\documentclass[aps,prb,twocolumn,superscriptaddress]{revtex4-2}

\usepackage{lineno}

\usepackage{amsmath, amssymb,  graphicx}

\usepackage[colorlinks=true ,urlcolor=blue,urlbordercolor={0 1 1}]{hyperref}

\usepackage{color}
\usepackage{xcolor}
\usepackage{braket}
\usepackage{booktabs}

\usepackage[utf8]{inputenc}
\usepackage{bm}
\usepackage{dsfont}
\usepackage{verbatim}
\usepackage{graphicx}
\usepackage{amsmath}
\usepackage{color}
\usepackage{xr}
\usepackage{hyperref}
\usepackage{float}
\usepackage{bbm}
\usepackage{amssymb}
\usepackage{slashed}
\usepackage[normalem]{ulem}
\usepackage{wasysym,bm,bbm,dsfont}

\newcommand{\bfk}{\mathbf{k}}

\newcommand{\bfK}{\mathbf{K}}
\newcommand{\bfG}{\mathbf{G}}
\newcommand{\bfr}{\mathbf{r}}
\newcommand{\bfR}{\mathbf{R}}

\begin{document}

\title{Ancilla theory of twisted bilayer graphene I: topological Mott localization and  pseudogap metal in twisted bilayer graphene}

\author{Jing-Yu Zhao}
\thanks{These two authors contributed equally}
\affiliation{Department of Physics and Astronomy, Johns Hopkins University, Baltimore, Maryland 21218, USA}

\author{Boran Zhou}
\thanks{These two authors contributed equally}
\affiliation{Department of Physics and Astronomy, Johns Hopkins University, Baltimore, Maryland 21218, USA}

\author{Ya-Hui Zhang}
\affiliation{Department of Physics and Astronomy, Johns Hopkins University, Baltimore, Maryland 21218, USA}

\date{\today}

\begin{abstract}
The recent experimental studies of twisted bilayer graphene (TBG) raise a fundamental question: how do we understand Mott localization in a topological band?   
In this work, we offer a new perspective of Mott physics, which can be generalized to TBG directly in momentum space. In our theory, the Mott gap is understood as from an exciton-like hybridization $\Phi(\mathbf k) c^\dagger(\mathbf k)\psi(\mathbf k)$ between the physical electron $c$ and an ancilla fermion $\psi$. In the conventional Mott insulator of trivial band, the hybridization is $s$-wave with $\Phi(\mathbf k)=\frac{U}{2}$, where $U$ is the on-site Hubbard interaction. On the other hand, the band topology in TBG enforces a topological Mott hybridization with $\Phi(\mathbf k)\sim k_x \pm i k_y$ in a small region around $\bfk=0$.   We dub this new Mott state as topological Mott localization because of the $p\pm ip$ order parameter analogous to the topological superconductor. At $\nu=0$, we find a topological Mott semimetal with a low energy effective theory resembling that of the untwisted bilayer graphene. For $\nu=\pm 1, \pm 2, \pm 3$, we show transitions from correlated insulators to Mott semimetals at smaller $U$.  In the most intriguing density region 
$\nu=-2-x$, we propose a symmetric pseudogap metal at small $x$, which hosts a small Fermi surface and violates the perturbative Luttinger theorem.   Interestingly, the quasiparticle is primarily formed by ancilla fermion, which we interpret as a composite fermion formed by a hole bound to a particle-hole pair.  Our theory offers a unified language to describe the Mott localization in both trivial and topological bands in momentum space, and we anticipate applications in other moir\'e systems with topological Wannier obstruction, such as the twisted transition-metal dichalcogenide (TMD) homobilayer.
\end{abstract}

\maketitle

\section{Introduction} \label{sec:Intro}

Discoveries of correlated insulators and superconductivity in the magic angle twisted bilayer graphene (TBG) \cite{cao2018unconventional,Cao2018mott,Yankowitz2019layer,Lu2019moreSC,Stepanov2020,Cao2021nematic,Liu2021Coulomb,arora2020superconductivity} and twisted multilayer graphene\cite{park2021tunable,hao2021electric,park2022robust,cao2021pauli}
have attracted a lot of attentions in recent years and initiated a whole new area of moir\'e materials\cite{andrei2020graphene,andrei2021marvels,nuckolls2024microscopic}. 
Although the phase diagram of TBG shares certain similarity with that of the high-temperature superconducting cuprates, theoretical studies indicate that the physics is different from a conventional Hubbard model due to the fragile band topology 
\cite{Po2018IVC,Tarnopolsky2019,Ahn2019fgtop,Ledwith2021Peda,Song2021symano}.
Previous theoretical studies have focused on quantum Hall physics and various symmetry breaking phases (such as inter-valley coherent orders)\cite{bultinck2020ground,kwan2021kekule,parker2021strain,wagner2022global}. 
Indeed experiments observed integer quantum anomalous Hall effects\cite{sharpe2019emergent,serlin2020intrinsic} when TBG is aligned with a hexagonal boron nitride (hBN)\cite{bultinck2020mechanism,zhang2019twisted} and even fractional Chern insulator (FCI) phases under a finite magnetic field\cite{xie2021fractional}. 
Hence one may conclude that the TBG physics is far away from high-Tc cuprate physics. However, there are also experimental evidences of local moments  from entropy measurements\cite{rozen2021entropic,saito2021isospin}, indicating Mott localization.

The main theoretical challenge lies in describing Mott localization in the topological bands. 
There have been attempts to construct single-particle basis with Wannier orbitals on the AA site.  The first approach is through the topological heavy fermion model (THFM)
\cite{Song2022THFM,Calugaru2023THFM,Yu2023THFM,Hu2023THFM,Vafek2024THFM,Zhou2024THFM,Hu2023THFM2,Chou2023THFM,Wang2024THFM,Lau2023THFM,Rai2024THFM,Youn2024DMFT}, which includes both the active bands and the remote bands.  
In the second approach, one can construct non-local Wannier orbitals only for the active bands\cite{Ledwith2024}, but the interaction may not be easily truncated to a simple on-site Hubbard $U$. 
In this work, we follow a different approach. Instead of searching for a good single-particle basis, we propose a \textit{new many-body theory} of Mott physics formulated directly in momentum space.
In our framework, the Mott gap is understood as from an exciton-like order parameter $\Phi(\bfk)\sim c^\dagger(\bfk)\psi(\bfk)$ between the physical electron $c$ and an ancilla fermion $\psi$. 
In the standard spin-1/2 Hubbard model\cite{Zhang2020,Zhou2024}, the hybridization is a uniform $s$-wave form with $\Phi(\bfk)=U/2$. 
While in TBG, the band topology enforces the hybridization to be  $p$-wave-like form $\Phi(\bfk)\sim k_x\pm ik_y$ around $\bfk=0$. We dub this new Mott state in topological band as topological Mott localization because its `order parameter' $\Phi(\bfk)$ is similar to $p\pm i p$ pairing of topological chiral superconductor. To our best knowledge, the topological structure of this new Mott state has not been revealed before.

\begin{figure}[ht]
    \centering
    \includegraphics[width=0.95\linewidth]{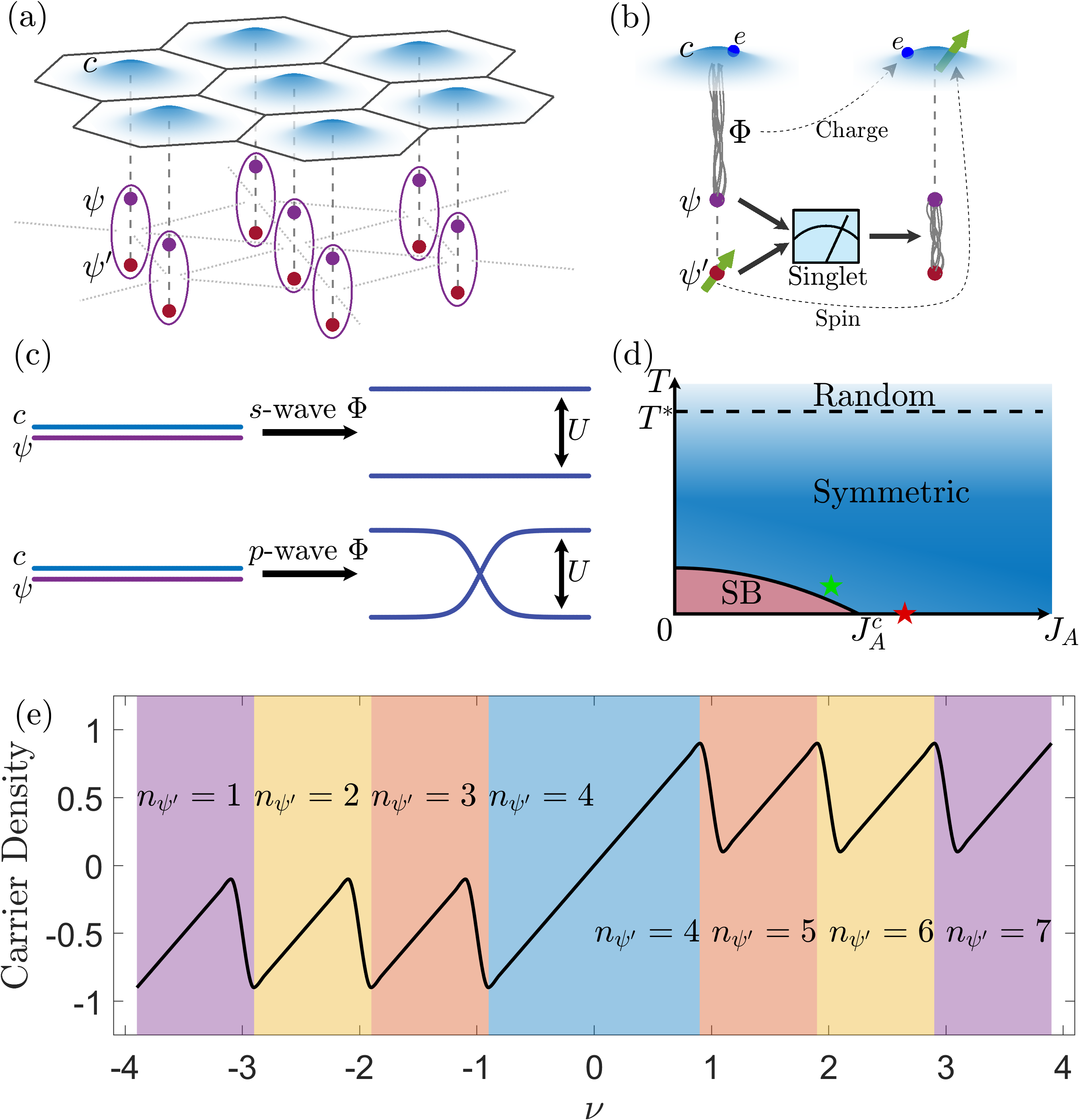}
    \caption{(a) Illustration of the ancilla wavefunction. The top layer is the physical TBG layer with the active band density centered at AA positions (blue wave packet). 
    The bottom two layers are the first ancilla $\psi$ (purple) and the second ancilla $\psi'$ (orange), respectively.
    (b) At each AA site, the wavefunction is projected back to the physical Hilbert space by enforcing the two ancilla fermions to form an SU(8) spin singlet. 
    (c) Different from a convention Hubbard model, which can be simulated with an $s$-wave exciton-like hybridization $\Phi(\bfk)=\frac{U}{2}$, here $c$ and $\psi$ are hybridized through a $p$-wave-like $\Phi(\bfk)\sim k_x\pm ik_y$ around $\Gamma_M$ point. 
    (d) An illustration of the phase diagram of the inter-valley interaction $J_A$ and temperature $T$. 
    Our symmetric ansatz applies to all of the blue regimes.
    (e) Illustration of the charge carrier density with physical density $\nu$ in our ansatz with $\Phi \neq 0$. We assume the second ancilla layer has density $n_{i;\psi'}=n$ with $n$ an integer. $n=4-\nu$ for integer $\nu$, and is pinned to the same integer when doping the Mott state.  The reset of the carrier density is assumed to be from a first-order transition with $n$ jumping by $1$ in our theory.}
    \label{fig:ancilla}
\end{figure}

In more detail, in the ancilla framework, we introduce two additional layers of ancilla fermions $\psi_{i;\alpha}$ and $\psi'_{i;\alpha}$, where $i$ labels a real space lattice site where the `local moment' lives. 
In TBG, we will put $\psi_i$ and $\psi'_i$ both on the AA site. 
$\alpha$ is the flavor index of a SU(N) fermion ($N=8$ in TBG). Then a physical state can be written as:
\begin{equation}\label{eqn:ancilla}
    |\Psi_c\rangle = P_S|\Psi_0[c,\psi,\psi']\rangle. 
\end{equation}
Here, $|\Psi_0[c,\psi,\psi']\rangle$ is a variational state in the enlarged Hilbert space  and $P_S$ is a projection operator that enforces the two ancilla fermions to form an SU(N) spin-singlet at each site $i$, as illustrated in Figs.~\ref{fig:ancilla}(a) and (b).
The physical layer $c$ and the first ancilla layer $\psi$ are coupled through a hybridization $\Phi(\mathbf k)c^\dagger_{\bfk;\alpha} \psi_{\bfk;\alpha}$ to open a charge gap, 
while the spin state is encoded in the second ancilla layer $\psi'$ which represents the local moments.  The validity of this seemingly exotic wavefunction for the Mott insulator of spin-1/2 Hubbard model at half-filling has been demonstrated both analytically and numerically \cite{Zhou2024}.  Especially, there is an interesting connection to quantum teleportation as the projection can be viewed as a Bell measurement with post-selection\cite{Zhou2024}.  Variational Monte Carlo studies\cite{shackleton2024emergent,muller2024polaronic} also demonstrated that the ancilla wavefunction can correctly capture the polaronic correlations in doped Mott insulators. At mean-field level the ancilla theory can also explain the `Fermi arcs' in underdoped cuprates\cite{Zhang2020}.

In the ancilla formalism,  we can describe the Mott physics in TBG  directly in the continuum model, though the same results can be obtained in the basis of THFM.  At $\nu=0$, we obtain a topological Mott semimetal with quadratic band touching at the $\Gamma_M$ point. This state has previously been discussed in Ref.~\cite{Ledwith2024} based on self-energy calculation, but now we can provide an effective theory to capture the band topology and also write down model wavefunction  at $T=0$. At $\nu=\pm 1, \pm 2, \pm 3$, we predict transitions from correlated insulators to  semimetals by reducing $U$ or increasing the twist angle. To the best of our knowledge, the Mott semimetal at non-zero integer filling is a new result. 

The most exciting discovery is  a symmetric pseudogap metal at filling $\nu=-2 -x$ with small $x$. We find a small Fermi surface state similar to  the pseudogap metal in the ancilla theory of the underdoped cuprate\cite{Zhang2020}. But  due to the band topology of the TBG, the quasiparticle of the small Fermi surface is dominated by the ancilla fermion $\psi$ at small $x$, and thus has vanishing spectral weight.  We interpret the ancilla fermion as a composite polaron or composite trion, formed by a spin moment on an AA site bound to a hole in the nearest-neighbor AA site.  We suggest that the normal state of the superconductor is this pseudogap metal formed by the composite polarons. This implies that a successful theory of superconductivity  must be formulated in terms of these composite fermions instead of single electrons. When there is a sizable anti-Hund's coupling $J_A$ mediated by optical phonons, the symmetric pseudogap metal can be a ground state.  In this case we can write down a model wavefunction, which, to the best of our knowledge, is impossible in any other theoretical framework.

\section{General formalism} 
The ancilla wavefunction in Eq.~\eqref{eqn:ancilla} can be directly generalized to the TBG system.  There are 8 active bands labeled by the valley $\tau = K,K'$, the spin $s=\uparrow,\downarrow$, and the orbital $a = +,-$.  
Then we introduce  SU(8) ancilla  fermions $\psi_{i;a\tau s}$ and $\psi'_{i;a\tau s}$  living on the AA site labeled by $i$. The projection operator $P_S$ enforces $\psi$ and $\psi'$ to form an SU(8) singlet at each site $i$. 
There are three constraints: (I) We fix $n_{i;\psi'}=n$ with $n$ being an integer at each site $i$. So the second ancilla $\psi'$ just represents a fully localized spin state in the $n_T=n$ representation of the SU(8) spin. 
(II) We fix $n_{i;\psi}=8-n$ at each site $i$. (III) We fix $S^{\alpha\beta}_{i;\psi}+S^{\alpha\beta}_{i;\psi'}=0$ at each site $i$. Here, $S^{\alpha\beta}$ with $\alpha,\beta=1,2,...,8$ labels each spin operator of SU(8). These three constraints give a $U(1)\times U(1) \times SU(8)$ gauge symmetry. 
But we will only consider the ansatz where all of the gauge fields are higgsed\footnote{Strictly speaking, the $U(1)$ gauge field corresponding to $n_{i;\psi'}$ may not be higgsed. But this is just the usual Gutzwiller projection to represent a spin state.}, so we can ignore the gauge symmetry in this work.  At integer fillings $\nu$ with total density $n_c=4+\nu$, we set $n_{i;\psi'}=4+\nu$ and $n_{i;\psi}=4-\nu$ to capture a Mott state. Away from integer fillings, we expect $n_{i;\psi'}$ to stay in the integer associated with the parent Mott state. In this picture the reset of the carrier density close to an integer $\nu$ is associated with a jump of $n_{i;\psi'}$ by $1$ (see Fig.~\ref{fig:ancilla}(e)).

In the ancilla description of a Mott state, we always have the charge sector formed by $c,\psi$ and the neutral spin sector from $\psi'$. 
In this work, we assume the spin state of $\psi'$ is symmetric.  
In reality there should be an effective ferromagnetic spin-spin coupling between the local moments $J<0$, which favors isospin polarization.  Our following discussion is  justified for two regimes (see Fig.~\ref{fig:ancilla}(b)): 
(I) We consider finite temperature with $|J|\ll T\ll U$, where the local moments in the $\psi'$ layer are just thermally fluctuating, but  the charge sector of $c,\psi$ still behaves like a low-temperature state. 
(II) We can add an additional spin-spin coupling term, for example, an anti-Hund's coupling $J_A \sum_{\mathbf q} \vec S_{\mathbf K}(\mathbf q) \cdot \vec S_{\mathbf K'}(-\mathbf q)$ \cite{zhang2020spin,wang2024molecular}, which enforces inter-valley spin-singlet  when $n_{\psi'}$ is even. 
Such a $J_A$ coupling can be mediated by optical phonons at momentum $K$\cite{chen2024strong}. In this case, we can have a symmetric spin state as a ground state and then  we can provide variational wavefunctions for symmetric Mott semimetal/insulator at $\nu=-2$ and a symmetric pseudogap metal at $\nu=-2-x$.  In the following we will mainly focus on the charge sector and provide mean-field theory of $c,\psi$. Our formulation has the advantage of being basis independent and we can obtain equivalent results in the continuum model and in the THFM.

\subsection{Continuum model}  
We can construct the Wannier orbital of $\psi_{i;a \tau s}$ as a $\delta$-function located at AA positions $i$ of TBG for each valley $\tau$, spin $s$, and orbital $a$. 
The mean-field Hamiltonian takes the form:
\begin{equation}\label{eqn:BM_psi}
    {H}_{\mathrm{MF}}^{(c\psi)} = {H}_{\mathrm{BM}} -\mu_c N_c - \mu_\psi N_\psi
    +{H}_{\mathrm{hyb}}^{(c\psi)}~.
\end{equation}
Here the physical electron is described by the continuum Bistritzer-MacDonald (BM) model\cite{Bistritzer2011}, denoted as ${H}_{\mathrm{BM}}$, 
which is briefly reviewed in Appendix~\ref{app:BM}. 
The ancilla $\psi$ is assumed to be completely flat with only a chemical potential term. 
The hybridization term is given by:
\begin{equation}
    {H}_{\mathrm{hyb}}^{(c\psi)} = \Phi_0\sum_{l,\bfk\in\mathrm{MBZ},\bfG_M} e^{i\frac{\pi}{4}l a}c^\dagger_{0,\bfk + \bfG_M;la\tau s} \psi^{}_{\bfk;a\tau s} +\mathrm{h.c.}~,
\end{equation}
where $\bfG_M$ is the moir\'{e} reciprocal vector, and $\Phi_0$ is the strength of the hybridization, which should scale with $U$.  
Note that $l$ is the layer index and $c_{0, \bfk; l a\tau s}$ labels the single-electron state in the original graphene lattice scale, which will be folded into the mini Brillouin zone (MBZ) by $H_{\mathrm{BM}}$.  We also need to emphasize that $\psi_{i;a=\pm}$ has angular momentum $L=\pm 1$ under $C_{3z}$ rotation around the AA site.

\subsection{Topological heavy fermion model} 
In the THFM, for each flavor $\alpha=a\tau s$, there is a flat and localized $f$ band and  itinerant $c_1,c_2$ bands. $c_1$ and $c_2$ form an effective model similar to the K valley of the AB-stacked bilayer graphene.  There is a large hybridization $\gamma c^\dagger_{1,\bfk} f_\bfk$, which opens the remote gap. In the end, the active bands are mainly from the localized $f$ states except around the $\Gamma_M$ point where they are dominated by $c_2(\bf k)$  with angular momentum $L=0$. 
The $f$ state has angular momentum $L=\pm 1$ around the AA site. Assuming the Mott localization happens due to Hubbard $U$ on the $f$ orbital, we introduce $\psi_i$ and $\psi'_i$ with the same Wannier orbital as $f$. The mean-field theory of the charge sector is:
\begin{equation}
    \begin{split}
    H_{\mathrm{MF}}^{(cf\psi)}=&H_\mathrm{THFM}+\Phi\sum_{i,a,\tau,s}\left(f^\dagger_{i;a,\tau,s}\psi_{i;a,\tau,s}+\mathrm{H.c.}\right)\\
    &-\mu (N_{c_1}+N_{c_2}+N_f)-\mu_\psi N_\psi,
    \end{split}
    \label{eq:THFM_mean_field}
\end{equation}
where $N_{c_1}, N_{c_2},N_f$ and $N_\psi$ represent the total density of $c_1,c_2,f$ and $\psi$ electrons, respectively. $\mu_{\psi}$ is introduced to fix the gauge constraint $\langle n_{i;\psi} \rangle=n$.

\begin{figure}[tbp]
    \includegraphics[width=0.95\linewidth]{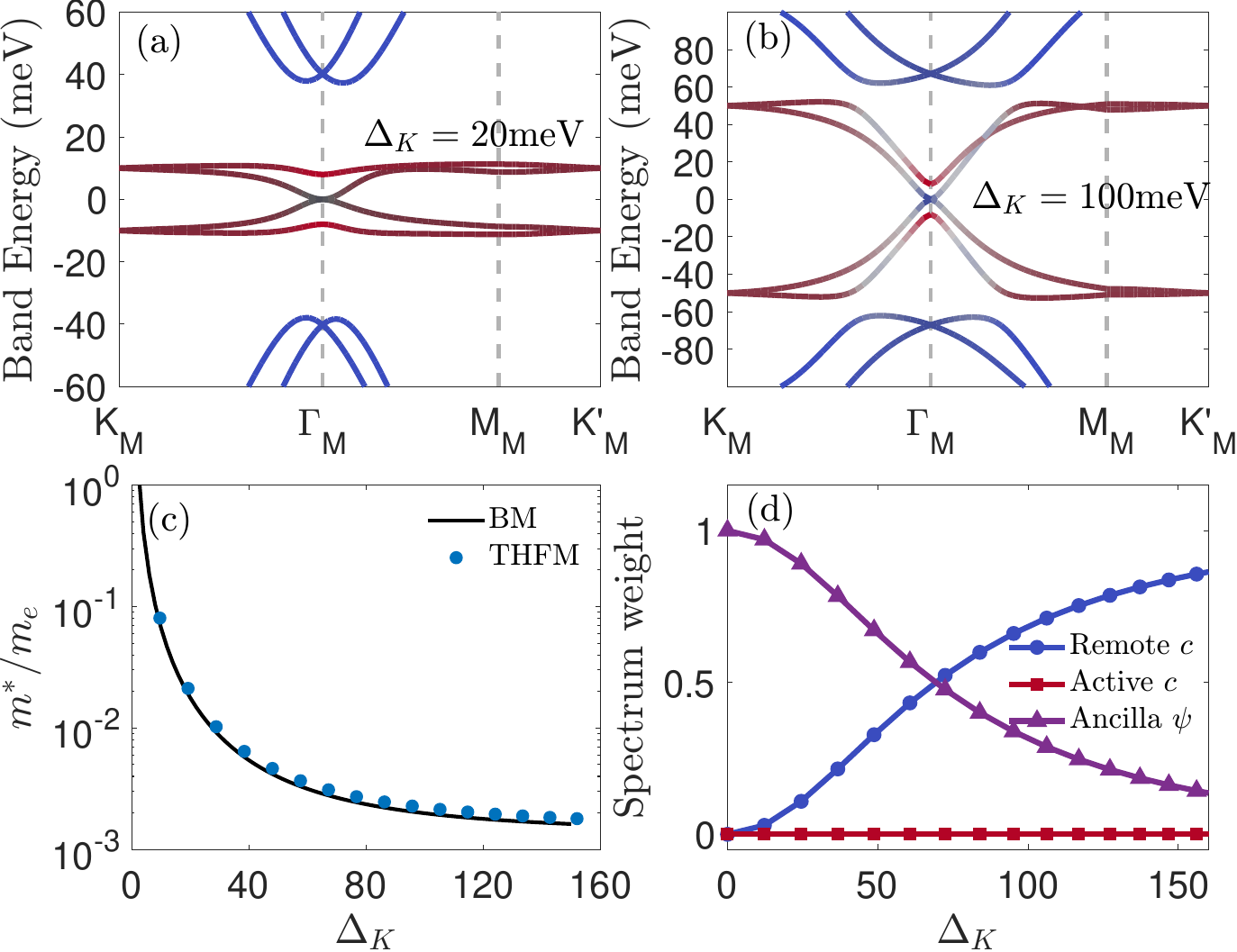}
    \caption{
        Mott bands calculated by the BM model at $\nu = 0, \theta=1.14^\circ,w_0/w_1=0.8$, for different hybridization strengths at (a) $\Delta_K = 20$ meV and (b) $\Delta_K=100$ meV, 
        where $\Delta_K$ is the $K_M$ point Mott gap. 
        The red, blue, and dark colors of the dispersion lines represent the weight of the active band, the remote bands and the ancilla bands, respectively. 
        In (c), we show the effective mass $m^*/m_e$ of the gapless $\Gamma_M$ point as a function of the $K_M$ point gap $\Delta_K$. We also show the results from the THFM.
        (d) The $\Gamma_M$ point spectral weights coming from the active band, remote bands, and ancilla band as a function of $\Delta_K$.
    }
    \label{fig:filling=0}
\end{figure}

\begin{figure}[ht]
    \centering
    \includegraphics[width=0.9\linewidth]{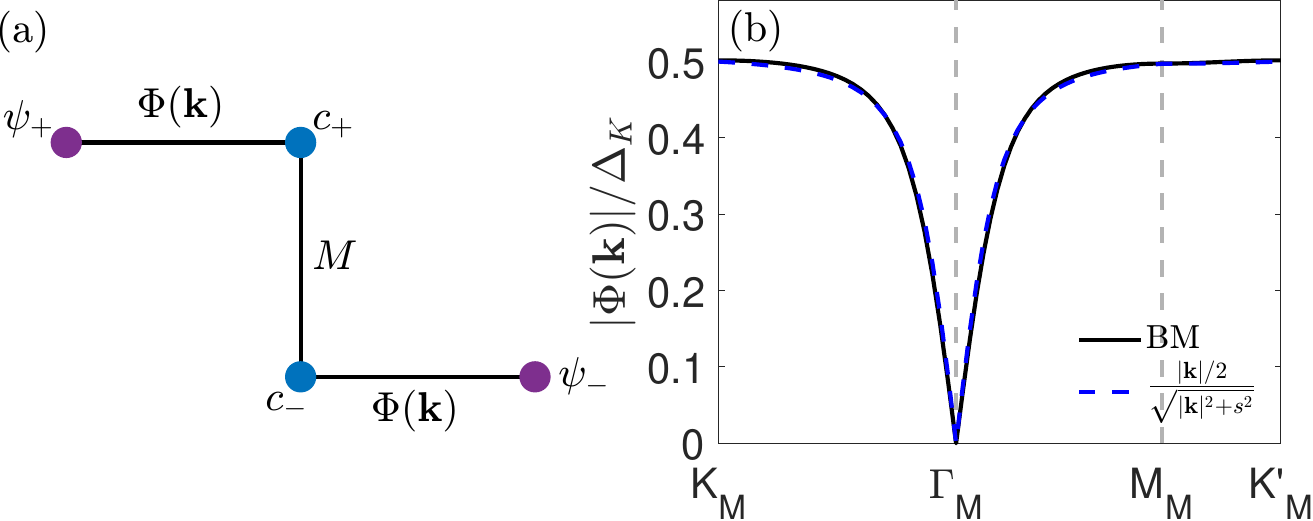}
    \caption{ (a)  Illustration of the effective model Eq.~\eqref{eq:effective_theory} near the $\Gamma_M$ point, which resembles the K corner of AB-stacked bilayer graphene.
    (b) Plot of $|\Phi(\bfk)|/\Delta_K$ along a high-symmetry path, computed at $\theta = 1.14^\circ$, $w_0/w_1 = 0.8$, and $\Delta_K = 30$ meV.
    The full $\Phi(\bfk)$ can be fitted by the function $|\Phi(\bfk)|/\Delta_K \sim |\bfk| / (2\sqrt{|\bfk|^2 + s^2})$, with the fitting parameter $s = 0.19 |\bfK_M|$.
    }
    \label{fig:ABG_illu_main}
\end{figure}

\section{Topological Mott semimetal at $\nu=0$} 
At $\nu=0$, the mean-field spectrum is shown in Figs.~\ref{fig:filling=0} (a)-(b) for different hybridization strengths, corresponding to different values of $U$.  Here we use the gap at the $K_M$ point, $\Delta_K$ to characterize the hybridization. $\Delta_K$ is just a rescaling of $\Phi_0$ and is the only variational parameter to be optimized for each $U$. In the large $U$ limit we expect $\Delta_K=U$, but $\Delta_K$ can be smaller than $U$ close to the Mott transition.
All the results are calculated at a twist angle $\theta = 1.14^\circ$ and $w_0/w_1 = 0.8$. 
Away from the $\Gamma_M$ point, we find upper and lower Hubbard bands just as in a conventional Hubbard model.  In the large $U$ limit, we find that the four bands around $\Gamma_M$ are basically the same as the $c_1$ and $c_2$ bands in the THFM (see the supplementary). Thus our ancilla theory can recover the result in the $\gamma \ll U$ limit of the THFM directly in the continuum model. When decreasing  $U$ (or equivalently $\Delta_K$), the quadratic touching crossovers from the $c_1$ band in THFM to the ancilla $\psi$. One can see that the effective mass $m^*$ increases by two orders of magnitude (see Fig.~\ref{fig:filling=0}(c)) and the physical spectral weight $Z$ vanishes (see Fig.~\ref{fig:filling=0}(d)) when we reach the limit $U \ll \gamma$.    The calculations in the continuum model and in the THFM basis give the same results (see the supplementary).

\subsection{Effective theory around $\bfk=0$} 

When $U \ll \gamma$,  we can perform the calculation projected into the active band. The active bands can be decomposed into two Chern bands $C = \pm 1$ labeled by $c_{\bfk;\pm}$\cite{Ledwith2021Peda}.
The hybridization between $c_{\bfk}$ and $\psi_{\bfk}$ is then constrained by symmetry to have a node $\Phi(\mathbf k)\sim k_x\pm i k_y$ around the $\Gamma_M$ point, while $\Phi(\bfk)=\frac{U}{2}$ away from $\Gamma_M$.  
As shown in Fig.~\ref{fig:ABG_illu_main}(b), the full momentum dependence of $\Phi(\bfk)$ is well captured by the fit $|\Phi(\bfk)|/\Delta_K \sim |\bfk| / (2\sqrt{|\bfk|^2 + s^2})$.
Within each spin-valley flavor, we can then easily write down an effective four-band model close to $\bfk=0$ in terms of $c=(c_+,c_-)$ and $\psi=(\psi_+,\psi_-)$, as illustrated in Fig.~\ref{fig:ABG_illu_main} (a):
\begin{equation}
\begin{aligned}
    {H}^{(c\psi)}_{\Gamma_M} = &\sum_\bfk c^\dagger_\bfk M\sigma_xc_\bfk+
    \sum_\bfk c^\dagger_\bfk \Phi_{\Gamma} (k_x\sigma_0-ik_y\sigma_z)\psi_\bfk+\mathrm{h.c.}\\
\end{aligned}
\label{eq:effective_theory}
\end{equation}
where $\sigma_a$ labels the Pauli matrix in orbital space, and $W=2M$ gives the bandwidth of the active bands. $\Phi_\Gamma$ is proportional to $\Delta_K$. The effective model is equivalent to the $K$ corner of the AB-stacked bilayer graphene if we treat $a=+,-$ as `layers'. When $M$ is finite, there is a quadratic band crossing mainly from $\psi$ with dispersion $\epsilon_\pm(\bfk)\approx \frac{\Phi_\Gamma^2}{M}k^2$.    Although $\psi$ has vanishing spectral weight $Z$, it responds to electromagnetic field just like an electron because we are in a Higgs phase\cite{Zhang2020}. So we expect the Landau level degeneracy sequence of $\pm 4, \pm 8, \pm 12, ...$ at energy scales below $M$. In the chiral limit with $M=0$, we have two decoupled Dirac cones in each spin-valley.  We note a Mott semimetal was also reported in a Monte Carlo study of a spin-valley polarized model of TBG\cite{huang2024QMC}.

\begin{figure}[tbp]
    \includegraphics[width=0.95\linewidth]{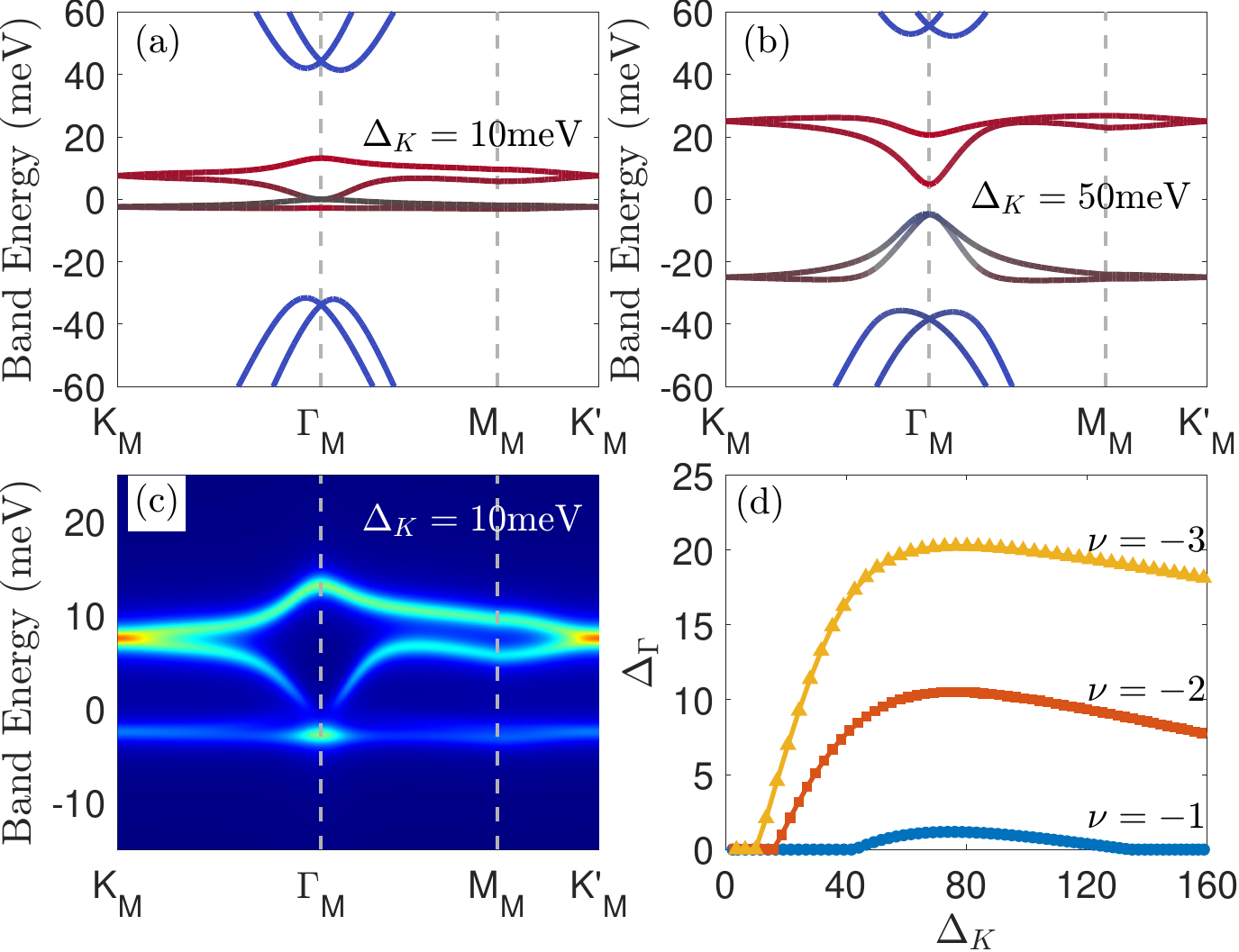}
    \caption{
        Mott bands calculated by the BM model at $\nu = -2, \theta = 1.14^\circ,w_0/w_1 = 0.8$ for different hybridization strengths at (a) $\Delta_K = 10$ meV, and (b) $50$ meV, respectively,
        where $\Delta_K$ is the $K_M$ point band gap. 
        The colors have the same meaning as in Figs.~\ref{fig:filling=0} (a) and (b). 
        (c) shows the corresponding physical spectrum function $A_c(\bfk,\omega)$ at $\Delta_K=10$ meV. 
        A broadening of $\eta = 1$ meV is used in getting the result. 
        And (d) shows the $\Gamma_M$ point gap $\Delta_\Gamma$ as a function of the $K_M$ point gap $\Delta_K$ and filling $\nu$, 
        which characterizes the semimetal to insulator transition. 
    }
    \label{fig:filling=-2}
\end{figure}

\section{Correlated insulators and semimetals at non-zero integer fillings} 

For other integer fillings, we obtain a gapped insulator for intermediate $U$. Then when decreasing $U$, the correlated insulator transits to a semimetal at a critical  $U_c$ (or equivalently $\Delta^c_K$). We show the Hubbard bands for $\nu=-2$ in  Fig.~\ref{fig:filling=-2} (a)-(b). For $\Delta_K<\Delta_K^c \approx 16$ meV, there is a quadratic band crossing at $\bfk=0$, though it is invisible in single-electron spectral function $A_c(\omega,\mathbf k)$, as shown in Fig.~\ref{fig:filling=-2}(c).    We note that spectral function $A_c(\omega,\mathbf k)$ may be measured by the twisting microscopy technique\cite{inbar2023quantum} in TBG, but we emphasize that single-electron spectroscopy may not be able to fully resolve all of the states. The insulator to semimetal transition also happens for $\nu=\pm 1, \pm 3$, as we show in  Fig.~\ref{fig:filling=-2}(d).  The gap of the correlated insulator arises from the  energy shift of $\psi$  relative to the physical band, as we need to adjust $\mu_\psi$ to fix $n_{\psi}=4-\nu$.  There is a gap-closing transition in both the small $\Delta_K$ and large $\Delta_K$ limits (see the supplementary).


\begin{figure}[ht]
    \includegraphics[width=0.95\linewidth]{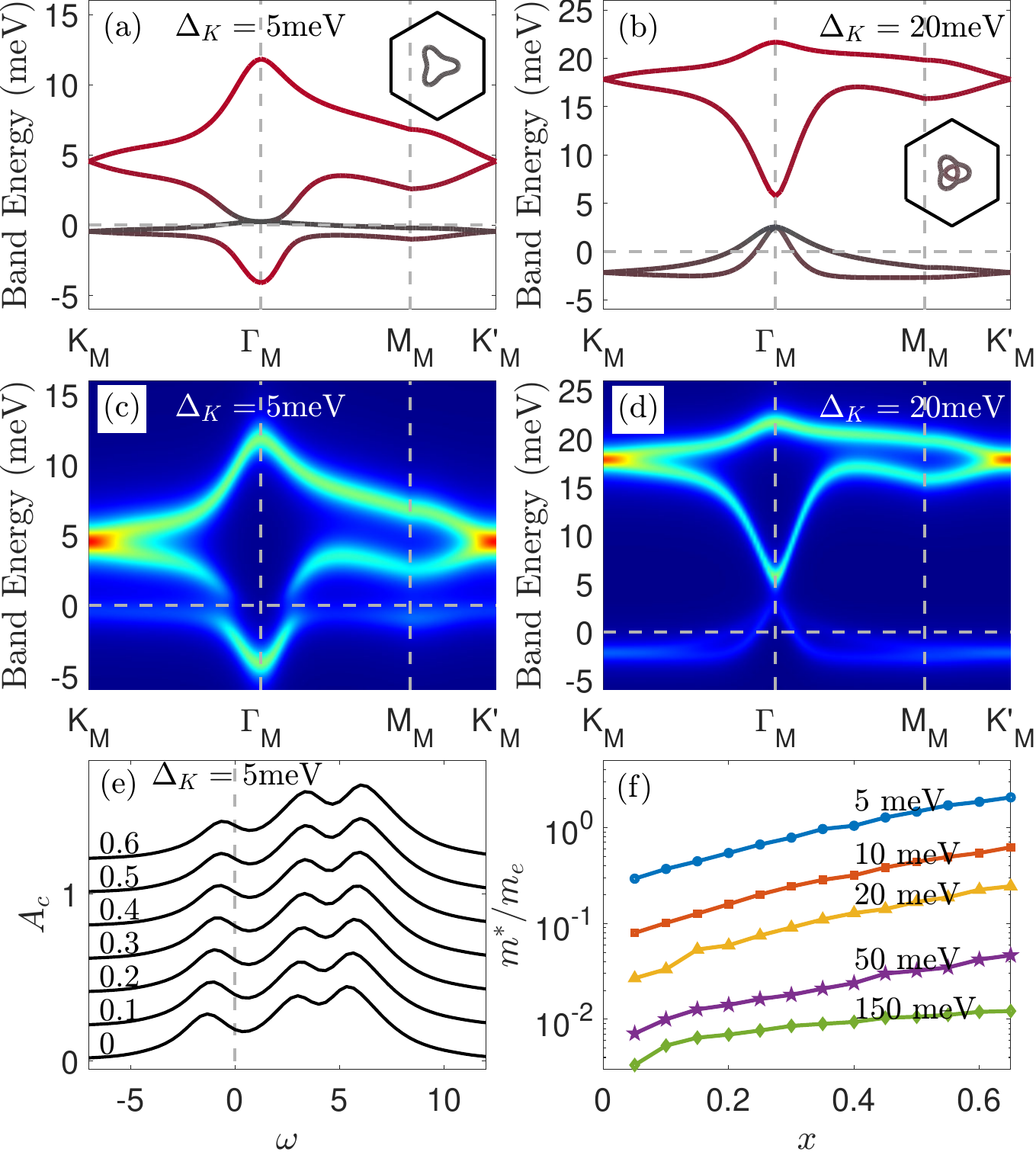}
    \caption{
        Mott bands calculated by the BM model with doping $\nu = -2-0.4$ and $\theta=1.14^\circ,w_0/w_1=0.8$, for different hybridization strengths at (a) $\Delta_K = 5$ meV and (b) $\Delta_K=20$ meV, where $\Delta_K$ is defined as the $K_M$ point gap. 
        The corresponding Fermi surfaces are also shown in the insets. 
        (c) and (d) show the physical electron spectrum function $A_c(\bfk,\omega)$ with the same parameters as in (a) and (b). 
        In (e), we show the STM spectrum $A_c(\omega)=\sum_\bfk A_c(\bfk,\omega)$ at $\Delta_K=5$ meV for different dopings $\nu=-2-x$, $x = 0-0.6$, with a slight vertical shift for each $x$. 
        A broadening $\eta = 1$ meV is used to obtain the spectrum functions $A_c(\bfk,\omega)$ and $A_c(\omega)$.
        (f) shows the Fermi surface effective mass $m^*/m_e$ as a function of doping $\nu=-2-0.4$ for different hybridization strengths $\Delta_K$. Here the effective mass $m^*$ is estimated by the Fermi surface density of states. 
        }
    \label{fig:pseudogap}
\end{figure}

\textbf{Symmetric pseudogap metal at $\nu=-2-x$} Now we try to dope  the $\nu=-2$ insulator or semimetal.  In the ancilla theory, we still keep $n_{i;\psi}=6$ and $n_{i;\psi'}=2$ at small $x$ . In this case, we find a small Fermi surface from holes doped into the lower Hubbard band at $\nu=-2$, which coexists with the local moments in the $\psi'$ layer. The Fermi surface volume per spin-valley flavor is $\frac{x}{4}$, and the state violates the perturbative Luttinger theorem.  We show the plots of the band and physical spectral function $A_c(\omega,\mathbf k)$ for two values of $\Delta_K$ in Fig.~\ref{fig:pseudogap}(a-d). When doping the correlated insulator, we get two Fermi surfaces within each spin-valley flavor. On the other hand, there is only one Fermi surface when slightly doping the semimetal.  We note that $\Delta_K$ likely decreases with the doping $x$, so we should use a smaller $\Delta_K$ even if $\Delta_K$ is large at $\nu=-2$.  For $\Delta_K=5$ meV, we provide the momentum-integrated spectral function $A_c(\omega)$ for a range of $x$ in Fig.~\ref{fig:pseudogap}(e). We can see that the spectral weight on the $\omega<0$ side is smaller than the $\omega>0$ side, which actually has also been observed in the scanning tunneling microscope (STM) experiment in the pseudogap region\cite{oh2021evidence}.  For the same value of $\Delta_K$, the effective mass $m^*$ reaches the order of electron mass $m_e$ (see Fig.~\ref{fig:pseudogap}(f)), which also agrees with the experimental estimation from quantum oscillations\cite{cao2018unconventional}.

Our current theory predicts a four-fold Fermi surface degeneracy, which may be further reduced to two-fold degeneracy by a small symmetry breaking order at low temperature. Alternatively, when there is a large anti-Hund's coupling $J_A$, the symmetric pseudogap metal can be a ground state and corresponds to an intrinsically strongly interacting Fermi liquid fixed point allowed by the non-perturbative Oshikawa-Luttinger theorem for the symmetry $U(1) \times U(1) \times SU(2)$\cite{zhang2020spin,yang2024strong}.

\section{Physical meaning of the ancilla fermion} 

The quasiparticle at the Fermi surface is dominated by the ancilla fermion $\psi$ at small $x$.  Actually the spectral function $A_c(\omega,\bfk)$ almost looks like there is a gap (see Fig.~\ref{fig:pseudogap}(c)).   In the supplementary, we show that $\psi$ is real and represents a many-body excitation. At $\bfk=0$, at leading order it corresponds to a trion:

\begin{align}
    O_{\bfk;\alpha}&\sim\frac{1}{N_\mathrm{site}} \sum_{\bfk',\mathbf q}Z(\mathbf k',\mathbf q)\mu_{f;\beta}(\bfk+\mathbf q)\mu_{f;\alpha}(\bfk')\mu_{f;\beta}^*(\mathbf k'+\mathbf q) \notag \\ 
    & c_{\beta;\bfk+\mathbf q}c^\dagger_{\beta;\bfk'+\mathbf q}c_{\bfk';\alpha}
\end{align}
where $\mu_{f;\beta}(\mathbf k)$ is the Bloch wavefunction of the $f$ fermion in the THFM and we expect $\mu_{f;\beta}(\mathbf k)\sim k_x \pm i k_y$ around the $\Gamma_M$ point. 
As a result, the trion is formed by electron and holes in the active band away from the Gamma point. $Z(\mathbf k',\mathbf q)\sim \sum_{\delta \mathbf r_1, \delta \mathbf r_2} e^{-i \mathbf k'(\delta \mathbf r_1+\delta \mathbf r_2)}e^{i \mathbf q \cdot (\delta \mathbf r_1-\delta \mathbf r_2)}Z(\delta \mathbf r_1, \delta \mathbf r_2)$ is the Fourier transformation of the wavefunction $Z(\delta \mathbf r_1, \delta \mathbf r_2)$ of the trion. Symmetry constrains $Z(\delta \mathbf r_1, \delta \mathbf r_2)$ to be s-wave under the $C_3$ rotation around the AA site. Here $\delta \mathbf r_1, \delta \mathbf r_2$ are  the displacement vectors of the first two hole/electron operators relative to the center of mass coordinate $\mathbf R_i$ of the AA site $i$, while the third hole is displaced by $-\delta \mathbf r_1-\delta \mathbf r_2$.  Such a composite fermion is orthogonal to the single-particle state at $\bfk=0$ due to different angular momenta under $C_{3z}$ rotation. Ref.~\onlinecite{ledwith2025exotic} shows that $Z(\mathbf k',\mathbf q)$ should be a constant deep inside the Mott regime with flat band and very concentrated Berry curvature, so there is a mapping to a trivial band Hubbard model with $t/U=0$.  This corresponds to a $Z(\delta \mathbf r_1, \delta \mathbf r_2)$ concentrating on $\delta \mathbf r_1=\delta \mathbf r_2=0$. However, with finite bandwidth, the trion wavefunction may be more extensive in the weak Mott regime. Also, upon hole doping, the extent of the trion can deform and may not be rigid. A fat trion can reduce repulsion between them. For the purpose of understanding superconductivity, it may be important to develop a method to estimate the trion wavefunction at finite hole doping.

There are a few remarks about the composite fermion here.  (I) We emphasize that the composite fermion emerges once the Mott localization (a non-zero $\Phi$) is developed below $T^*$ and does not depend on any specific spin state. Therefore it is very different from previous discussions\cite{khalaf2022baby,davydova2023itinerant,tao2024observation} of spin polarons on top of specific magnetic orders.  Hence we should dub it as Mott trion. Actually in our formalism the Mott trion always come together with the local moment represented by $\psi'$. Once $\Phi \neq 0$, both the Mott trion ($\psi$) and local moments emerge. In the Fermi liquid with $\Phi=0$, they confine each other and disappear together. Therefore it is clear that the Mott trion is associated with the local moment formation, but does not depend on specific spin state. (II) The existence of the Mott trion does not rely on the band topology. It also exists in the trivial band Hubbard model. Just there in the strong Mott insulator $\psi$ is always mixed with the electron.  But quasi-particle dominated by the composite fermion $\psi$ is still possible when the energy difference $\delta \epsilon(\mathbf k)$ between $c(\mathbf k)$ and $\psi(\mathbf k)$ is large relative to $\Phi(\mathbf k)$. This can happen in the weak Mott regime at integer filling or in the underdoped regime. Actually, in the ancilla theory of the pseudogap metal of underdoped cuprate\cite{Zhang2020}, the composite fermion $\psi$ dominates in the backside of a hole pocket and explains the `Fermi arcs' in the experiment.  What makes the topological band in TBG special is that now  $\Phi(\mathbf k)$ has a node at $\bf k=0$, so the band of the composite fermion $\psi$ is split from the electron $c(\mathbf k)$ around $\Gamma$ point even in the large U regime at integer filling. Also at integer filling $\nu \neq 0$, there is a large contribution to $\delta \epsilon(\mathbf k)$ by the relative chemical potential shift of the ancilla fermion to fix its density.  At $\nu=-2-x$, in the symmetric pseudogap metal, the entire hole pocket is dominated by the composite fermion at small $x$.  This is different from ancilla theory of the underdoped cuprate where the composite fermion and single electron form different sides of the pocket\cite{Zhang2020}. However, the existence of this composite quasi-particle  is non-topological and universal in Mott state or pseudogap metal state with local moment formation. Our ancilla theory provides a unified framework to capture it in both the trivial and topological band. (III) It is worthwhile to point out that $b\sim \psi^\dagger_{\alpha} \psi'_{\alpha}$, the bound state of the Mott trion and the spinon $\psi'$, represents a spinless charged boson when $\Phi \neq 0$,  which is equivalent to the famous slave boson in the conventional treatment of the trivial band Hubbard model or t-J model\cite{lee2006doping}. Slave boson theory has played important roles in studying metallic and superconducting states in doped Mott insulator\cite{lee2006doping}. Our framework thus provides a way to perform similar descriptions also for topological band.


\section{Conclusion} 

In summary, we propose a new theoretical framework to describe the Mott physics of the twisted bilayer graphene directly in momentum space.  With the help of ancilla fermions $\psi$ and $\psi'$, we can capture correlated insulators and Mott semimetals at integer $\nu$. More importantly, we propose a symmetric pseudogap metal at $\nu=-2-x$. This exotic metal has a small Fermi surface formed by composite fermions with vanishing spectral weight at small $x$.  Our symmetric states may be unstable to symmetry breaking at lower temperatures, which we plan to include in a future study. On the other hand, we propose to add an anti-Hund's coupling $J_A$ in theoretical and numerical studies
\cite{hofmann2022fermionic,pan2022dynamical,zhang2021QMC,huang2024QMC,huang2024angle,soejima2020efficient}, 
which may suppress the various generalized ferromagnetic orders at $\nu=0,-2,-2-x$  and reveal the true essential physics.  Our theory of the pseudogap metal shares the same spirit as the fractional Fermi liquid (FL*) candidate for underdoped cuprates\cite{Zhang2020}. Therefore, the ancilla theory may provide a unified language to bridge the TBG physics and the high-Tc cuprate physics. We also anticipate applications of this framework to other moir\'e systems, such as the twisted WSe$_2$ system\cite{xia2025superconductivity,guo2025superconductivity}.

\textit{Note added:} Upon finishing the manuscript, we became aware of another preprint\cite{Hu2025THFM} which calculated self energy and spectral function $A_c(\omega,\bfk)$ for integer fillings of TBG, mainly in the flat band limit.

\textbf{Acknowledge} YHZ thanks Patrick Ledwith for useful discussions. This work was supported by the National Science Foundation under Grant No. DMR-2237031.

\bibliography{refs.bib}

\appendix

\section{Review of the Ancilla wavefunction method}\label{app:ancilla}

In this section, we provide more details about the ancilla wavefunction in Eq.~\eqref{eqn:ancilla} for the SU(N) Hubbard model, with $N=8$ relevant for TBG. 
Specifically, we give a quantum teleportation interpretation of the ancilla wavefunction in Eq.~\eqref{eqn:ancilla} and demonstrate its validation by comparing it with the perturbative ground state of a Hubbard model. 
Consider an SU(N) Hubbard model: 
 \begin{equation}\label{eqn:hubbardSUn} 
    H = -\sum_{i,j,\alpha}t_{ij}c_{i;\alpha}^\dagger c_{j;\alpha}+U\sum_{i,\alpha\beta}n_{i;\alpha}n_{i;\beta}-\mu \sum_{i,\alpha} n_{i;\alpha}~,
\end{equation}
where $\alpha = 1,\cdots, N$ is the flavor index. 
Here, we focus on the insulator integer fillings with $\nu\equiv n_T-N/2=-N/2+1,\cdots,N/2-1$, where $n_T\in \mathbb{Z}$ is the average particle number per site.

As already discussed in the main text, we can introduce two ancilla layers $\psi_{i;\alpha}$ and $\psi'_{i;\alpha}$ and consider the following wavefunction ansatz:
\begin{equation}
    \lvert \Psi_c\rangle =
    P_S|\Psi_0[c,\psi,\psi']\rangle =
    P_S \left( \lvert\mathrm{Slater}[c,\psi]\rangle\otimes \lvert\Psi_{\psi^\prime}\rangle\right),
\end{equation}
where $P_S$ is a projection operator enforcing (I) $n_{i;\psi}=N/2-\nu$; (II) $n_{i;\psi'}=N/2+\nu$; and (III) the two local ancilla qubits $\psi_i,\psi_i^\prime$ form an SU(N) singlet on each site $i$. 

Here, we attribute the charge degree of freedom to the physical layer $c$ and the first ancilla layer $\psi$, 
which is taken as a Slater determinant $\lvert\mathrm{Slater}[c,\psi]\rangle $ as a ground state of a parent mean-field Hamiltonian:
\begin{equation}\label{eqn:HMecf}
\begin{split}
H_M^{(c\psi)}=&-\sum_{i,j,\alpha}t_{ij}c^\dagger_{i;\alpha}c^{}_{j;\alpha}+\Phi\sum_{i,\alpha}\left(c^\dagger_{i;\alpha}\psi_{i;\alpha}+\mathrm{H.c.}\right)\\
&+\frac{\Delta_\psi}{2}\sum_i (n_{i;\psi}-n_{i;c}) - \mu \sum_i(n_{i;\psi}+n_{i;c}),
\end{split}
\end{equation}
where $\Delta_\psi$ and $\mu$ are chosen to make $\langle n_{i;c}\rangle=N/2+\nu$ and $\langle n_{i;\psi}\rangle=N/2-\nu$. 

On the other hand, the spin degree of freedom is encoded in the second ancilla $\psi'$, which forms a local spin moment in the representation of $N/2+\nu$ of SU(N). 
Its generic wavefunction takes the form:
\begin{equation}\label{eqn:psiprime}
    \lvert\Psi_{\psi^\prime}\rangle=\sum_{\{\bm{\alpha}_i\}}w(\{\bm{\alpha}_i\})\prod_{i=1}^{N_\mathrm{site}}\prod_{\alpha_i} {\psi^\prime}^\dagger_{i;\alpha_{i}}\lvert 0\rangle,
\end{equation}
where $\bm{\alpha}_i=(\alpha_{i;1},\alpha_{i;2},...,\alpha_{i;N/2+\nu})^T$ 
labels the occupied flavors on site $i$. 
The precise form of the spin state depends on the induced Heisenberg spin interactions and lattice geometry, whose details are not important for our current discussion.

\subsection{Infinite $U$ limit}

First consider the infinite $U$ limit, where we can take $\Phi\rightarrow \infty $ and $t_{ij}=0$ in Eq.~\eqref{eqn:HMecf}. 
The physical layer $c$ then strongly hybridizes with the first ancilla $\psi$ and forms a spin-singlet EPR pair on each site $i$: 
\begin{equation}\label{eqn:slatercpsit=0}
   \lvert\mathrm{Slater}[c,\psi]\rangle_0= \prod_{i=1}^{N_\mathrm{site}} \prod_{\alpha=1}^{N} 
   \left(\cos\frac{\phi}{2}c^\dagger_{i;\alpha} +\sin\frac{\phi}{2}\psi^\dagger_{i;\alpha}\right)
   \lvert 0\rangle,
\end{equation}
where $\tan(\phi/2)=\Delta_\psi/2\Phi-\sqrt{1+(\Delta_\psi/2\Phi)^2}$. 
We use a subscript 0 under $|\mathrm{Slater}[c,\psi]\rangle_0$ to denote the infinite $U$ limit wavefunction. 
The relative chemical potential $\Delta_\psi$ is determined by 
fixing the average particle number of the $c$ layer to be $N/2 + \nu$ per site, which gives 
\begin{equation}\label{eqn:deltanu}
    \Delta_\psi = \frac{2\nu}{N} \Delta_{\mathrm{Mott}}, \quad
    \Phi = \frac{\sqrt{N^2/4-\nu^2}}{N} \Delta_{\mathrm{Mott}},
\end{equation}
where $\Delta_{\mathrm{Mott}} \equiv 2\sqrt{\Phi^2+(\Delta_\psi/2)^2}$ is the Mott gap opened in the mean-field Hamiltonian Eq.~\eqref{eqn:HMecf}. 

To obtain the real physical wavefunction, we need another EPR pair between the first ancilla $\psi$ and the second ancilla $\psi'$, 
which is chosen as the SU(N) singlet state on each site $i$:
\begin{equation}
\begin{aligned}
    \lvert s_i\rangle=&
    \frac{1}{\sqrt{\mathcal{N}_s}} \sum_p (-1)^{p} \psi^\dagger_{i;p_1} \cdots\psi^\dagger_{i;p_{N/2-\nu}}\\
    &\times {\psi^\prime}^\dagger_{i;p_{N/2-\nu+1}}...{\psi^\prime}^\dagger_{i;p_{N}}\lvert 0 \rangle~.
\end{aligned}
\end{equation}
Here, the sum is over permutations $p$ of the flavor index $\{1,2,...,N\}$ modulo the permutation within each layer $\{1,\cdots,N/2\pm\nu\}$, 
$(-1)^p$ is the inversion number of $p$, and $\mathcal{N}_s = N!/(N/2-\nu)!/(N/2+\nu)!$ is a normalization factor. 

We can now measure the two ancilla layers with a post-selection on $|s_i\rangle$, as illustrated in Fig.~\ref{fig:ancilla} (c) in the main text. 
The measurement procedure is equivalent to the projection
\begin{equation}
    P_S=\prod_i \ket{s_i}\bra{s_i}~.
\end{equation}
With straightforward calculation, we see that the spin degrees of freedom in the second ancilla $\psi'$ are perfectly teleported to the physical layer: 
\begin{equation}
\begin{aligned}
    |\Psi_c\rangle_0 =& P_s|\mathrm{Slater}[c,\psi]\rangle_0\otimes|\Psi_{\psi'}\rangle \\
    =&A_s
    \sum_{\{\bm{\alpha}_i\}}w(\{\bm{\alpha}_i\})\prod_{i=1}^{N_{\mathrm{site}}}\prod_{\alpha_i}^{n_T}c^\dagger_{i;\alpha_{i}}\lvert 0\rangle \otimes |s_i\rangle,
\end{aligned}
\end{equation}
which is equivalent to $\lvert\Psi_{\psi^\prime}\rangle$ in  Eq.~\eqref{eqn:psiprime} up to a factor $A_s = \frac{1}{\sqrt{\mathcal{N}_s}}\cos(\phi/2)^{N/2+\nu}\sin(\phi/2)^{N/2-\nu}$. 
Again, we use the subscript 0 under $|\Psi_c\rangle_0$ to denote the infinite $U$ limit wavefunction. 
If we assume that $\lvert\Psi_{\psi^\prime}\rangle$ is already the ground state of the Hubbard model in Eq.~\eqref{eqn:hubbardSUn} for $U\rightarrow \infty$, 
the ancilla wavefunction $\lvert\Psi_c\rangle$ is also the true ground state. 

In addition to the ground state,  the operators can also be teleported to the physical Hilbert space. For example, one can verify that
\begin{equation}\label{eqn:cpsirelation}
    \begin{split}
        P_S c^\dagger_{i;\alpha} \psi_{j;\beta}\lvert
        \mathrm{Slater}[c,\psi]\rangle_0\otimes|\Psi_{\psi'}\rangle 
         = & \tan\frac{\phi}{2}c^\dagger_{i;\alpha}c_{j;\beta}\lvert\Psi_c\rangle_0,\\
        P_S \psi^\dagger_{i;\alpha} c_{j;\beta}| \mathrm{Slater}[c,\psi]\rangle_0\otimes|\Psi_{\psi'}\rangle  = & -\cot\frac{\phi}{2}c^\dagger_{i;\alpha}c_{j;\beta}\lvert\Psi_c\rangle_0.\\
    \end{split}
\end{equation}

\subsection{Finte $U$ but $U \gg t_{ij}$}

The above analysis can be easily generalized to the more physically relevant regime with finite $U$ and $ U \gg t_{ij}$. 
The EPR pairs between $c$ and $\psi$ would now have a finite spread in real space, 
but are still well-defined in momentum space. 
In the small $t$ limit, we can write it down by slightly rotating the $t=0$ wavefunction in Eq.~\eqref{eqn:slatercpsit=0} at each momentum $\bfk$: 
\begin{equation}
    \lvert\mathrm{Slater}[c,\psi]\rangle= R^{(c\psi)}
    \lvert\mathrm{Slater}[c,\psi]\rangle_0~,
\end{equation}
where
\begin{equation}
    R^{(c\psi)}=\prod_\bfk\exp\left(-\frac{\mathrm{1}}{2}\delta\phi\left(\mathbf{k}\right)\sum_{\alpha}\left(c^\dagger_{\bfk;\alpha} \psi_{\bfk;\alpha}-\psi^\dagger_{\bfk;\alpha} c_{\bfk;\alpha}\right)\right),
\end{equation}
the rotation angle $\delta\phi(\mathbf{k})$ is approximately
\begin{equation}
    \delta\phi(\mathbf{k})=\frac{t(\mathbf{k})}{2}\frac{\Phi}{\Phi^2+\Delta_\psi^2/4}+O\left(\frac{t^2(\mathbf{k})}{\Phi^2}\right),
\end{equation}
and $t(\mathbf{k}) = \sum_jt_{ij}e^{i\bfk\cdot \bfR_j}$. 


All the remaining calculations are similar to what we did in the infinite $U$ limit. 
We perform the measurement on $\psi$ and $\psi'$ on each site and post-select the singlet states $|s_i\rangle$. 
Making use of the relations in Eq.~\eqref{eqn:cpsirelation}, 
we obtain the finite $t$ physical wavefunction as: 
\begin{equation}\label{eqn:ancilla_SW}
\begin{split}
    |\Psi_c\rangle =& P_S R^{(c\psi)} |\mathrm{Slater}[c,\psi]\rangle_0\otimes|\Psi_{\psi'}\rangle\\
    =&\left(1+\frac{1}{2\sqrt{\Phi^2+\Delta_\psi^2/4}}\sum_{i,j,\alpha}t_{ij}c^\dagger_{i;\alpha}c_{j;\alpha}\right)\lvert\Psi_c\rangle_0~.
\end{split}
\end{equation}

One can also utilize an inverse Schrieffer-Wolff transformation to obtain the ground state of the Hubbard model in Eq.~\eqref{eqn:hubbardSUn} \cite{Zhou2024}:
\begin{equation}\label{eqn:Hubbard_SW}
    \lvert\mathrm{GS,Hubbrd}\rangle \approx \left(1+\frac{1}{U}\sum_{i,j,\alpha}t_{ij}c^\dagger_{i;\alpha}c^{}_{j;\alpha}\right)\lvert\Psi_c\rangle_0.
\end{equation}

Comparing Eq.~\eqref{eqn:ancilla_SW} and Eq.~\eqref{eqn:Hubbard_SW}, we conclude that the two wavefunctions are equivalent as long as we choose the parameter such that $\Delta_{\mathrm{Mott}}\equiv 2\sqrt{\Phi^2+\Delta_\psi^2/4}=U$. 
In the more general case with a smaller $U$, we should treat $\Phi$ (and thus $\Delta_{\mathrm{Mott}}$) as a variational parameter to be optimized for each $U$.  
The relation $\Delta_{\mathrm{Mott}}=U$ is only excepted to be precise in the large $U$ limit. 
At smaller $U$, $\Delta_{\mathrm{Mott}}$ may decrease faster than $U$.

For the TBG system, we should use $N=8$. Then we have the relations $\Delta_{\psi}=\frac{\nu}{4} \Delta_{\mathrm{Mott}}$, and $\Delta_{\mathrm{Mott}}=U$ in the large $U$ limit. 
Although TBG differs from the lattice Hubbard model described here due to the fragile topology,  
most of the densities in the active band of TBG are localized on AA sites, and only a small region around the $\Gamma_M$ point contributes to the topological obstruction.  
So we expect most of our analysis here still apply to the TBG system away from the $\Gamma_M$ point.

\section{Ancilla theory in the continuum model}\label{app:BM}

\subsection{BM model and its symmetries}

\begin{table}[tb]
    \centering
    \begin{tabular}{cc}
    \toprule
    Symmetry & Transformation of $c_{0,\bfk}$ \\
    \midrule
    $C_{2z}T$ & $c^{}_{0,\bfk} \rightarrow \sigma_x\mathcal{K}c^{}_{0,\bfk}$ \\
    $P$ & $c^{}_{0,\bfk} \rightarrow i\mu_y \mathcal{K}c^\dagger_{0,-\bfk}$ \\
    $S$ & $c^{}_{0,\bfk} \rightarrow \sigma_z\mathcal{K}c^\dagger_{0,\bfk}$ \\
    $C_{3z}$ & $c^{}_{0,\bfk} \rightarrow e^{-i2\pi/3\sigma_z}c^{}_{0,C_{3z}\bfk}$ \\
    $C_{2x}$ & $c^{}_{0,\bfk} \rightarrow \sigma_x\mu_xc^{}_{0,C_{2x}\bfk}$ \\
    \bottomrule
    \end{tabular}
    \caption{Symmetry action of the original basis $gc_{0,\bfk} = D(g)\mathcal{K}^{0,1}c^\pm_{0,\tilde{g}\bfk}$, 
    where $D(g)$ is a $4\times 4$ matrix including both the layer index $l$ and orbital index $a$, and $\mathcal{K}$ is the complex conjugation. 
    We use $\sigma$ and $\mu$ to denote the Pauli matrices of the orbital and layer indexes, respectively.}
    \label{tab:symmetry_detail}
\end{table}

\begin{table}[tb]
    \centering
    \begin{tabular}{lccccc}
        \toprule
        &$C_{2z}T$ & $P$ & $S$ & $C_{3z}$ & $C_{2x}$\\
        \midrule
        $c_\bfk$ & $\sigma_x\mathcal{K}c_\bfk$& $i\sigma_z\mathcal{K}c_{-\bfk}^\dagger$
        & $\sigma_z\mathcal{K} c^\dagger_\bfk$ &
        $e^{i\alpha(\bfk)\sigma_z}c_{C_{3z}\bfk}$& $\sigma_xc_{C_{2x}\bfk}$ \\
        $\psi_\bfk$ & $\sigma_x\mathcal{K}\psi_\bfk$& $i\sigma_z\mathcal{K}\psi^\dagger_{-\bfk}$ & $-\sigma_z\mathcal{K}\psi^\dagger_\bfk$
        & $e^{-i2\pi/3\sigma_z}\psi_{C_{3z}\bfk}$ & $\sigma_x\psi_{C_{2x}\bfk}$ \\
        \bottomrule
    \end{tabular}
    \caption{Symmetry action in the active bands and the ancilla bands. Here, $\alpha(\bfk)$ satisfies $\alpha(\bfk) + \alpha(C_{3z} \bfk) + \alpha(C_{3z}^{-1}\bfk) = 2\pi$. 
    It can be fixed at high-symmetry points as and $\alpha(\Gamma_M) = 0, \alpha(K_M)=\alpha(K_M') = 2\pi/3$.}
    \label{tab:symmetry_act}
\end{table}

Here, we briefly review the Bistritzer-MacDonald (BM) model\cite{Bistritzer2011} for the twisted bilayer graphene (TBG) system, with a special focus on the symmetries of both the original model and the active bands.
For a given valley, say $\tau =K$, we consider the following Hamiltonian: 
\begin{equation}\label{eqn:HBM}
\begin{aligned}
    H_{\mathrm{BM}} =& \sum_{l,\mathbf{k}} c^\dagger_{0,\mathbf{k};l} v_F(\bfk-\bfK_M^l)\cdot\boldsymbol{\sigma} c^{}_{0,\bfk;l}\\
    &+ \sum_{\bfk,j=1,2,3}(c^\dagger_{0,\bfk+\bfG_{M,j};T} T_j c^{}_{0,\bfk;B}+\mathrm{h.c.})~.
\end{aligned}
\end{equation}
Here,  $c_{0,\bfk;l}^\dagger = (c_{0,\bfk;l+}^\dagger,c_{0,\bfk;l-}^\dagger)$
is the original TBG electron from layer $l=T,B$ with the orbital index $a=\pm $ suppressed and the valley and spin indices omitted. 

The Hamiltonian in Eq.~\eqref{eqn:HBM}  consists of an intra-layer kinetic energy term and an inter-layer hopping term. 
For the intra-layer term, we keep only the linear Dirac cone dispersion of each layer and omit the small $\theta$ rotation of the momentum. 
Note that the momentum $\bfk$ of $c_{0,\bfk;l}$ is related to the physical momentum  $\bar{\bfk}$ through 
\begin{equation}\label{eqn:momshift_BM}
    \bar{\bfk} = \mathbf{K}^{T/B}+\bfk-\bfK_M^{T/B}~,  
\end{equation}
with $\mathbf{K}^{T} = M_\theta \mathbf{K}^B$ being the $K$ point momentum of the original top and bottom  graphene layers, respectively. 
The momentum shift $\bfk\rightarrow\bfk-\mathbf{K}^{T/B}_M$ is introduced such that the Dirac cone is located at the $\bfK^T_M = \bfK_M$ and $\bfK^B_M=\bfK'_M$ points of the MBZ,
which centers the MBZ $\Gamma_M$ point at 0. 

For the inter-layer term, hopping occurs between the top and bottom layers with a momentum shift $\bfG_{M,j} = -\mathbf{q}_{j}-\bfK^T_M+\bfK^B_M$, 
where $\mathbf{q}_1 = \bfK^T-\bfK^B = 2|\bfK|\sin(\theta/2)\mathbf{e}_y$ and $\mathbf{q}_2 = C_{3z}\mathbf{q}_1$,$\mathbf{q}_3 = C_{3z}^{-1}\mathbf{q}_1$. 
Finally, the $T_j$ hopping matrix takes the form 
\begin{equation}\label{eqn:BM_Tjs}
    T_j = w_0 \mathds{1} + w_1 \sigma_x e^{-i\frac{2\pi}{3}(j-1)\sigma_z}~. 
\end{equation}
In all of our calculations, we use the parameters 
\begin{equation}
\begin{aligned}
    &w_1 = 110\ \mathrm{meV}, 
    \quad  v_F|\bfK| = 10.16\ \mathrm{eV}~,
\end{aligned}
\end{equation}
The specific twisted angle $\theta$ and the ratio $w_0/w_1$ are denoted in the caption of each figure.

The TBG system hosts various symmetries, including 
point group symmetries, time-reversal symmetry, and an approximating particle-hole symmetry. 
The actions of these symmetries on the electron annihilation operator 
can be generally written as 
\begin{equation}
    gc_{0,\bfk} = D(g)\mathcal{K}^{0,1}c^\pm_{0,\tilde{g}\bfk}~,
\end{equation}
where we further suppress the layer index $l$, so that $D(g)$ is a $4\times 4$ matrix in layer and orbital space, and $\mathcal{K}$ is the complex conjugate operator. 
Note that due to the momentum shift in Eq.~\eqref{eqn:momshift_BM}, the modified action of $g$ on the momentum $\tilde{g}\bfk$ is such that $\tilde{g}\bfk-\bfK^{l'} = g(\bfk-\bfK^l)$, 
which will be omitted in our notation for simplicity. 
The transformation rules of all the symmetries on the $c_{0,\bfk}$ electrons are summarized in Table.~\ref{tab:symmetry_detail}, 
where we use $\sigma$ and $\mu$ to denote the Pauli matrices for the orbital and layer indices, respectively. 

Here we only consider symmetries within each valley and spin. 
First, there are point group symmetries including the threefold rotation $C_{3z}c_{0,\bfk}=e^{-i2\pi/3\sigma_z}c_{0,C_{3z}\bfk}$ and the twofold rotation $ C_{2x}c_{0,\bfk}=\sigma_xc_{0,C_{2x}\bfk}$. 
Time reversal symmetry sends the physical momentum from $\bar{\bfk}$ to $-\bar{\bfk}$ and therefore change the valley from $K$ to $K'$. 
But we can consider the combination of a twofold rotation and time reversal $C_{2z}Tc_{0,\bfk} = \sigma_x\mathcal{K}c_{0,\bfk}$. 
When the twist angle $\theta\ll 1$ and the nonlinear dispersion of single-layer graphene is ignored, as we have down, there is also a particle-hole symmetry, which becomes approximate in real materials. 
Here we use the particle-hole symmetry that sends $c_{0,\bfk}$ to $i\mu_y\mathcal{K} c^\dagger_{0,-\bfk}$. 
Finally, when $w_0=0$ in Eq.~\eqref{eqn:BM_Tjs}, there will be another particle-hole like symmetry called the chiral symmetry $S$, which sends $c_{0,\bfk}$ to $\sigma_z\mathcal{K}c^\dagger_{0,\bfk}$. 
In this limit, the two flat bands form topological bands carrying Chern number $C = \pm 1$, 
but this is only an approximate symmetry when $w_0$ takes a finite value, as in our calculation. 

As the flat bands are of particular importance, 
here we work out the symmetry representations within the active bands for later convenience. 
We choose the so-called sublattice basis such that the matrix $\Gamma_{mn}=\langle u_m(\bfk)|\sigma_z|u_n(\bfk)\rangle$ is diagonal. 
We therefore continue to use index $a$ to label the band basis. 
The sublattice basis has the merit that the chiral symmetry $S\sim\sigma_z$ is nearly diagonalized so that each of the basis states carries a non-trivial Chern number $\pm 1$. 

Working in the sublattice basis where $\sigma_z$ is diagonal and noting that $\{C_{2z}T,\sigma_z\}=0$, we can choose a gauge such that $C_{2z}T c_{\bfk}= \sigma_x e^{-i\theta_T(\bfk)}\mathcal{K}c_{\bfk}$. 
The phase $\theta_T(\bfk)$ can always be set to zero by appropriately choosing the overall phase of the states $|u_a(\bf{k})\rangle$. 
Similarly, we can always choose a gauge such that the action of particle-hole symmetry is $Pc_{\bfk} = i\sigma_zc^\dagger_{-\bfk}\mathcal{K}$, and the $x$-axis rotation is $C_{2x}c^{}_{\bfk} = \sigma_xc^{}_{C_{2x}\bfk}$, where we usedthe commutation relations $[S,P]=[C_{2z}T,P] = \{C_{2x},S\} = \{C_{2x},P\} = [C_{2x},C_{2z}T]=0$ to constrain the possible forms of the symmetries. 

The only remaining unsettled symmetry is the 3-fold rotation $C_{3z}$, which generally takes the form $C_{3z}c_{\bfk} = e^{i\alpha(\bfk)\sigma_z}c_\bfk$ and 
$\alpha(\bfk)+\alpha(C_{3z}\bfk)+\alpha(C_{3z}^{-1}\bfk) = 0$. 
The precise form of the $\alpha(\bfk)$ is gauge dependent except at the high symmetry points, 
where numerical calculations show that $\alpha(\Gamma_M) = 0$, and $\alpha(K_M)=\alpha(K'_M) = 2\pi/3$. 
Generally, $\alpha(\bfk)$ cannot be taken as a continuous and periodic function at the same time due to nontrivial Chern numbers \cite{Fang2012}. 
Here we will choose a gauge such that $\alpha(\bfk)=0$ near the $\Gamma_M$ point. 
All the symmetry actions within the active bands are summarized in Table~\ref{tab:symmetry_act}. 

\subsection{Ancilla orbital in BM model}

It is known that the active band has a fragile topology  and cannot be Wannierized to a local orbital. 
On the other hand, the ancilla bands $\psi_i$ and $\psi'_i$  introduced in the main text only live on the AA lattice sites labeled by $i$. As an auxiliary degree of freedom, the Wannier orbital of the ancilla fermions seems to be meaningless. However, in our final ansatz with finite hybridization $\Phi(\bfk)c^\dagger_\bfk\psi_\bfk^{}$, the ancilla fermion will represent a physical operator and the ansatz of the ancilla fermion $\psi_i$ influences the final physical state after projection $P_s$.  Or in other words, we need to fix the hybridization $\Phi(\bfk)$ as a variational ansatz. Equivalently we can imagine $\psi_{i}$ has a `Wannier orbital' and couples to the physical state $c_{0;\bfk}$ through local hopping. In principle, the `Wannier orbital' of the ancilla $\psi$ is arbitrary and should be determined by optimizing the energy of the final physical state. Here we simply construct an ansats based on the intuition that the ancilla fermion $\psi$ represents a target of a fully Mott localized state.  With the expectation of opening a Mott gap for states at momenta away from the $\Gamma_M$ point, we require $\psi_{i;a=\pm}$ to have angular momentum $L=\pm 1$ under $C_3$ rotation around an AA site.

We will thus place the first ancilla layer on a ``copy'' of the original TBG lattice, and consider the following Wannier orbital localized at AA sites denoted as $\{\mathbf{R}_i\}$: 
\begin{equation}\label{eqn:Wannier_I}
\begin{aligned}
    \psi^\dagger_{i;a\tau s} =&\sum_le^{i\frac{\pi}{4}la }
    \int \mathrm{d}^2r e^{i \bfK_M^l\cdot(\bfr- \mathbf{R}_i) }\\
    &\times W_\psi(\mathbf{r}-\mathbf{R}_i)\psi^\dagger_{0,la\tau s}(\bfr)~.
\end{aligned}
\end{equation}
Here, the real space ancilla $\psi_{0,la\tau s}^\dagger(\bfr)$ in layer $l$ 
is defined as $\psi_{0,la\tau s}^\dagger(\bfr) = \frac{1}{\sqrt{V}}\sum_\bfk e^{-i\bfk\cdot\bfr} \psi_{0,\bfk;la\tau s}^\dagger$, 
where $V$ is the total volume of the space. 
The overall phase factors $e^{i\frac{\pi}{4}l a}$ and $e^{i\Delta\mathbf{K}_l\cdot(\bfr-\bfR_I)}$ are introduced to make the Wannier ancilla $\psi_{i;a\tau s}^\dagger$ respect the particle-hole and translation symmetries. 
Note that here we assume $\psi_{0,\bfk;la\tau s}$ has the same symmetry properties of $c_{0,\bfk;la\tau s}$. 
$W_\psi(\mathbf{r}-\mathbf{R}_i)$ is a Gaussian-like function centered around $\mathbf{R}_{i}$, whose exact form will be shown to be unimportant as long as it respects the symmetries $W_\psi(g\bfr)=W_\psi(\bfr)$.
More generally, $W_\psi$ should also be layer and sublattice dependent but here we only consider the simplest form with 
\begin{equation}\label{eqn:Wgauss}
    W_\psi(\bfr)\propto e^{-|\bfr|^2/2\lambda_0^2}~. 
\end{equation}

Transforming to the momentum space, we get: 
\begin{equation}\label{eqn:Wannier_K}
\begin{aligned}
    \psi_{\bfk;a\tau s}^\dagger = &\frac{1}{\sqrt{N_{\mathrm{site}}}}\sum_i e^{i\bfk\cdot\bfR_i}\psi_{i;a\tau s}^\dagger  \\
    =& \sum_{l,\bfG_M}e^{i\frac{\pi}{4}l a} 
    \tilde{W}_\psi(\mathbf{k}-\bfK_M^l+\bfG_M)
    \psi^\dagger_{0,\bfk+\bfG_M;la\tau s}~,
\end{aligned}
\end{equation}
where $N_{\mathrm{site}}$ is the number of Moir\'{e} unit cells of the system and 
$\tilde{W}_\psi(\mathbf{k}) = \sqrt{\frac{N_{\mathrm{site}}}{V}}\int \mathrm{d}^2rW_\psi(\mathbf{r})e^{-i\mathbf{k}\cdot\mathbf{r}}\propto e^{-\lambda_0^2|\bfk|^2/2}$ is the Fourier transformation of $W_\psi(\bfr)$. 

We now check that the symmetries of the Wannier orbital constructed in Eq.~\eqref{eqn:Wannier_I} or \eqref{eqn:Wannier_K}.  
We have assumed that the ancilla ``copy'' layer $\psi_{0,\bfk}$ transforms the same as the physical electrons $c_{0,\bfk}$ as in Table~\ref{tab:symmetry_detail}.  
The calculation of the transformation rules of all the symmetries then becomes straightforward and the results are listed in Table~\ref{tab:symmetry_act}, which indeed coresponds to a $p_x\pm ip_y$ orbital. 
The only subtle calculation is the particle-hole symmetry, where one needs to use the relation $ i\sigma_ze^{-i\frac{\pi}{4}\sigma_z} = e^{i\frac{\pi}{4}\sigma_z}$. 

\subsection{Hybridization matrix between $c$ and $\psi$}

\begin{figure}[ht]
    \centering
    \includegraphics[width=0.95\linewidth]{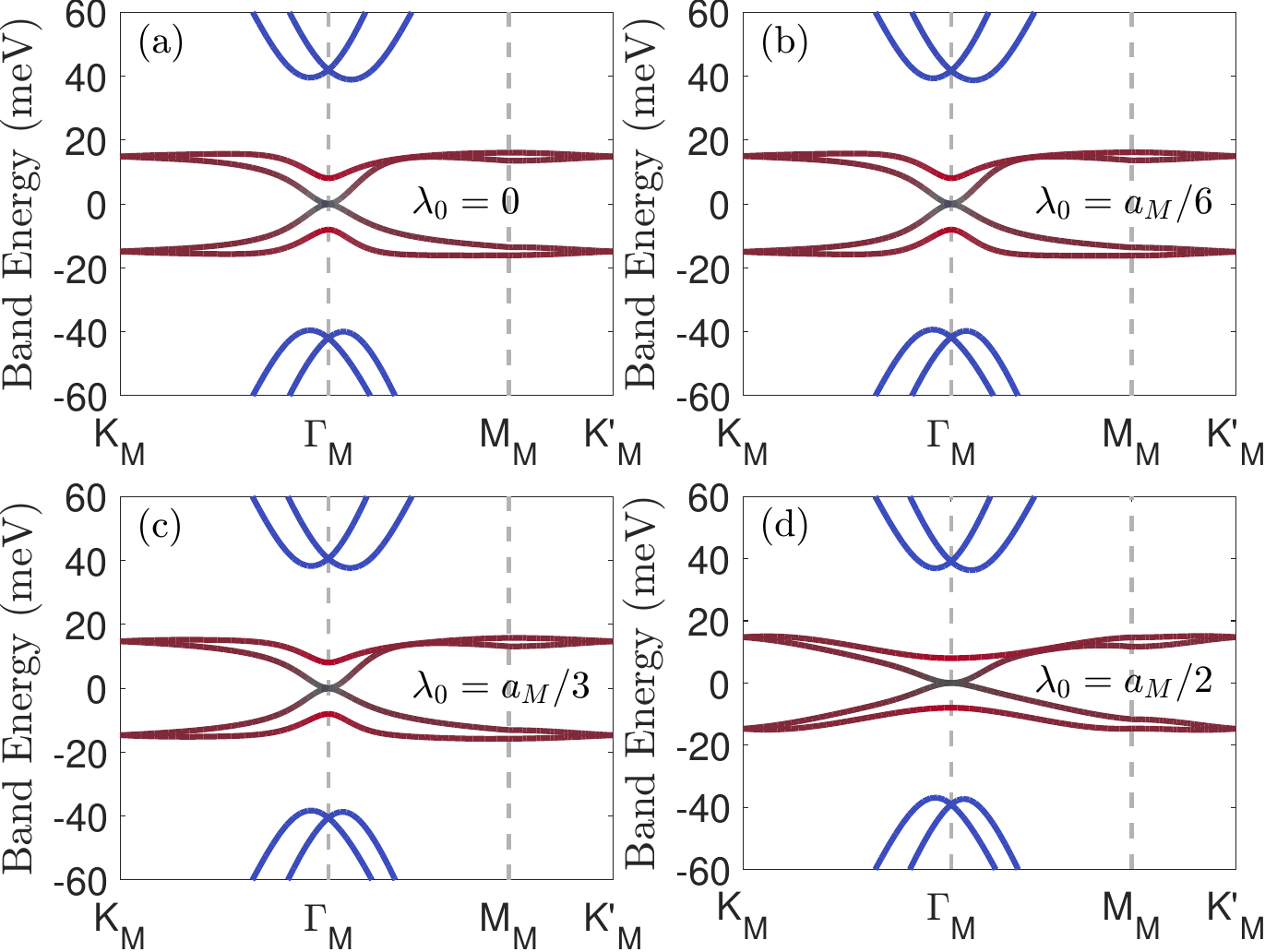}
    \caption{Spectra calculated for different choices of $\psi$ Wannier orbitals Eq.~\eqref{eqn:Wgauss} at the charge neutral point $\nu=4$ and $\Delta_K\approx 30$ meV
    for (a) $\lambda_0 = 0$, (b) $\lambda_0=a_M/6$, (c) $\lambda_0 = a_M/3$ and (d) $\lambda_0=a_M/2$. 
    The unit of length is chosen as $a_M$ where which is the distance between two nearest AA positions. 
    No obvious change of the spectrum is observed as long as $\lambda_0 < a_M/2$. }
    \label{fig:spec_lambda0}
\end{figure}

To obtain the hybridization Hamiltonian between the physical layer and the ancilla layer, 
we can assume that the hybridization happens only when the physical electron and its ancilla ``copy'' coincide at the same lattice in real space, i.e.: 
\begin{equation}
    H_{0,\mathrm{hyb}}^{(c\psi)} = \Phi_0\sum_{l,\bfR, a}
    c_{0;\bfR,\mathbf{t}_{a},l}^\dagger \psi_{0;\bfR,\mathbf{t}_a,l}^{} + \mathrm{h.c.}~.
\end{equation}
Here, $\mathbf{R}$ is the position vector of the original graphene lattice unit cell in layer $l$, and $\mathbf{t}_a$ is the relative position vector of sublattice $a$ within a unit cell. $\Phi_0$ is the microscopic hybridization strength. 
One can assume a more general hybridization form as long as it respects the symmetry, but the simple form used here is sufficient to capture the essential physics. 
The hybridization matrix is then projected onto the Wannier orbitals of the ancilla layer as: 
\begin{equation}
\begin{aligned}
    H^{(c\psi)}_{\mathrm{hyb}} =& P_W^\psi H_{0,\mathrm{hyb}} P_W^\psi \\
    =& \Phi_0\sum_{la\tau s}\sum_{\bfk\in \mathrm{MBZ},\bfG_M} 
    e^{i\frac{\pi}{4}la} \tilde{W}_\psi(\bfk-\bfK_M^{l}+\bfG_M)\\
    &c^\dagger_{0,\bfk+\bfG_M;la\tau s} \psi^{}_{\bfk;a\tau s}+\mathrm{h.c.}~,
\end{aligned}
\end{equation}
where $P^\psi_W = \sum_{\bfk\in \mathrm{MBZ},a\tau s}\psi^\dagger_{\bfk;a\tau s}\psi^{}_{\bfk;a\tau s}$ is a projection of $\psi$ to the subspace spanned by the Wannier orbitals $\psi^\dagger_{\bfk;a\tau s}, \bfk \in\mathrm{MBZ}$. 

In realistic calculations, we can choose a Gaussion like $W_\psi(\bfr)$ as in Eq.~\eqref{eqn:Wgauss} with a correlation length scale $\lambda_0$.  
In Fig.~\ref{fig:spec_lambda0}, we show the band structure at $\nu = 0$ for several different values of $\lambda_0 = 0, \lambda_0=a_M/6,\lambda_0=a_M/3$ and $\lambda_0 = a_M/2$,
where $a_M$ is the distance between two AA sites. 
No obvious change of the spectra is observed as long as $\lambda_0$ is smaller than $a_M/2$, i.e. when two Wannier orbitals have small overlap. 
We therefore choose a delta-function Wannier orbital and use a constant $\tilde{W}(\bfk) = 1$ unless otherwise mentioned. 
In this limit, the hybridization Hamiltonian projected onto the ancilla Wannier orbitals can be written as:
\begin{equation}\label{eqn:MFBM_hyb}
    H^{(c\psi)}_{\mathrm{hyb}} = \Phi_0\sum_{la\tau s}\sum_{\bfk\in \mathrm{MBZ},\bfG_M} c^\dagger_{0,\bfk+\bfG_M;la\tau s} e^{i\frac{\pi}{4}la}\psi_{\bfk;a\tau s}+\mathrm{h.c.}
\end{equation}

Finally, we summarize the total mean-field Hamiltonian model as: 
\begin{equation}\label{eqn:MFBM_app}
    H_{\mathrm{MF}}^{(c\psi)} = H_{\mathrm{BM}} + H^{(c\psi)}_{\mathrm{hyb}} - \mu(N_c+N_\psi) + \Delta_\psi(N_\psi-N_c)/2 ~,
\end{equation}
with $H_{\mathrm{BM}}$ defined in Eq.~\eqref{eqn:HBM} and $H_{\mathrm{hyb}}^{(c\psi)}$ defined in Eq.~\eqref{eqn:MFBM_hyb}. 

\section{Analytical treatment projected to the active bands}

\begin{figure}[t]
    \centering
    \includegraphics[width=0.9\linewidth]{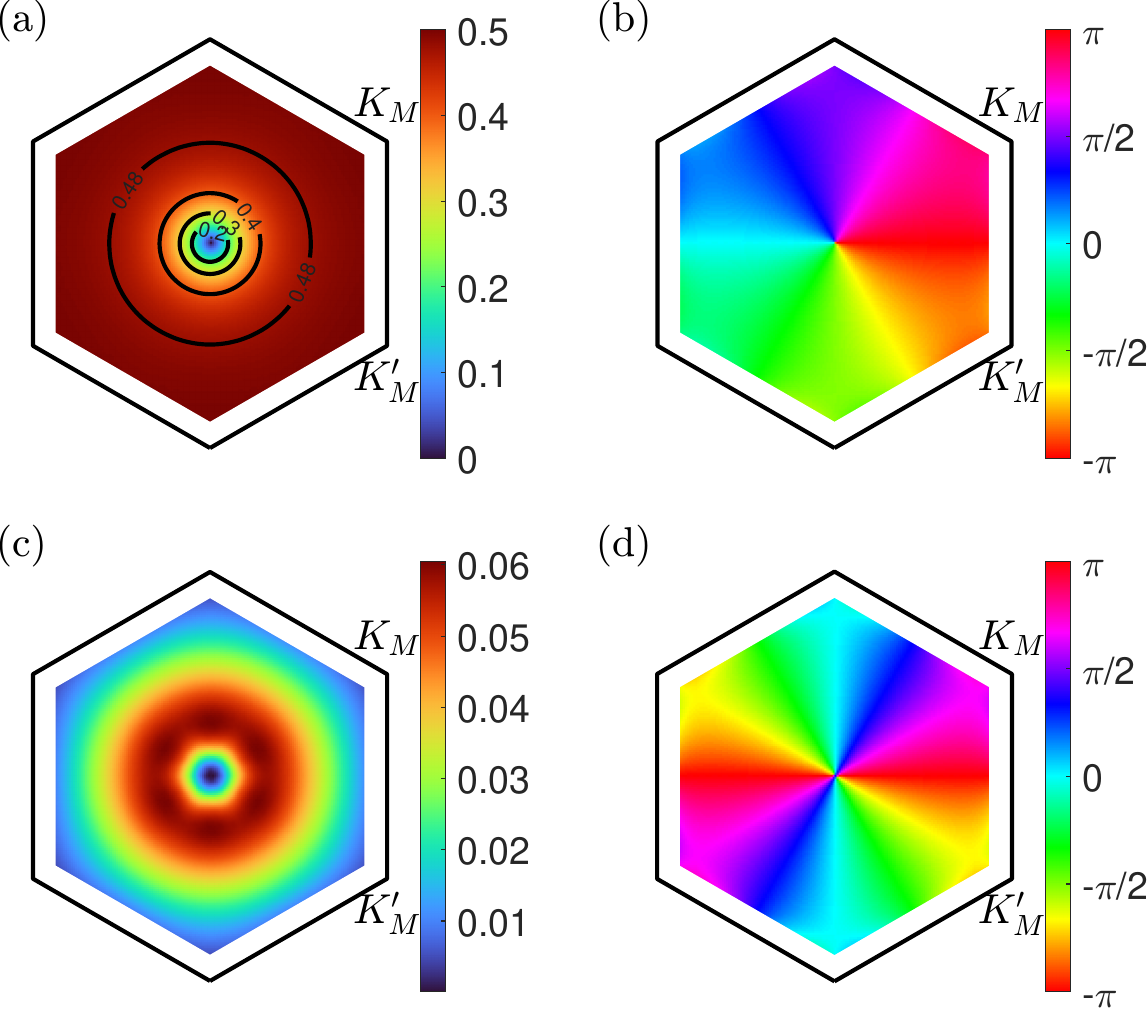}
    \caption{(a) The absolute value and (b) the phase of $\Phi(\bfk)/\Delta_K$; 
    and (c) the absolute value and (d) the phase of $\delta\Phi(\bfk)/\Delta_K$, calculated at $\theta=1.14^\circ,w_0/w_1=0.8,\Delta_K=30$ meV. 
    }
    \label{fig:Phik_BM}
\end{figure}

\begin{figure}
    \centering
    \includegraphics[width=0.9\linewidth]{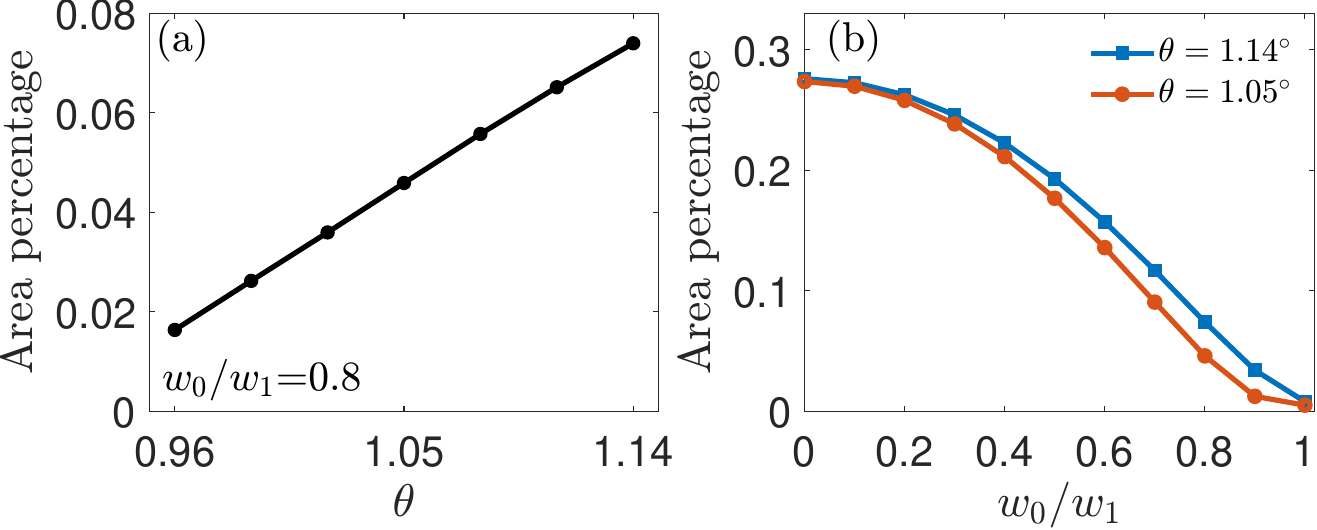}
    \caption{(a) Evolution of the area percentage of MBZ enclosed by the contour $|\Phi(\bfk)|=0.4\Delta_K$ as a function of the twisted angle $\theta$ for $w_0/w_1=0.8$. 
    And (b) evolution of the area percentage of MBZ enclosed by the contour $|\Phi(\bfk)|=0.4\Delta_K$ as a function of $w_0/w_1$ for twisted angle $\theta=1.14^\circ$.  
    }
    \label{fig:Phik_BM_area}
\end{figure}

\begin{figure}
    \centering
    \includegraphics[width=0.99\linewidth]{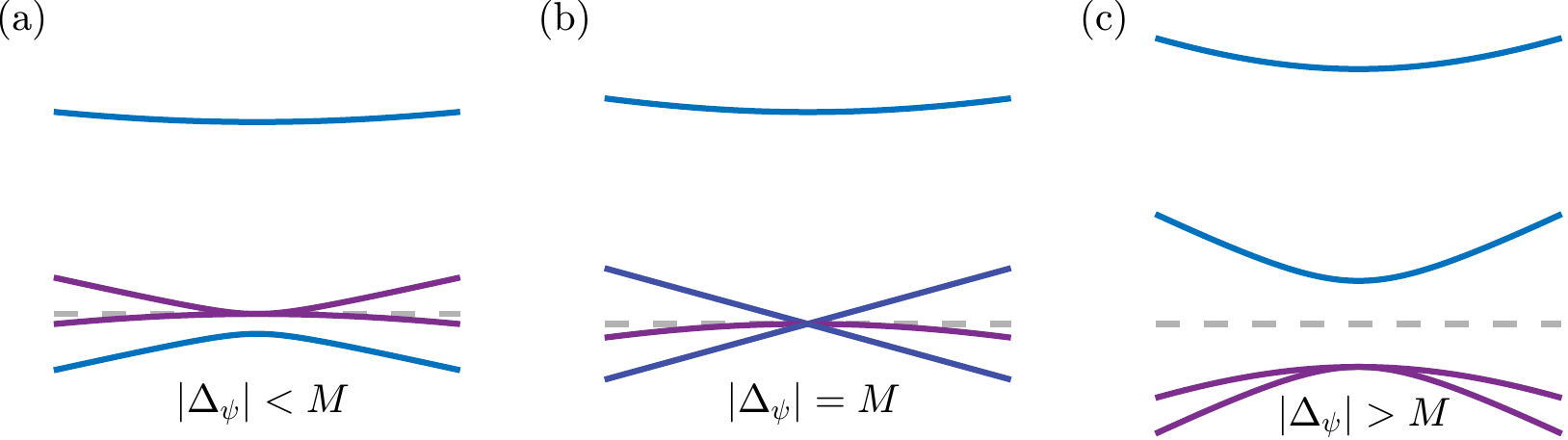}
    \caption{
    Illustration of the band structure evolution around the $\Gamma_M$ point across the semimetal to insulator transition for 
    (a) $|\Delta_\psi|<M$, (b) $|\Delta_\psi|=M$ and (c) $|\Delta_\psi|>M$. 
    Here, the blue and purple colors represent the bands of $c$ and $\psi$, respectively, and the dark blue color in (b) represents mixing of the two bands. 
    The gray shaded line marks the chemical potential $\mu$. 
    }
    \label{fig:ABG_illu}
\end{figure}


When $U$ is much smaller than the remote band energies, we can further project the hybridization Hamiltonian into the active flat bands of the BM model in Eq.~\eqref{eqn:HBM}: 
\begin{equation}
    c^{}_{0,\bfk+\bfG_M;la\tau s} = \sum_{a'=\pm} u^{\mathrm{BM}}_{\bfG_M la\tau s;a'\tau s}(\bfk)c^{}_{\bfk;a'\tau s}+\cdots~,
\end{equation}
where the original electron operator $c^{}_{0,\bfk+\bfG_M;la\tau s}$ is expressed in terms of the active bands $c^{}_{\bfk;\pm \tau s}$ and some small contributions from the remote bands. 
The exact form of the wavefunctions $u^{\mathrm{BM}}_{\bfG_M la\tau s;\pm \tau s}(\bfk)$ is gauge dependent. 
Here we choose a symmetric gauge in the aforementioned sublattice basis, such that $c_{\bfk;a\tau s}$ transforms as in Table~\ref{tab:symmetry_act} and $u^{\mathrm{BM}}_{\bfG_M la\tau s;\pm \tau s}(\bfk)$ is continuous around the $\Gamma_M$ point. 

Before proceeding with numerical calculations of the hybridization matrix, 
we can guess the form of the hybridization matrix using the symmetries listed in Table.~\ref{tab:symmetry_act}. 
First, the $C_{2z}T$ symmetry restricts the hybridization Hamiltonian to be 
\begin{equation}\label{eqn:hybrid_proj_BM}
    H^{(c\psi)} = \sum_{\bfk\in \mathrm{MBZ}} c_{\bfk}^\dagger\left[ \begin{matrix} 
    \Phi(\bfk) & \delta\Phi(\bfk) \\ \delta\Phi^*(\bfk) & \Phi^*(\bfk)
    \end{matrix}\right]\psi_\bfk + \mathrm{h.c.}~.
\end{equation}
Here $\delta\Phi(\bfk)$ is generally a small value compared to $\Phi(\bfk)$ in the sublattice basis, 
which can be proved to be exactly zero when the chiral symmetry $S$ is exact. 

We pay a particular attention to the $\Gamma_M$ point, 
where the active bands are $s$-wave under $C_{3z}$ and the ancilla bands are $p_x\pm ip_y$-wave, 
so that the hybridization matrix is forced to be zero. 
Around the $\Gamma_M$ point, we choose the gauge such that $\alpha(\bfk) = 0$ in Table~\ref{tab:symmetry_act}. 
Then the hybridization Hamiltonian can be linked by a $C_{3z}$ rotation: 
$\Phi(C_{3z}\bfk) = \Phi(\bfk) e^{-i2\pi/3} $, $\delta\Phi(C_{3z}\bfk) = \delta\Phi(\bfk) e^{i2\pi/3}$, 
a $C_{2x}$ rotation: $\Phi(C_{2x}\bfk) = \Phi^*(\bfk)$,  $\delta\Phi(C_{2x}\bfk) = \delta\Phi^*(\bfk)$ and 
a particle-hole transformation : $\Phi(-\bfk) = -\Phi(\bfk), \delta\Phi(-\bfk) = \delta\Phi(\bfk)$. 
Expanding the Hamiltonian to the second order of $k=k_x+ik_y$, a Hamiltonian respecting all the symmetries takes the form: 
\begin{equation}\label{eq:cpsi_app}
    H^{(c\psi)}_{\Gamma_M}  \approx \Phi_{\Gamma_M}\sum_\bfk c^\dagger_\bfk\left[ \begin{matrix} 
    \bar{k} & \delta \bar{k}^2 \\ \delta {k}^2 & {k}
    \end{matrix}\right]\psi_\bfk, 
\end{equation}
where $\Phi_{\Gamma_M}$ and $\delta$ are parameters that need to be determined numerically. 
Deviating from the $\Gamma_M$ point, there is no longer obstruction of the hybridization and we need to recover the Mott physics $|\Phi(\bfk)|\sim\Delta_K/2$ away from $\Gamma_M$.

Numerical calculations of both $\Phi(\bfk)/\Delta_K$ and $\delta\Phi(\bfk)/\Delta_K$ in the aforementioned gauge is shown in  Fig.~\ref{fig:Phik_BM} (a)-(d) for $\Delta_K=30$ meV. 
Near the $\Gamma_M$ point, we find  $\Phi(\bfk)\sim k_x- ik_y$. 
and $\delta\Phi(\bfk)\sim (k_x-ik_y)^2$ consistent with Eq.~\eqref{eq:cpsi_app}. 
Away from the $\Gamma_M$ point, we find that $|\Phi(\bfk)|\sim \Delta_K/2$ is roughly a constant proportional to the Mott gap.  
An overall ansatz in the full MBZ would be 
\begin{equation}
    \frac{2\Phi(\bfk)}{\Delta_K}=\frac{k_x-ik_y}{\sqrt{|\bfk|^2+s^2}}~, 
\end{equation}
which is found to be indeed a good approximation as shown in the main text Fig.~\ref{fig:ABG_illu_main} (b). 
On the other hand, $|\delta\Phi(\bfk)| \ll \Delta_K$ for the entire MBZ, and can be neglected in the following analysis. 

Here we show the area percentage $\alpha\approx 1.07\pi (s/|\bfK_M|)^2$ of the MBZ enclosed by a contour $|\Phi(\bfk)|=0.4\Delta_K$ in Fig.~\ref{fig:Phik_BM_area}, 
which characterizes the concentration of the active band berry curvature. 
It decreases as a function of $w_0/w_1$, reaching its maximum value at the chiral limit $w_0/w_1=0$, and almost vanish at $w_0/w_1=1$. 
On the other hand, it increases almost linearly as a function of the twisted angle $\theta$.   When $w_0/w_1=0.8$ at magic angle, $\alpha$ is around $5\%$. This means that at least 95$\%$ of the states in the  MBZ acquire a Mott gap equal to $\Delta_K$, and only less than 5$\%$ of the MBZ around the $\Gamma$ point has a significantly smaller gap. 

We can interpret our ansatz for $H^{(c\psi)}$ as an analog to that of an excitonic insulator. For conventional Mott insulator, the exciton order parameter $\Phi(\bfk)$ is $s$-wave, while in TBG we need to have a vortex for the exciton order parameter at the $\Gamma_M$ point. The parameter $\alpha$ is the vortex core size. In this work we restrict to symmetric ansatz, but one can imagine a symmetry breaking order emerges at lower temperature to further gap out the small region around the $\Gamma_M$ point.  Compared to the symmetry breaking ansatz, our symmetric Mott state has an energy cost $\delta E \sim \alpha U$ from the vortex core.  The symmetry breaking order is likely a generalized ferromagnetic order, thus we estimate the ferromagnetic exchange $|J|\sim \alpha U$ and is at least one order of magnitude smaller than $U$. Our symmetric Mott state then is justified in the temperature regime $|J|<T<<U$.  Alternatively, for even integer filling, we can add an anti-Hund's coupling $J_A$ which penalizes the ferromagnetic phase. When $J_A>|J|\sim \alpha U$, the symmetric Mott state can be the ground state.  If $U=30$ meV, $\alpha U=1.5$ meV if $\alpha=5\%$.  Interestingly $J_A$ from electron-phonon coupling is also estimated to be at order of $1-3$ meV\cite{wang2024molecular}. Therefore it is conceivable that in real system $J_A$ is close to the critical value $J_{A;c}$. 

\begin{figure*}[ht]
    \centering
    \includegraphics[width=0.95\linewidth]{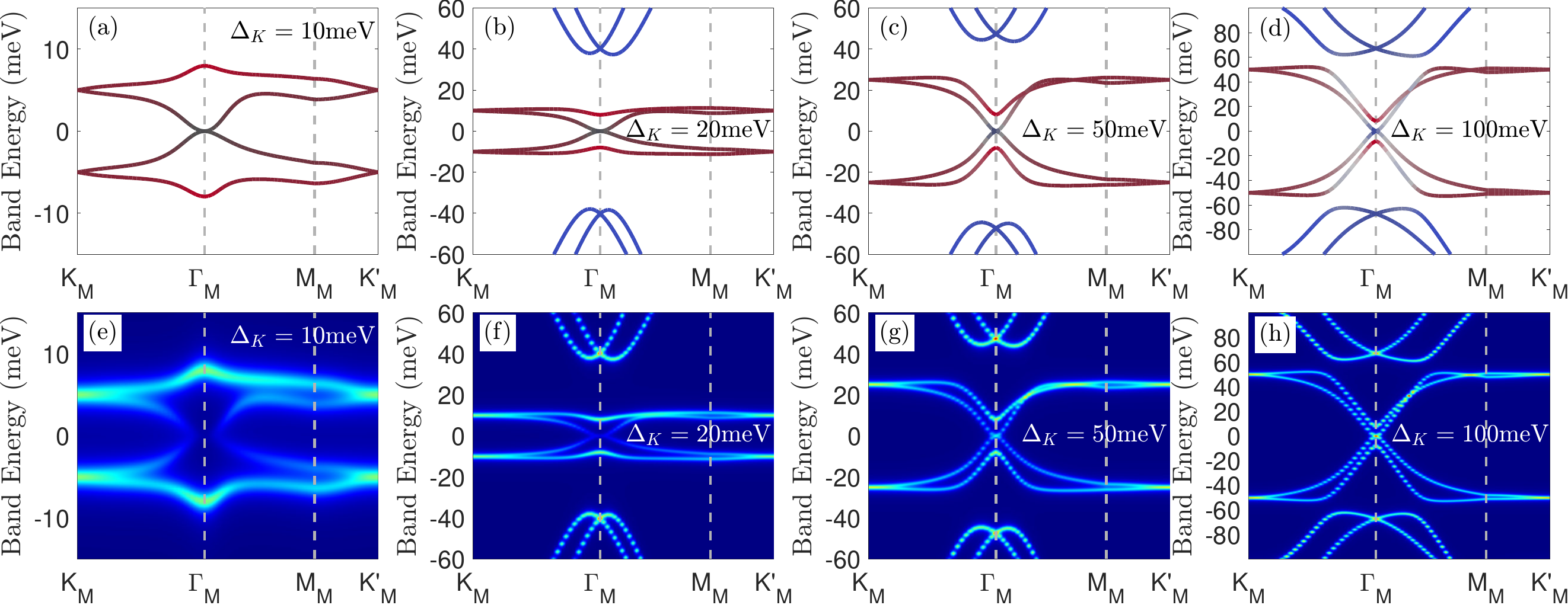}
    \caption{Mott bands calculated by BM model for $\nu = 0,\theta= 1.14^\circ,w_0/w_1=0.8$, with different hybridization strength at (a) $\Delta_K = 10$ meV, (b) $\Delta_K=20$ meV (c) $\Delta_K=50$ meV and (d) $\Delta_K=100$ meV,
    where $\Delta_K$ is defined as the $K_M$ point gap. 
    The blue, red, and dark colors of the dispersion line represent the weight of the active band, the remote bands and the ancilla bands, respectively. 
    The corresponding physical electron spectrum function $A_c(\bfk,\omega)$ is shown in (e)-(h), with a broadening chosen as $\eta=1$ meV. }
    \label{fig:nu0sBM_app}
\end{figure*}

\begin{figure*}[ht]
    \centering
    \includegraphics[width=0.95\linewidth]{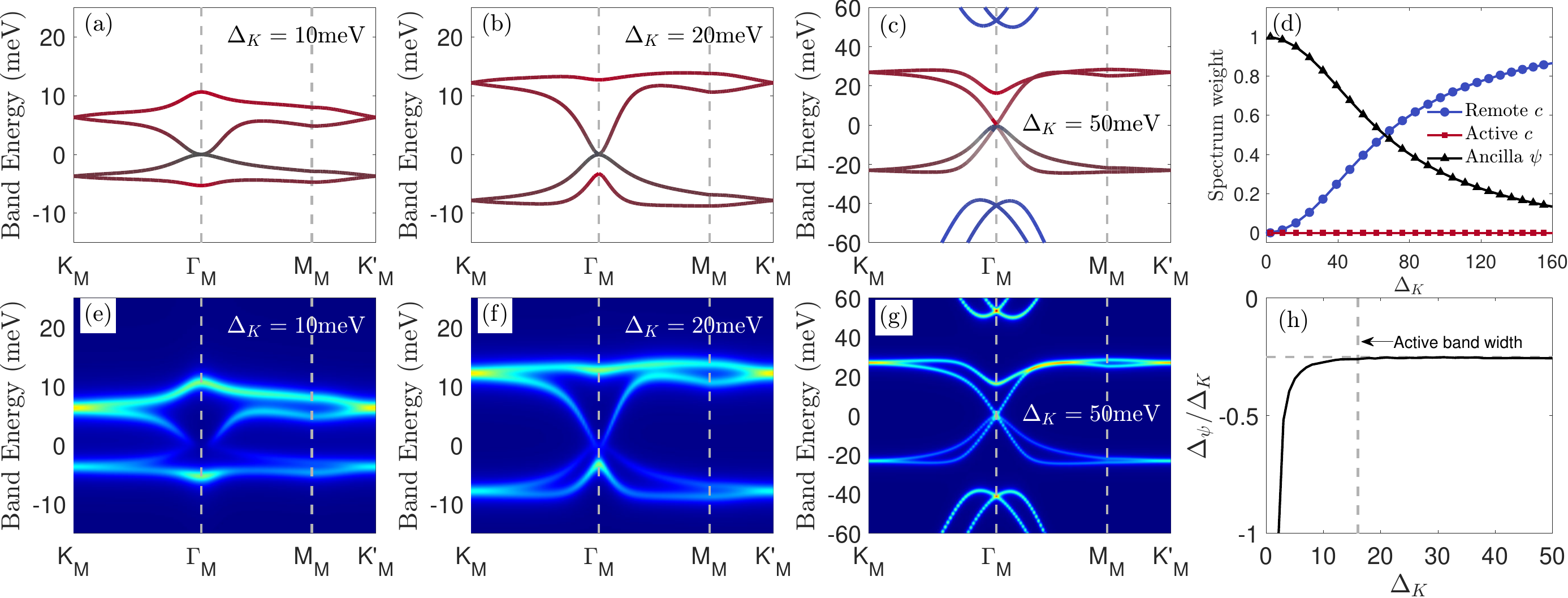}
    \caption{Mott bands calculated by BM model for $\nu = -1,\theta= 1.14^\circ,w_0/w_1=0.8$, with different hybridization strengths at (a) $\Delta_K = 10$ meV, (b) $\Delta_K=20$ meV and (c) $\Delta_K=50$ meV,
    where $\Delta_K$ is defined as the $K_M$ point gap. 
    The blue, red and dark colors of the dispersion line represent the weight of the active band, the remote bands and the ancilla bands, respectively. 
    The corresponding physical electron spectrum function $A_c(\bfk,\omega)$ is shown in (e)-(g), with a broadening chosen as $\eta=1$ meV.  
    (d) shows the $\Gamma_M$ point spectrum weight coming from the active, remote and ancilla bands as a function of $\Delta_K$. 
    And (h) shows the relative chemical potential shift $\Delta_\psi/\Delta_K$ as a function of $\Delta_K$, which goes to a constant $-1/4$ at large $\Delta_K$. }
    \label{fig:nu-1sBM_app}
\end{figure*}

\begin{figure*}[ht]
    \centering
    \includegraphics[width=0.95\linewidth]{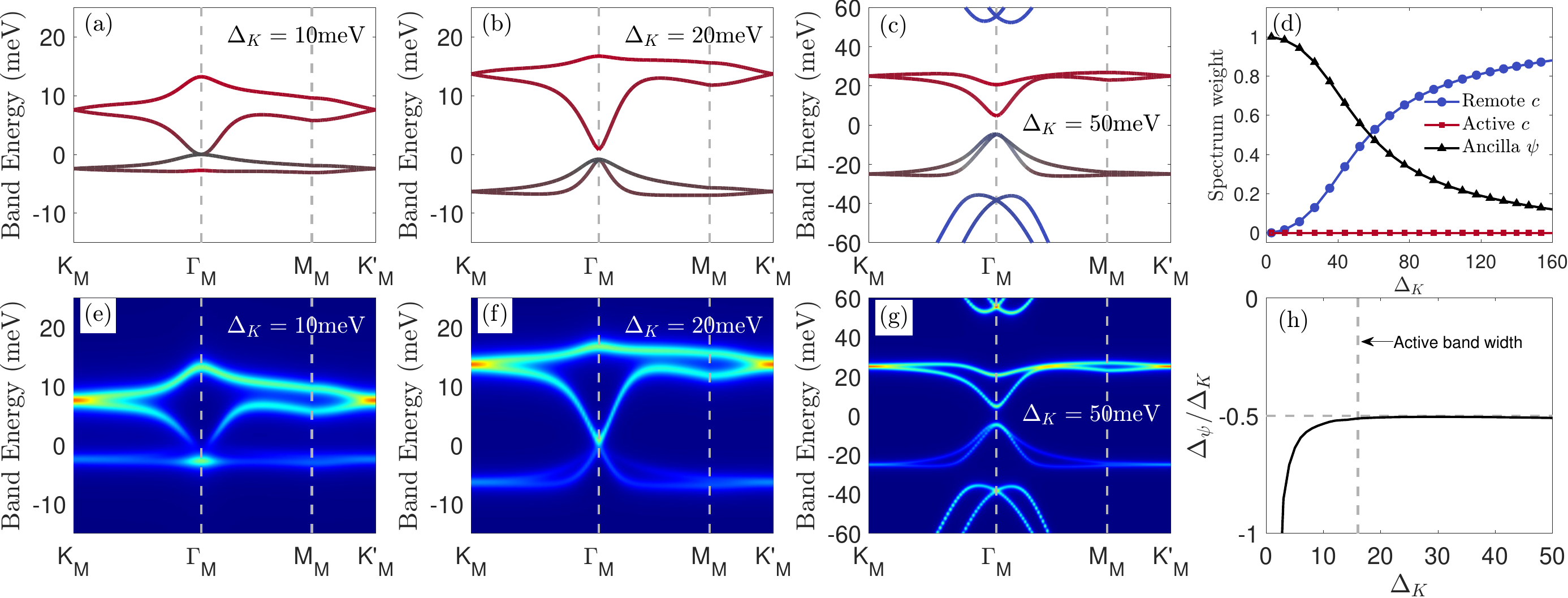}
    \caption{Mott bands calculated by BM model for $\nu = -2,\theta= 1.14^\circ,w_0/w_1=0.8$, with different hybridization strength at (a) $\Delta_K = 10$ meV, (b) $\Delta_K=20$ meV and (c) $\Delta_K=50$ meV,
    where $\Delta_K$ is defined as the $K_M$ point gap. 
    The blue, red, and dark colors of the dispersion line represent the weight of the active band, the remote bands and the ancilla bands, respectively. 
    The corresponding physical electron spectrum function $A_c(\bfk,\omega)$ is shown in (e)-(g), with a broadening chosen as $\eta=1$ meV. 
    (d) shows the $\Gamma_M$ point spectrum weight coming from the active, remote and ancilla bands as a function of $\Delta_K$. 
    And (h) shows the relative chemical potential shift $\Delta_\psi/\Delta_K$ as a function of $\Delta_K$, which goes to a constant $-1/2$ at large $\Delta_K$. }
    \label{fig:nu-2sBM_app}
\end{figure*}

\begin{figure*}[ht]
    \centering
    \includegraphics[width=0.95\linewidth]{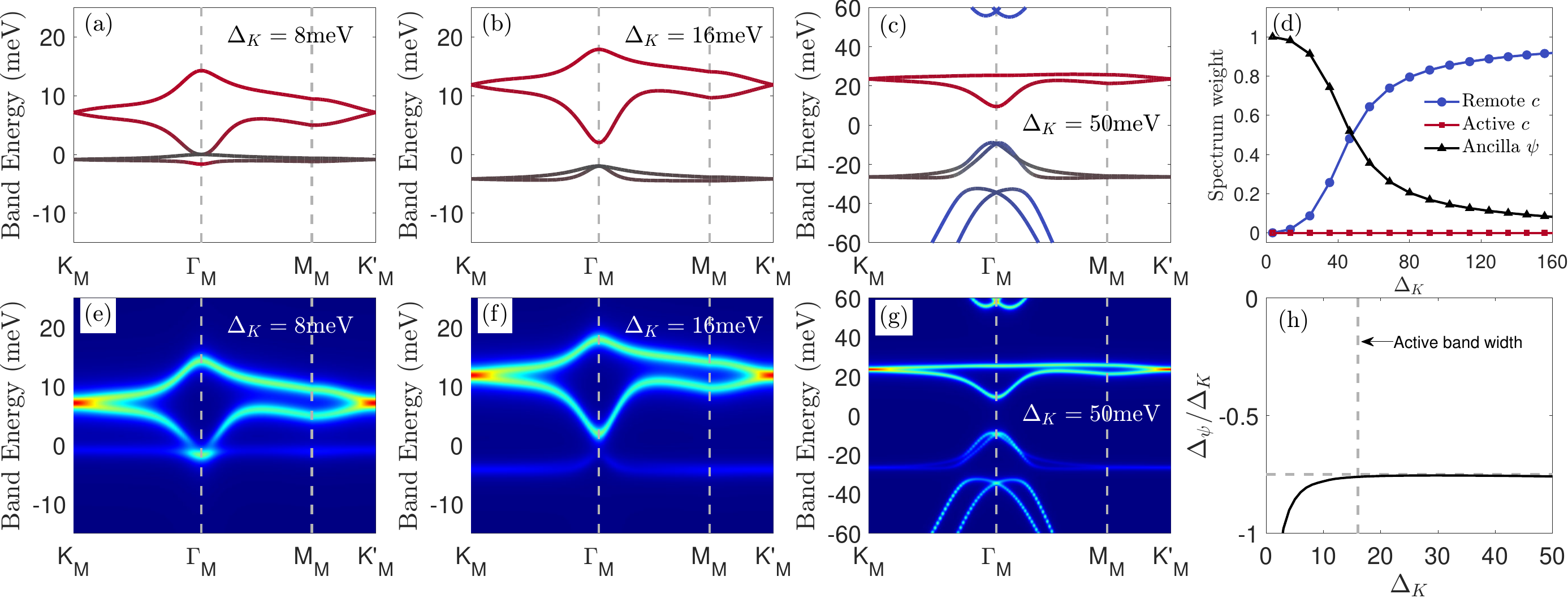}
    \caption{Mott bands calculated by BM model for $\nu = -3,\theta= 1.14^\circ,w_0/w_1=0.8$, with different hybridization strengths at (a) $\Delta_K = 8$ meV, (b) $\Delta_K=16$ meV and (c) $\Delta_K=50$ meV, 
    where $\Delta_K$ is defined as the $K_M$ point gap. 
    The blue, red, and dark colors of the dispersion line represent the weight of the active band, the remote bands and the ancilla bands, respectively. 
    The corresponding physical electron spectrum function $A_c(\bfk,\omega)$ are shown in (e)-(g), with a broadening chosen as $\eta=1$ meV. 
    (d) shows the $\Gamma_M$ point spectrum weight coming from the active, remote and ancilla bands as a function of $\Delta_K$. 
    And (h) shows the relative chemical potential shift $\Delta_\psi/\Delta_K$ as a function of $\Delta_K$, which goes to a constant $-3/4$ at large $\Delta_K$. }
    \label{fig:nu-3sBM_app}
\end{figure*}

\begin{figure*}[ht]
    \centering
    \includegraphics[width=0.95\linewidth]{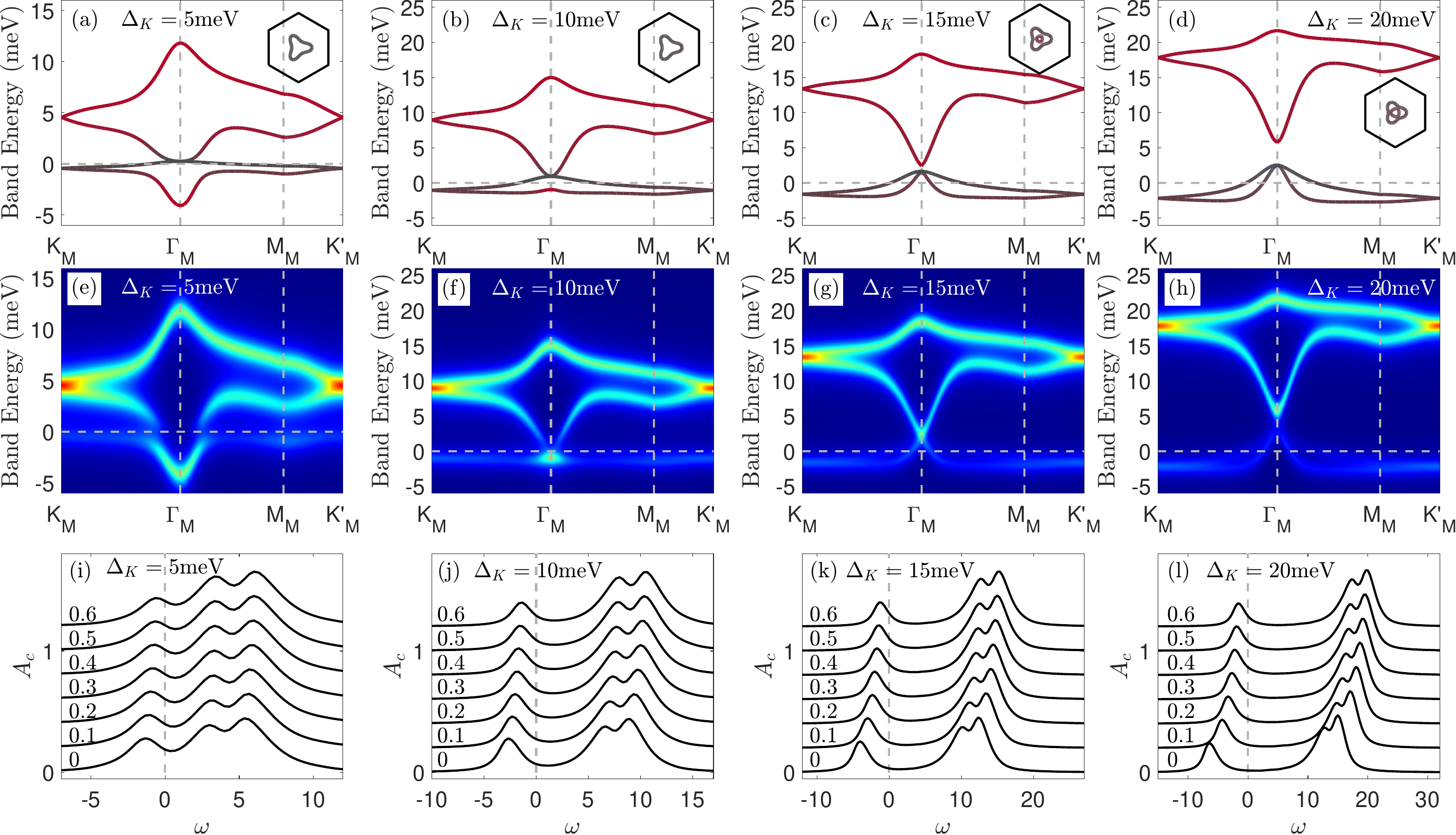}
    \caption{Mott bands calculated by BM model for $\nu = -2-0.4, \theta= 1.14^\circ,w_0/w_1=0.8$, with different hybridization strengths at (a) $\Delta_K = 5$ meV, (b) $\Delta_K=10$ meV (c) $\Delta_K=15$ meV and (d) $\Delta_K=20$ meV, 
    where $\Delta_K$ is defined as the $K_M$ point gap. 
    The blue, red and dark colors of the dispersion line represent the weight of the active band, the remote bands and the ancilla bands, respectively. 
    The corresponding Fermi surface are shown in the insets of (a)-(d), where the color of the Fermi surface has the same meaning of the bands.
    In (e)-(h), we show the physical electron spectrum function $A_c(\bfk,\omega)$ in correspondence with (a)-(d), 
    and in (i)-(l), we show the corresponding STM spectrum $A_c(\omega)=\sum_\bfk A_c(\bfk,\omega)$
    , with a broadening chosen as $\eta=1$ meV. }
    \label{fig:nu-2-xBM_app}
\end{figure*}

\begin{figure*}[ht]
    \centering
    \includegraphics[width=0.95\linewidth]{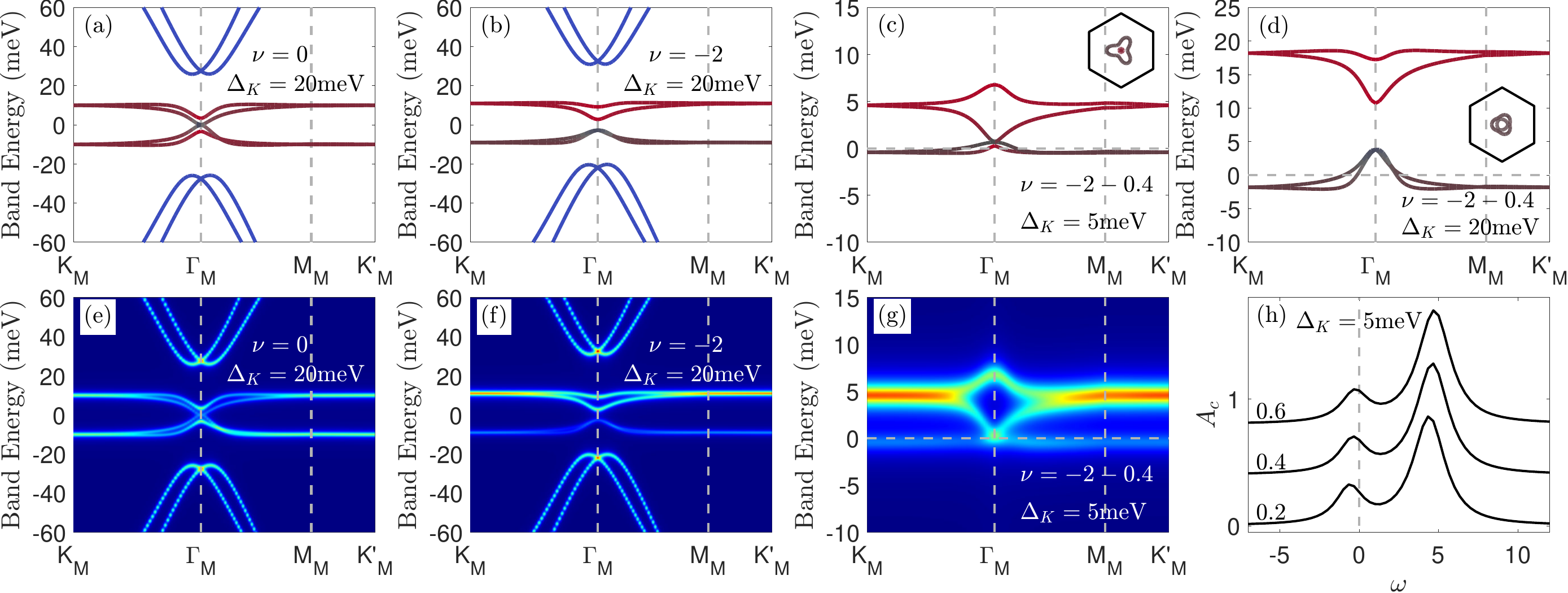}
    \caption{Mott bands calculated by BM model for the magic angle $\theta= 1.05^\circ$ and $w_0/w_1=0.8$ for 
    (a) and (e) $\nu=0, \Delta_K=20$ meV, 
    (b) and (f) $\nu=-2,\Delta_K=20$ meV, 
    and for the doped case 
    (c) and (g) $\nu=-2-0.4,\Delta_K=5$ meV
    and (d) $\nu=-2-0.4, \Delta_K = 20$ meV, with the Fermi surface shown in the insets of (c) and (d). 
    Again $\Delta_K$ is defined as the $K_M$ point gap.  
    The colors in the dispersion have the same meaning as in Fig.~\ref{fig:filling=0} (a) and the broadening is chosen as $\eta = 1$ meV. 
    In (h), we show the STM spectrum $A_c(\omega)=\sum_\bfk A_c(\bfk,\omega)$ for the doped system $\nu = -2-x$, and $x=0.2,0.4,0.6$, with a slightly vertical shift for each $x$. }
    \label{fig:magicBM_app}
\end{figure*}
\begin{figure*}[ht]
    \centering
    \includegraphics[width=0.95\linewidth]{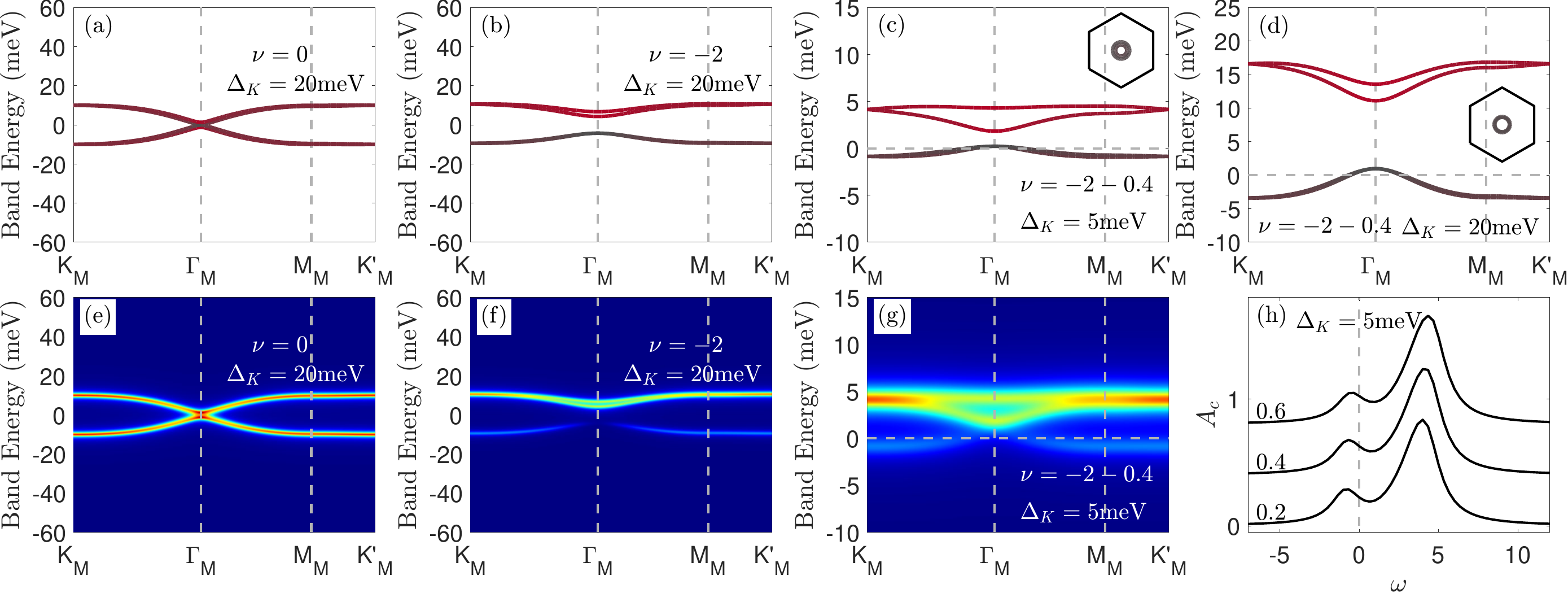}
    \caption{Mott bands calculated by BM model for the magic angle $\theta= 1.05^\circ$ and chiral limit $w_0/w_1=0$ for 
    (a) and (e) $\nu=0, \Delta_K=20$ meV, 
    (b) and (f) $\nu=-2,\Delta_K=20$ meV, 
    and for the doped case 
    (c) and (g) $\nu=-2-0.4,\Delta_K=5$ meV
    and (d) $\nu=-2-0.4, \Delta_K = 20$ meV, with the Fermi surface shown in the insets of (c) and (d). 
    Again $\Delta_K$ is defined as the $K_M$ point gap.  
    The colors in the dispersion have the same meaning as in Fig.~\ref{fig:filling=0} (a) and the broadening is chosen as $\eta = 1$ meV. 
    In (h), we show the STM spectrum $A_c(\omega)=\sum_\bfk A_c(\bfk,\omega)$ for the doped system $\nu = -2-x$, and $x=0.2,0.4,0.6$, with a slightly vertical shift for each $x$. }
    \label{fig:chiralBM_app}
\end{figure*}

Writing together the active band $c$ and ancilla band $\psi$ as $\Psi^\dagger_\bfk = (c_{\bfk;+}^\dagger, \psi_{\bfk;+}^\dagger, c_{\bfk;-}^\dagger, \psi_{\bfk;-}^\dagger)$ 
and discarding $\delta\Phi(\bfk)$ in Eq.~\eqref{eqn:hybrid_proj_BM}, 
the full low-energy Hamiltonian near the $\Gamma_M$ point then takes the following form: 
\begin{equation}\label{eqn:ABgraphene}
\begin{aligned}
    H_{\Psi} = &\sum_\bfk \Psi^\dagger_\bfk\left[ \begin{matrix} 
    -\Delta_\psi/2 & \Phi(\bfk) & M & 0 \\  
    \Phi^*(\bfk) &  \Delta_\psi/2 &0 & 0 \\  
    M & 0 & -\Delta_\psi/2 & \Phi^*(\bfk)\\
    0 & 0 & \Phi(\bfk) &\Delta_\psi/2\\
    \end{matrix}\right]\Psi_\bfk \\
    &-\mu\sum_\bfk \Psi^\dagger_\bfk\Psi_\bfk^{}~,
\end{aligned}
\end{equation}
which can be seen as an AB-stacking graphene model at the $K$ point, with $c_+$ and $\psi_+$ forming the first layer and $c_-$, $\psi_-$ forming the second layer, as already illustrated in the main text Fig.~\ref{fig:ABG_illu_main} (a). 
Dispersion of this model can be directly solved as
\begin{equation}
     \epsilon(\bfk) =\left\{ \begin{aligned}
         &\pm (M/2 + \sqrt{(M\mp\Delta_\psi)^2/4+|\Phi(\bfk)|^2})\\
        &\pm (M/2 - \sqrt{(M\mp\Delta_\psi)^2/4+|\Phi(\bfk)|^2})
     \end{aligned}
    \right.
\end{equation}

Given the above analytical form, we 
can calculate the self-energy of the active bands near the $\Gamma_M$ point.  
Without the ancilla bands, the Green's function is simply: 
\begin{equation}
    G_0^{-1} = \left(\begin{matrix}
        \omega+\mu & -M\\
        -M & \omega+\mu
    \end{matrix}\right)~.
\end{equation}
On the other hand, with ancilla, the Greens' function can be calculated as
\begin{equation}
    [G]_{cc}^{-1} =  \left(\begin{matrix}
    \frac{(\omega+\mu)^2 - |\Phi(\bfk)|^2 - \Delta_\psi^2/4}{\omega+\mu-\Delta_\psi/2}& -M\\
    -M & \frac{(\omega+\mu)^2 - |\Phi(\bfk)|^2 - \Delta_\psi^2/4}{\omega+\mu-\Delta_\psi/2}
    \end{matrix}\right)~,
\end{equation}
where $[G]_{cc}$ is the Green's function projected to the active $c$ channel. 
 The self-energy can then be calculated as
\begin{equation}
    \Sigma(\omega,\bfk) = G_0^{-1}-[G]_{cc}^{-1} = -\Delta_\psi/2+
     \frac{|\Phi(\bfk)|^2}{\omega+\mu-\Delta_\psi/2}~.
\end{equation}

\subsection{$\nu=0$}

At $\nu=0$, $\Delta_\psi=\mu=0$ and thus we can easily get the Hubbard bands around the $\Gamma_M$ point to be:
\begin{equation}
    \epsilon(\bfk) \approx
    \left\{\begin{aligned}
        &\pm( M+ |\Phi_{\Gamma_M}\bfk|^2/M)\\ &\pm |\Phi_{\Gamma_M}\bfk|^2/M
    \end{aligned}\right.~.
\end{equation}
The first line are the active bands separated by $2M$, 
and the second line are the ancilla bands lying between the active bands. 

The self energy in this case is simply $\Sigma(\omega,\bfk) = |\Phi(\bfk)|^2/\omega$, which gives a gap proportional to $|\Phi(\bfk)|\sim\Phi_{\Gamma_M}|\bfk|$ that vanishes at $\Gamma_M$ point. 
Deviating from $\Gamma_M$ point, we can use the fit in Fig.~\ref{fig:Phik_BM}, which gives 
$\Sigma(\omega,\bfk)\approx \frac{\Delta_K^2}{4\omega}\frac{|\bfk|^2}{|\bfk|^2+s^2}$, consistent with the result in \cite{Ledwith2024}.

\subsection{Other integers $\nu=\pm1,\pm2,\pm3$}

Deviating from the charge neutrality point, the chemical potentials $\mu$ and $\Delta_\psi$ are generally nonzero. 
If we assume that charge doping are most contributed by electrons away from $\Gamma_M$ point, 
we can approximate $\Delta_\psi\approx \nu\Delta_K/4$ as discussed in Appendix.~\ref{app:ancilla}. 
In this case, there will be a phase transition from semimetal to insulator as we increase the Hubbard interaction $\Delta_K\sim U$. 
The overall chemical potential $\mu$ needs to be determined self-consistently such that $c$ and $\psi$ together is half-filling. 

For small Hubbard interaction, $|\Delta_\psi| < M$, the relative shift of ancilla $\psi$ bands and active $c$ bands
is still small and the system remains a semimetal phase, as shown in Fig.~\ref{fig:ABG_illu}. 
On the other hand, as $\Delta_K$ increases, the chemical potential shift $\Delta_\psi$ will finally be larger than the active band width $|\Delta_\psi|>M$, and therefore goes to an insulator phase. 
For both case, the dispersions around $\Gamma_M$ point can be approximated as
\begin{equation}
\epsilon(\bfk) \approx \left\{
\begin{aligned}
    &\pm M -\Delta_\psi/2\pm\frac{|\Phi_{\Gamma_M}\bfk|^2}{M\mp\Delta_\psi}\\
    & \Delta_\psi/2 \pm \frac{|\Phi_{\Gamma_M}\bfk|^2}{M\pm \Delta_\psi}~,
\end{aligned}
\right.
\end{equation}


Of particular interest is the transition point, where $|\Delta_\psi|=M$, say $M+\Delta_\psi=0$. 
The above approximation becomes wrong due to zero in the denominator. 
Now near the $\Gamma_M$ point, 
\begin{equation}
\epsilon(\bfk) \approx \left\{
\begin{aligned}
    &3M/2+\frac{|\Phi_{\Gamma_M}\bfk|^2}{2M}\\
    &-M/2-\frac{|\Phi_{\Gamma_M}\bfk|^2}{2M}\\
    & -M/2 \pm |\Phi_{\Gamma_M}\bfk|
\end{aligned}
\right.
\end{equation}
Here three bands touch at the same point at $\bfk=0, \epsilon(0)=-M/2$, with two of them forming a linear touching 
and another band still having a quadratic dispersion. 

The linear dispersion can be understood by projecting the $c$ orbital in Eq.~\eqref{eq:cpsi_app} into $\frac{1}{\sqrt{2}}(c_+-c_-)$: 
\begin{equation}
\begin{aligned}
    H_{\tilde{\Psi}}  = &\sum_\bfk \tilde\Psi^\dagger_\bfk\left[ \begin{matrix} 
     -\frac{|\Phi(\bfk)|^2}{4M} &  \frac{\Phi^*(\bfk)}{\sqrt{2}} &-\frac{\Phi^*(\bfk)^2}{4M} \\  
     \frac{\Phi(\bfk)}{\sqrt{2}} & 0 & -\frac{\Phi^*(\bfk)}{\sqrt{2}}\\
    -\frac{\Phi(\bfk)^2}{4M} & -\frac{\Phi(\bfk)}{\sqrt{2}}  & -\frac{|\Phi(\bfk)|^2}{4M} 
    \end{matrix}\right]\tilde\Psi_\bfk~,
\end{aligned}
\end{equation}
where we set the chemical potential to be $\mu=\Delta_\psi/2$, and $\tilde\Psi_\bfk^\dagger = (\psi_+^\dagger,\frac{1}{\sqrt{2}}(c_+^\dagger-c_-^\dagger),\psi_-^\dagger)$. 
One of the two $\psi$ orbitals hybridizes with a single $\frac{1}{\sqrt{2}}(c_+-c_-)$ orbital and forms a Dirac cone, 
while the other orbital remains decoupled and only gets a quadratic correction after considering the band repulsion between the $\frac{1}{\sqrt{2}}(c_++c_-)$ band.

\section{Results from ancilla theory in continuum model}

In this appendix we show more results of the charge sector in ancilla theory based on the continuum model Eq.~\eqref{eqn:MFBM_app}. 
We show the Mott bands for the integer filings $\nu=0, -1, -2,-3$, the doped case $\nu=-2-x$, and the results for the magic angle $\theta=1.05^\circ$ and chiral limit $w_0/w_1=0$. 
For all of the results, we use the $K_M$ point band gap $\Delta_K$ (instead of the microscopic parameter $\Phi_0$ in Eq.~\eqref{eqn:MFBM_hyb}) to characterize the Mott gap size, 
which has similar physical meaning with the $\Delta_{\mathrm{Mott}}$ defined in Eq.~\eqref{eqn:deltanu} in Appendix.~\ref{app:ancilla} for conventional Hubbard model. 
Note also that we always define $\Delta_K$ as the band gap between two lowest $p$-wave orbital when band crossing happens, as the $s$-wave orbital has zero overlap with the $\psi$ orbital and does not contribute to the Mott physics.

\subsection{Integer fillings}

We plot the Mott bands and physical spectral functions $A_c(\bfk,\omega)$ for integer fillings 
$\nu=0$ in Fig.~\ref{fig:nu0sBM_app}, $\nu=-1$ in Fig.~\ref{fig:nu-1sBM_app}, $\nu=-2$ in Fig.~\ref{fig:nu-2sBM_app} and $\nu=-3$ in Fig.~\ref{fig:nu-3sBM_app}.
The $\nu> 0$ results are not shown here due to the particle-hole symmetry of our model. 

For the charge neutrality point, we show the semimetal bands for more interaction strength $\Delta_K = 10$ meV, $20$ meV, $50$ meV and $100$ meV. 
The $\Gamma_M$ point remain gapless regardless of the Mott interaction $\Delta_K$. 
The $\Gamma_M$ effective mass decreases and remote band weight increases as $\Delta_K$ increases, consistent with our conclusion in the main text. 

For other integer filling,  both the $\nu=-1$ and $\nu=-3$ case goes through a semimetal to insulator transition as the interaction $\Delta_K$ is gradually increased, similar to the $\nu=-2$ results. 
The gapless mode at $\Gamma_M$ point in semimetal phase is mainly contributed by $\psi$ band and  is almost invisible in the spectrum function. 
Upon increasing $\Delta_K$, the $\Gamma_M$ point spectrum weight gradually increases when the remote bands comes in. 
The main difference for different integer fillings is that the $\Gamma_M$ point gap is much weaker for $\nu=-1$ and much stronger for $\nu=-3$, as also shown in Fig.~\ref{fig:filling=-2} (d) in the main text.  
We therefore expect a more robust semimetal phase for $\nu=-1$. 

As discussed in the main text, the gap at $\Gamma_M$ point is not directly from the Mott interaction $U$, but from a Chemical potential shift of the ancilla and active band $\Delta_\psi$. 
Here we plot $\Delta_\psi/\Delta_K$ as a function of $\Delta_K$ for different dopings in Fig.~\ref{fig:nu-1sBM_app} (h), Fig.~\ref{fig:nu-2sBM_app} (h) and Fig.~\ref{fig:nu-3sBM_app} (h). 
When the Mott gap  $\Delta_K$ is larger than the active bandwidth $2M$, the chemical potential shift $\Delta_\psi$ satisfies the relation $\Delta_\psi=\nu\Delta_K/4$ for all the integer fillings $\nu=-1,-2,-3$. 
On the other hand, the above relation fails in the semimetal phase when $\Delta_K$ is comparable with $2M$.

\subsection{Pseudogap metal at $\nu=-2-x$}

More results of the pseudogap metal are shown in Fig.~\ref{fig:nu-2-xBM_app} for $\nu = -2-0.4$. 
Similar to the main text results Fig.~\ref{fig:pseudogap}, the Fermi surface are mainly contributed by the ancilla layer $\psi$, 
which almost disappear in the physical electron spectrum $A_c(\bfk,\omega)$. 
A transition from single Fermi surface to two Fermi surfaces happens between $\Delta_K=10$ meV and $\Delta_K=15$ meV, which is in correspondence with the semimetal to insulator transition at inter filling $\nu=-2$. 

\subsection{Magic angle and chiral limit}

Finally, we briefly show the results of magic angle $\theta=1.05^\circ, w_0/w_1=0.8$ in Fig.~\ref{fig:magicBM_app} and the chiral limit $\theta= 1.05^\circ,w_0/w_1=0$ in Fig.~\ref{fig:chiralBM_app}. 
Overall, the results are similar to our main results in $\theta=1.14^\circ, w_0/w_1=0.8$ case, 
but the semimetal phase shrinks due to the narrower bandwidth of the active bands. 
Specifically, the two lower Hubbard bands at $\nu=-2$ and $\nu=-2-x$ coincide with each other in the chiral limit $w_0/w_1=0$ in Figs.~\ref{fig:chiralBM_app} (c) and (d). 
And the double peaks on the positive bias side in Fig.~\ref{fig:pseudogap} (e) becomes a single one in Fig.~\ref{fig:magicBM_app} (h) and Fig.~\ref{fig:chiralBM_app} (h) due to a narrower bandwidth. 

\begin{table}[tb]
    \centering
    \begin{tabular}{lccccc}
        \toprule
        &$C_{2z}T$ & $P$ & $S$ & $C_{3z}$ & $C_{2x}$\\
        \midrule
        $c_{1,\mathbf{k}}$& $\sigma_x\mathcal{K}c_{1,\mathbf{k}}$& $-\mathrm{i}\sigma_z\mathcal{K}c^\dagger_{1,-\mathbf{k}}$ & $-\sigma_z\mathcal{K} c_{1,\mathbf{k}}$ &
        $e^{\mathrm{i}2\pi/3\sigma_z}c_{1,C_{3z}\mathbf{k}}$& $\sigma_x c_{1,C_{2x}\mathbf{k}}$\\
        $c_{2,\mathbf{k}}$& $\sigma_x\mathcal{K}c_{2,\mathbf{k}}$& $-\mathrm{i}\sigma_z\mathcal{K} c^\dagger_{2,-\mathbf{k}}$ & $\sigma_z\mathcal{K} c_{2,\mathbf{k}}$ &
        $\sigma_0c_{2,C_{3z}\mathbf{k}}$& $\sigma_x c_{2,C_{2x}\mathbf{k}}$\\
        $f_\mathbf{k}$ & $\sigma_x\mathcal{K}f_\mathbf{k}$& $\mathrm{i}\sigma_z \mathcal{K}f^\dagger_{-\mathbf{k}}$ & $\sigma_z \mathcal{K}f_\mathbf{k}$ &
        $e^{\mathrm{i}2\pi/3\sigma_z} f_{C_{3z}\mathbf{k}}$& $\sigma_x f_{C_{2x}\mathbf{k}}$ \\
        $\psi_\mathbf{k}$ & $\sigma_x\mathcal{K} \psi_\mathbf{k}$& $-\mathrm{i}\sigma_z \mathcal{K}\psi^\dagger_{-\mathbf{k}}$ & $\sigma_z \mathcal{K}\psi_\mathbf{k}$& $e^{\mathrm{i}2\pi/3\sigma_z} \psi_{C_{3z}\mathbf{k}}$ & $\sigma_x \psi_{C_{2x}\mathbf{k}}$ \\
        \bottomrule
    \end{tabular}
    \caption{Symmetry action on $c,f,\psi$ fermion.}
    \label{table:symmTHFM}
\end{table}

\section{Ancilla theory in the topological heavy fermion model}\label{app:THFM}
Here we review the topological heavy fermion model \cite{Song2022THFM} and the detail of its symmetry transformation. The THFM contains heavy fermion $f$ and two types of itinerant electrons $c_1, c_2$. The free Hamiltonian of THFM is written as:
\begin{equation}\label{eq:THFM_app}
\begin{split}
        H_\mathrm{THFM}=&\sum_{\mathbf{k},\tau,s}\left(v_\star\tau_z\left( k_x\sigma_0+\mathrm{i}k_y\sigma_z\right)c^\dagger_{1,\mathbf{k};\tau s}c_{2,\mathbf{k};\tau s}+\mathrm{H.c.}\right)\\
        &+\sum_{\mathbf{k},\tau,s}c^\dagger_{2,\mathbf{k};\tau s}M\sigma_x c_{2,\mathbf{k};\tau s}+H^{\left(cf\right)}_0,\\
        H^{(cf)}_0=&\frac{1}{\sqrt{N_\mathrm{site}}}\sum_{\mathbf{k},i,\tau,s}\Big(e^{\mathrm{i}\mathbf{k}\cdot\mathbf{R}_i-\frac{k^2\lambda^2}{2}}f^\dagger_{i;\tau s}\big(\gamma\sigma_0\\
        &+v^\prime_\star\left(k_x\sigma_x+k_y\tau_z\sigma_y\right)\big)c_{1,\mathbf{k};\tau s}+\mathrm{H.c.}\Big).
\end{split}
\end{equation}
In our calculation, we choose that $\theta=1.14^\circ$, $v_\star=-4.544$ eV $\overset{\circ}{\mathrm{A}}$, $M=-7.422$ meV, $\gamma=-38.037$ meV, $v^\prime_\star=1.702$ eV $\overset{\circ}{\mathrm{A}}$, $\lambda=0.3375a_M$. Here $i$ represents the AA sites of TBG and $\mathbf{k}$ is summed over the entire momentum space. $f_{i;\tau s}=(f_{i;+\tau s},f_{i;-\tau s})^T$, $c_{1;\tau s}(\mathbf{k})=(c_{1;+\tau s}(\mathbf{k}),c_{1;-\tau s}(\mathbf{k}))^T$, $c_{2;\tau s}(\mathbf{k})=(c_{2;+\tau s}(\mathbf{k}),c_{2;-\tau s}(\mathbf{k}))^T$. $f_\pm$ has angular momentum $L=\pm1$. $\sigma$ represents the Pauli matrix in orbital space. The symmetry action of $f,c_1,c_2$ are listed in Table.~\ref{table:symmTHFM}. The first ancilla fermion $\psi$ stays in the same orbital with $f$, hence they transform in the same way under symmetry action except particle-hole transformation. $f$ and $\psi$ are coupled as:
\begin{equation}
    H_\mathrm{hyb}=\Phi\sum_{i,a,\tau,s}\left(f^\dagger_{i;a\tau s}\psi_{i;a\tau s}+\mathrm{H.c.}\right).
\end{equation}
Under particle-hole transformation, we require that $H_\mathrm{hyb}$ remains invariant at charge neutrality $\nu=0$, then $\psi$ transforms as $\psi_\mathbf{k}\rightarrow-\mathrm{i}\sigma_z\mathcal{K}\psi^\dagger_{-\mathbf{k}}$. The total Hamiltonian can be written as:
\begin{equation}\label{eq:THFM_ancilla_app}
\begin{split}  H^{(cf\psi)}_\mathrm{MF}=&H_\mathrm{THFM}+H_\mathrm{hyb}-\mu(N_{c_1}+N_{c2}+N_f)\\
&-\mu_\psi N_\psi,
\end{split}
\end{equation}
which is Eq. 4 in the main text.

\subsection{Effective model at $\nu=0$}

For $\nu=0$ and in the limit $\gamma\gg M,\Phi$, $c$ and $f_1$ are gapped. The low energy part of  Eq.~\eqref{eq:THFM_ancilla_app} only contains $c_2$ and $\psi$. The effective model can be written as:
\begin{equation}
\begin{split}
    H^{(c_2 f)}_\mathrm{eff}\approx &\frac{\Phi v_\star}{\sqrt{2}\gamma}\sum_{\mathbf{k},\tau,s}\left(c^\dagger_{2,\mathbf{k};\tau s}
    \begin{bmatrix}
        - k & \frac{v^\prime_\star}{\gamma} k^2\\
        \frac{v^\prime_\star}{\gamma} \bar{k}^2 & - \bar{k}
    \end{bmatrix}
   \psi_{\mathbf{k};\tau s} +\mathrm{H.c.}\right)\\
   &+\sum_{\mathbf{k},\tau,s}c^\dagger_{2,\mathbf{k};\tau s} M\sigma_x c_{2,\mathbf{k};\tau s},
\end{split}
\end{equation}
where $k=k_x+\mathrm{i}k_y$. This equation is consistent with the result in BM model Eq.~\eqref{eq:cpsi_app}.

\subsection{Comparison with the calculation in BM model}
\begin{figure*}[ht]
    \centering
    \includegraphics[width=0.95\linewidth]{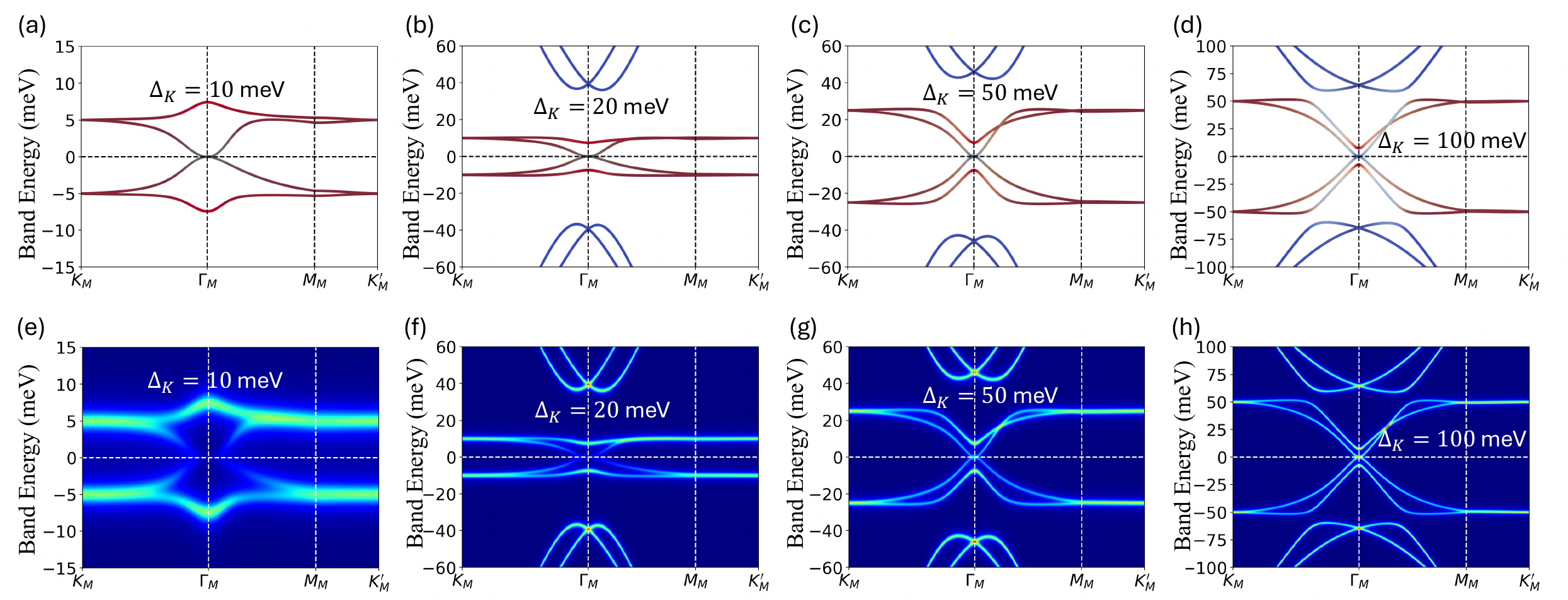}
    \caption{Mott bands calculated by THFM for $\theta=1.14^\circ,\nu = 0$ with different hybridization strength at (a) $\Delta_K = 10$ meV, (b) $\Delta_K=20$ meV and (c) $\Delta_K=50$ meV and (d) $\Delta_K = 100$ meV. The colors has the same meaning as in Figs.~\ref{fig:filling=0} (a) and (b). 
    The corresponding physical electron spectrum function $A_c(\bfk,\omega)$ are shown in (e)-(h). }
    \label{fig:nu0sTHFM_app}
\end{figure*}

\begin{figure*}[ht]
    \centering
    \includegraphics[width=0.95\linewidth]{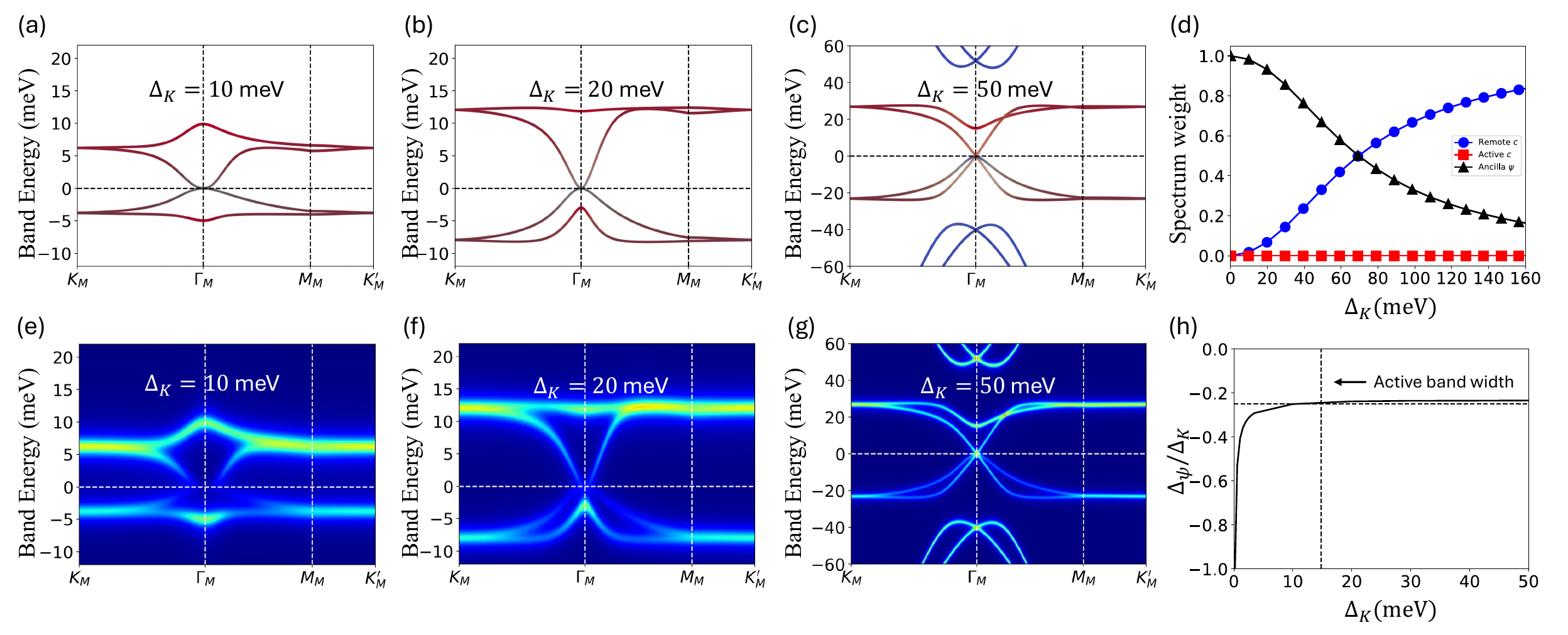}
    \caption{Mott band calculated by THFM for $\theta=1.14^\circ,\nu = -1$ with different hybridization strength at (a) $\Delta_K = 10$ meV, (b) $\Delta_K=20$ meV and (c) $\Delta_K=50$ meV. The colors has the same meaning as in Figs.~\ref{fig:filling=0} (a) and (b). 
    The corresponding physical electron spectrum function $A_c(\bfk,\omega)$ are shown in (e)-(g). 
    (d) shows the $\Gamma_M$ point spectrum weight coming from the active, remote and ancilla bands as a function of $\Delta_K$. 
    And (h) shows the relative chemical potential shift $\Delta_\psi/\Delta_K$ as a function of $\Delta_K$, which goes to a constant $-1/4$ at large $\Delta_K$. }
    \label{fig:nu-1sTHFM_app}
\end{figure*}

\begin{figure*}[ht]
    \centering
    \includegraphics[width=0.95\linewidth]{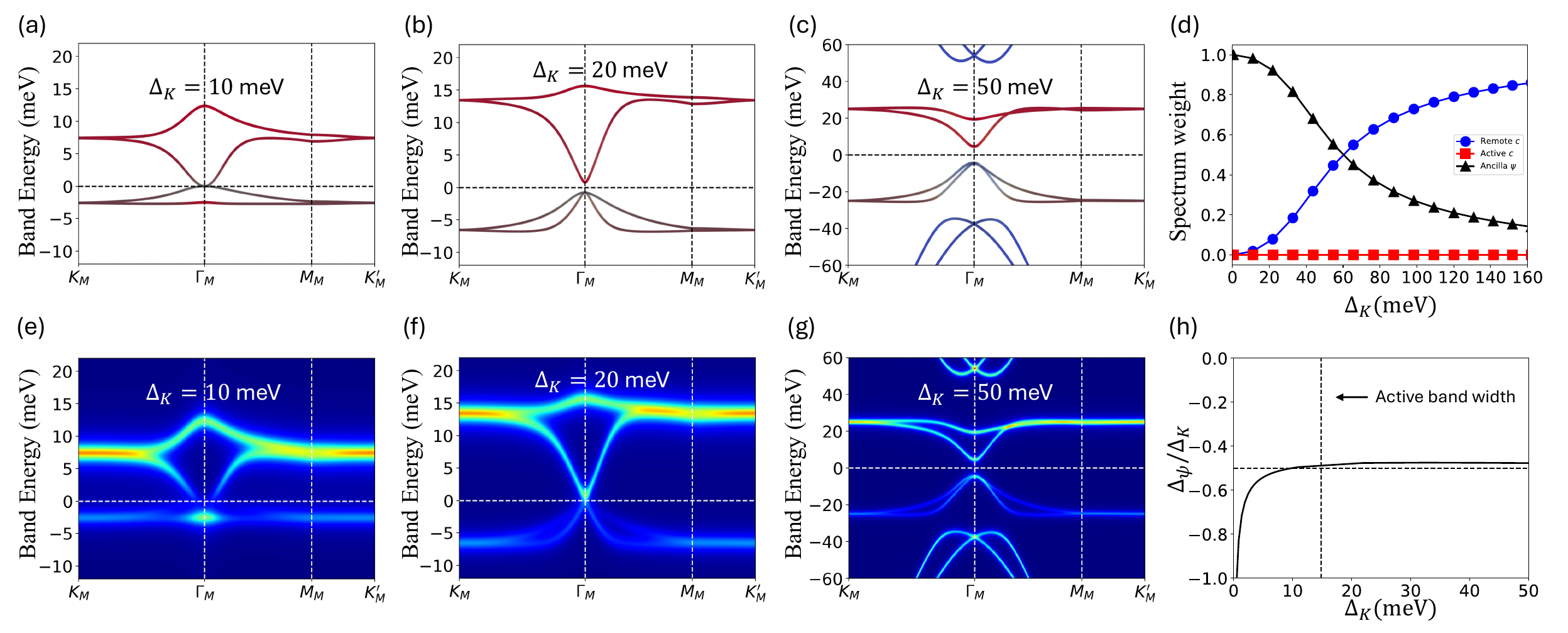}
    \caption{Mott bands calculated by THFM for $\theta=1.14^\circ,\nu = -2$ with different hybridization strength at (a) $\Delta_K = 10$ meV, (b) $\Delta_K=20$ meV and (c) $\Delta_K=50$ meV. The colors has the same meaning as in Figs.~\ref{fig:filling=0} (a) and (b). 
    The corresponding physical electron spectrum function $A_c(\bfk,\omega)$ are shown in (e)-(g). 
    (d) shows the $\Gamma_M$ point spectrum weight coming from the active, remote and ancilla bands as a function of $\Delta_K$. 
    And (h) shows the relative chemical potential shift $\Delta_\psi/\Delta_K$ as a function of $\Delta_K$, which goes to a constant $-1/2$ at large $\Delta_K$. }
    \label{fig:nu-2sTHFM_app}
\end{figure*}

\begin{figure*}[ht]
    \centering
    \includegraphics[width=0.95\linewidth]{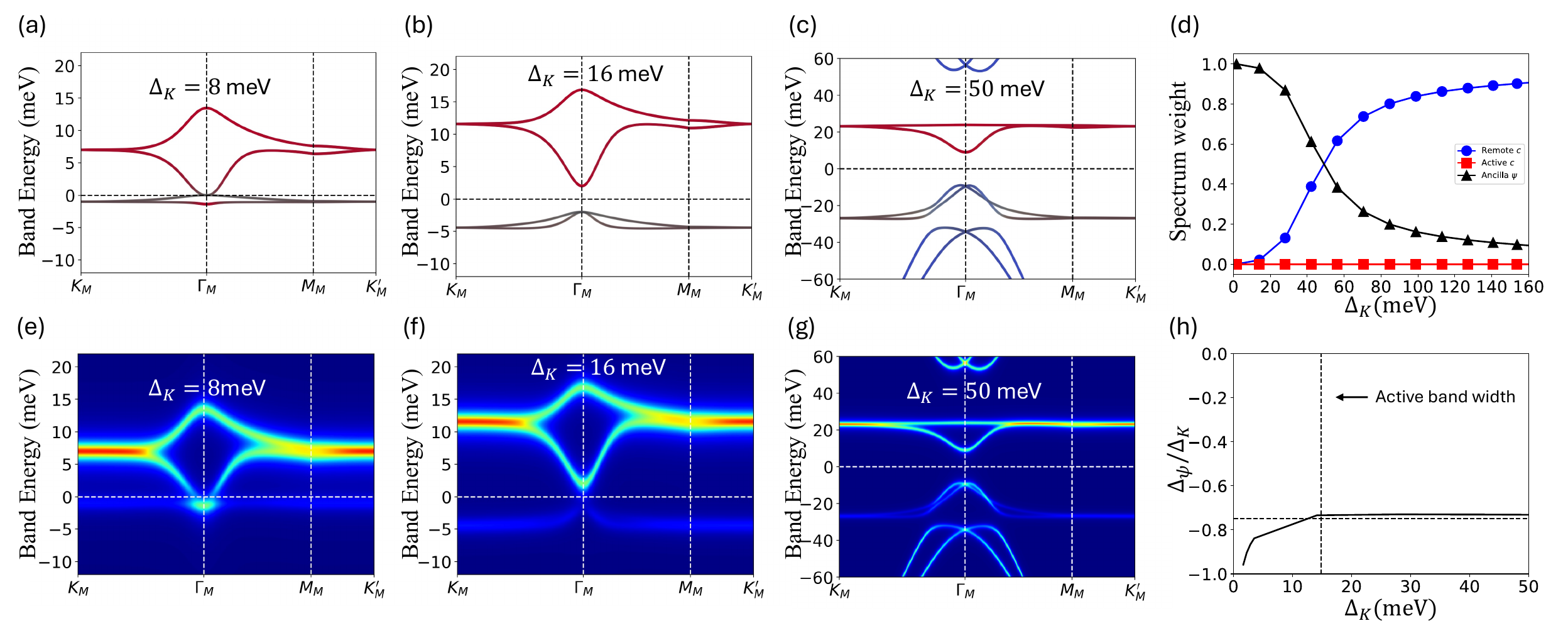}
    \caption{Mott bands calculated by THFM for $\theta=1.14^\circ,\nu = -3$ with different hybridization strength at (a) $\Delta_K = 8$ meV, (b) $\Delta_K=16$ meV and (c) $\Delta_K=50$ meV. The colors has the same meaning as in Figs.~\ref{fig:filling=0} (a) and (b). 
    The corresponding physical electron spectrum function $A_c(\bfk,\omega)$ are shown in (e)-(g). 
    (d) shows the $\Gamma_M$ point spectrum weight coming from the active, remote and ancilla bands as a function of $\Delta_K$. 
    And (h) shows the relative chemical potential shift $\Delta_\psi/\Delta_K$ as a function of $\Delta_K$, which goes to a constant $-3/4$ at large $\Delta_K$. }
    \label{fig:nu-3sTHFM_app}
\end{figure*}

\begin{figure*}[ht]
    \centering
    \includegraphics[width=0.95\linewidth]{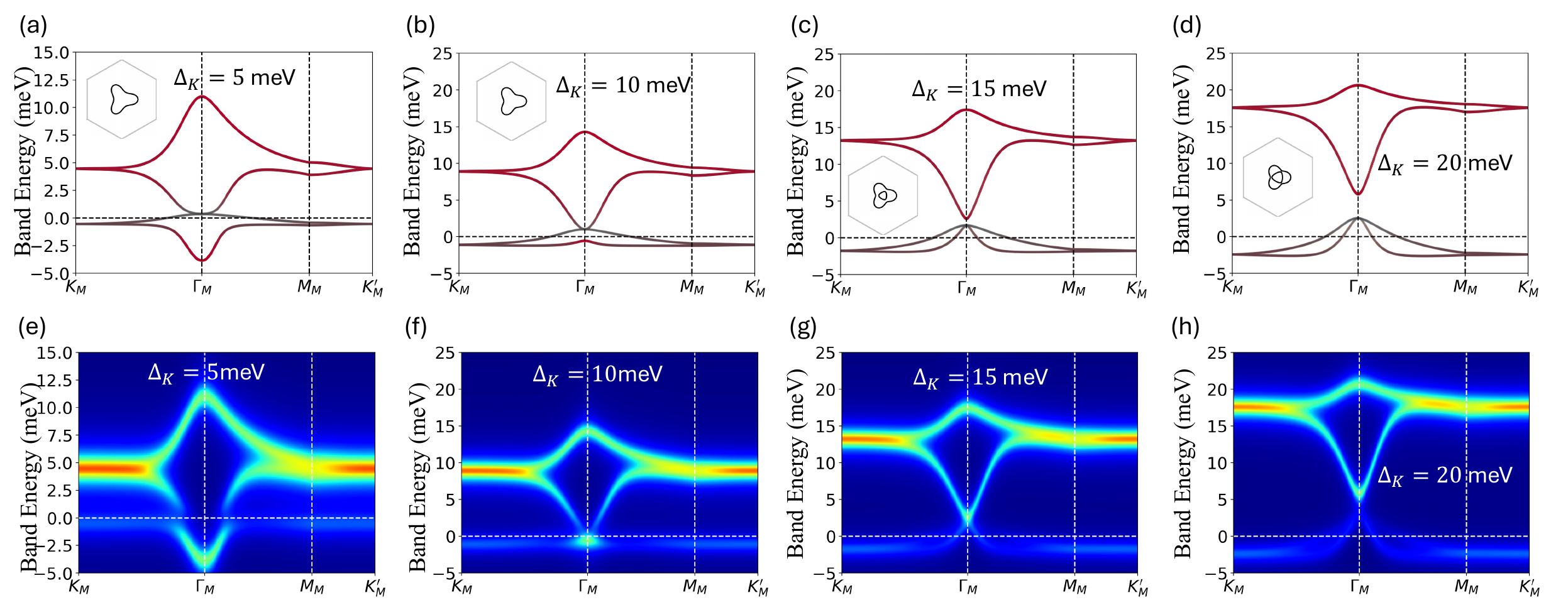}
    \caption{Mott bands calculated by THFM model for $ \theta= 1.14^\circ,\nu = -2-0.4$, with different hybridization strength at (a) $\Delta_K = 5$ meV, (b) $\Delta_K=10$ meV (c) $\Delta_K=15$ meV and (d) $\Delta_k=20$ meV, 
    where $\Delta_K$ is defined as the $K_M$ point gap. 
    The blue, red and dark colors of the dispersion line represent the weight of the active band, the remote bands and the ancilla bands, respectively. The corresponding Fermi surface are shown in the insets of (a)-(d).
    In (e)-(h), we further show the physical electron spectrum function $A_c(\bfk,\omega)$ in correspondence with (a)-(d), with a broading chosen as $\eta=1$ meV. 
  }
    \label{fig:nu-2-xTHFM_app}
\end{figure*}

\begin{figure*}[ht]
    \centering
    \includegraphics[width=0.95\linewidth]{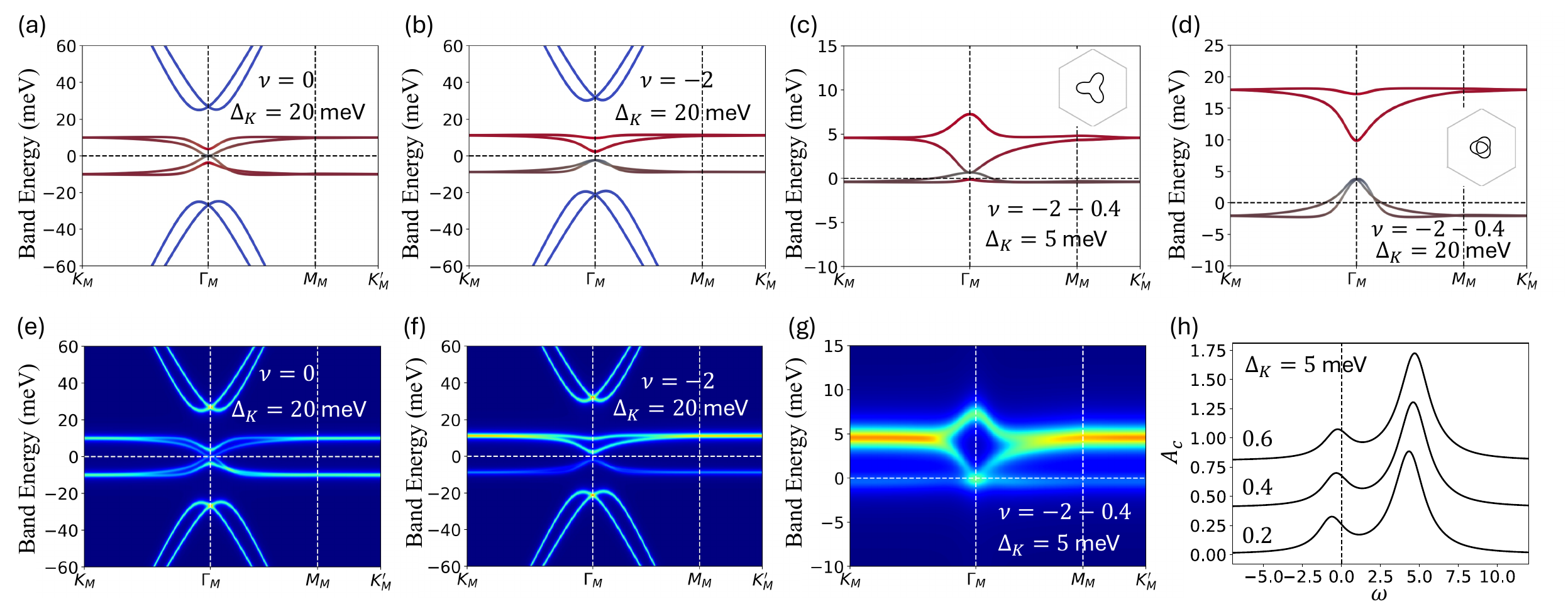}
    \caption{Mott bands calculated by THFM model for the magic angle $\theta= 1.05^\circ$ for 
    (a) and (e) $\nu=0, \Delta_K=20$ meV, 
    (b) and (f) $\nu=-2,\Delta_K=20$ meV, 
    and for the doped case 
    (c) and (g) $\nu=-2-0.4,\Delta_K=5$ meV
    and (d) $\nu=-2-0.4, \Delta_K = 20$ meV, with the Fermi surface shown in the insets of (c) and (d). 
    Again $\Delta_K$ is defined as the $K_M$ point gap. The broading is chosen as $\eta = 1$ meV. 
    In (h), we show the STM spectrum $A_c(\omega)=\sum_\bfk A_c(\bfk,\omega)$ for the doped system $\nu = -2-x$, and $x=0.2,0.4,0.6$, with a slightly vertical shift for each $x$. In this calculation, we choose that $v_\star=-4.303$ eV $\overset{\circ}{A}$, $M=3.697$ meV, $\gamma=-24.75$ meV, $v^\prime_\star=1.622$ eV $\overset{\circ}{A}$, $\lambda=0.3375a_M$. }
    \label{fig:magicTHFM_app}
\end{figure*}

\begin{figure*}[ht]
    \centering
    \includegraphics[width=0.95\linewidth]{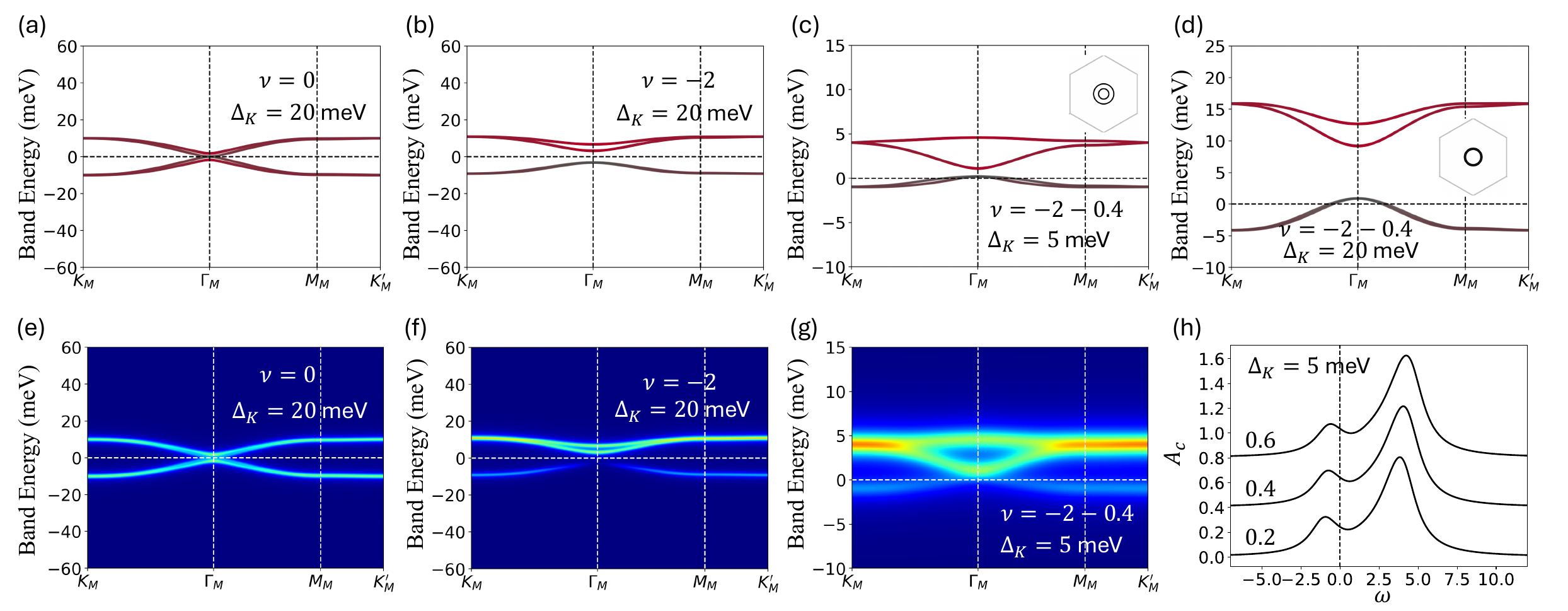}
    \caption{Mott bands calculated by THFM model for the magic angle $\theta= 1.05^\circ$ in the chiral limit for 
    (a) and (e) $\nu=0, \Delta_K=20$ meV, 
    (b) and (f) $\nu=-2,\Delta_K=20$ meV, 
    and for the doped case 
    (c) and (g) $\nu=-2-0.4,\Delta_K=5$ meV
    and (d) $\nu=-2-0.4, \Delta_K = 20$ meV, with the Fermi surface shown in the insets of (c) and (d). 
    Again $\Delta_K$ is defined as the $K_M$ point gap.  
   The broading is chosen as $\eta = 1$ meV. 
    In (h), we show the STM spectrum $A_c(\omega)=\sum_\bfk A_c(\bfk,\omega)$ for the doped system $\nu = -2-x$, and $x=0.2,0.4,0.6$, with a slightly vertical shift for each $x$. In this calculation, we choose that $v_\star=-5.144$ eV $\overset{\circ}{\mathrm{A}}$, $M=3.697$ meV, $\gamma=-100.79$ meV, $v^\prime_\star=0$, $\lambda=0.3375a_M$.}
    \label{fig:chiralTHFM_app}
\end{figure*}

We present the Mott bands and physical spectral functions $A_c(\omega,\bfk)$ for integer fillings $\nu=0$ in Fig.~\ref{fig:nu0sTHFM_app}, $\nu=-1$ in Fig.~\ref{fig:nu-1sTHFM_app}, $\nu=-2$ in Fig.~\ref{fig:nu-2sTHFM_app}, $\nu=-3$ in Fig.~\ref{fig:nu-3sTHFM_app}, for pseudogap metal at $\nu=-2-0.4$ in Fig.~\ref{fig:nu-2-xTHFM_app}, for the magic angle and chiral limit case in Fig.~\ref{fig:magicTHFM_app}, ~\ref{fig:chiralTHFM_app}. The results are similar to that from the BM model Fig.~\ref{fig:nu0sBM_app}$-$~\ref{fig:chiralBM_app}. This indicates that the ancilla theory is basis independent and we can obtain consistent results from different single particle basis.



\section{Correlated insulator to semimetal transition in $U\gg\gamma$ limit of THFM}

In the main text we discussed a semimetal to insulator transition at filling $\nu=-1,-2,-3$ by increasing $U$ in the regime $U<\gamma$ or $U\sim \gamma$.  It turns out that there is a similar transition in the $U\gg\gamma$ limit, where the THFM offers a good starting point.  Here we will discuss this transition based on simple $\frac{\gamma^2}{U}$ perturbation in THFM and show that the result matches the calculation in the ancilla theory. This supports the correctness of the ancilla theory in the $U\gg\gamma$ limit.

\subsection{$\frac{\gamma^2}{U}$ perturbation theory in THFM}
In the THFM, the total Hamiltonian can be written as:
\begin{equation}\label{eqn:totalHwithoutancilla}
    H=H_0+\frac{U}{2}\sum_{i}\left(n_{f,i}-4\right)^2-\mu_fN_f,
\end{equation}
where $H_0$ is the free term and $i$ represents AA sites. $\frac{U}{2}\left(n_{f,i}-4\right)^2$ is the $SU(8)$ Hubbard interaction of $f$ electrons. $\mu_f$ is tuned to fix $f$ electrons at integer filling $4+\nu$. 
If we only consider $c$ electrons at $\Gamma$ point, $H_0$ can be simplified as:
\begin{equation}\label{eqn:freec1gamma}
    H_0(\Gamma)=\frac{\gamma}{\sqrt{N_\mathrm{site}}}\sum_{i,a,\tau,s}(c^\dagger_{1,\Gamma;a\tau s}f_{i;a\tau s}+\mathrm{H.c.}),
\end{equation}
At integer filling $\nu$, we now perform a perturbation calculation to the order $o(\frac{\gamma^2}{U})$ in the large $U\gg\gamma$ limit. We assume $f$ orbital is Mott localized with total filling $n_f=4+\nu$. The contributions come from two kinds of processes: (1) $c_1\rightarrow f, f\rightarrow c_1$; (2) $f\rightarrow c_1, c_1\rightarrow f$. The total contribution can be written as:
\begin{equation}
\begin{split}
   &-\frac{1}{N_\mathrm{site}}\sum_{i,a,\tau,s}\big(J_{K;1}c^\dagger_{1,\Gamma;a\tau s}f_{i;a\tau s} f^\dagger_{i;a^\prime\tau^\prime s^\prime} c_{1,\Gamma;a^\prime\tau^\prime s^\prime}\\
   &+J_{K;2}f^\dagger_{i;a\tau s}c_{1,\Gamma;a\tau s}c^\dagger_{1,\Gamma;a^\prime\tau^\prime s^\prime}f_{i;a^\prime\tau^\prime s^\prime}\big)\\
   =&\frac{J_{K;1}+J_{K;2}}{N_\mathrm{site}}\sum_i\left(\sum_{\alpha,\beta}S_{c_1,\Gamma;\alpha\beta}S_{f,i;\beta\alpha}+\frac{n_{c_1,\Gamma} n_{f,i}}{8}\right)\\
    &-\left(J_{K;1}n_{c_1,\Gamma}+\frac{J_{K;2}}{N_\mathrm{site}}\sum_{i}n_{f,i}\right),
\end{split}
 \end{equation}
 where $J_{K;1}=\frac{\gamma^2}{U/2-\mu_f+\nu U}$ and $J_{k;2}=\frac{\gamma^2}{U/2+\mu_f-\nu U}$. $S_{c_1,\Gamma;\alpha\beta}=c^\dagger_{1,\Gamma;\alpha}c_{1,\Gamma;\beta}-\frac{\delta_{\alpha\beta}}{8}n_{c_1,\Gamma}$, $S_{f,i;\alpha\beta}=f^\dagger_{i;\alpha}f_{i;\beta}-\frac{\delta_{\alpha\beta}}{8}n_{f,i}$, where $\alpha,\beta=a\tau s$.
  
  The effective Hamiltonian consists of a $SU(8)$ Kondo coupling between $c$ electron and the localized spin moments from $f$ orbital.  But there is an additional energy shift for the $c_1$ state due to the virtual process.
 In the absence of any orbital/valley/spin order of the local $f$ spin moments, the energy shift of $c_1$ can be estimated as:
 \begin{equation}
 \begin{split}
     &\frac{J_{K;1}+J_{K;2}}{8}\left(n_{c_1,\Gamma}-\frac{8J_{K;2}}{J_{K;1}+J_{K;2}}\right)\\
     &\left(\frac{1}{N_\mathrm{site}}\sum_{i}n_{f,i}-\frac{8J_{K;1}}{J_{K;1}+J_{K;2}}\right)+\mathrm{Const.},
\end{split}
 \end{equation}
which contributes an energy shift to $c_1$ electrons at $\Gamma$ point. By substituting $J_{K;1}=\frac{\gamma^2}{U/2-\mu_f+\nu U}$ and $J_{K;2}=\frac{\gamma^2}{U/2+\mu_f-\nu U}$ into the above equation, we can obtain the shift term as:
\begin{equation}
    \frac{\gamma^2\left(\nu U-8(\mu_f-\nu U)\right)}{2\left(U^2-4(\mu_f-\nu U)^2\right)}n_{c_1,\Gamma}.
\end{equation}
In the following we set $\mu_f=\nu U$. At $\nu=0$,  the energy shift vanishes. This is also ensured by the particle hole symmetry at charge neutrality. At other integer fillings $\nu=\pm1,\pm2,\pm3$, The energy shift is $\frac{\gamma^2 \nu}{2U}$ for $c_1$.

$c_1$ and $c_2$ are kept at charge neutrality together. But now there is an energy shift for $c_1$, but not for $c_2$. Therefore, $c$ bands at $\Gamma$ point have four energies: $\{\frac{\gamma^2\nu}{2U},\frac{\gamma^2\nu}{2U},\pm M\}$. Here the first two energies correspond to $c_1$ electrons and the rest corresponds to $c_2$ electrons. When the $c$ electrons are half filling, the gap of the Hubbard bands at integer filling $\nu$ can be estimated as:
\begin{equation}
    \Delta_\Gamma=\left\{
    \begin{array}{rcl}
      & \left\vert\left\vert\frac{\gamma^2\nu}{2U}\right\vert-\left\vert M\right\vert\right\vert   & \left\vert\frac{\gamma^2\nu}{2U}\right\vert>\left\vert M \right\vert  \\
      &0   & \left\vert\frac{\gamma^2\nu}{2U}\right\vert<\left\vert M \right\vert 
    \end{array}\right. 
\end{equation}
One can see that there is a semimetal to insulator transition by reducing $U$ in the $U\gg\gamma$ limit. Its nature is actually quite similar to the transition at $U<\gamma$ limit presented in the main text. Just in the later case the $c_1,c_2,f$  basis is not a good staring point, as the semimetal is formed by the active band $c_{\bfk}$ and ancilla $\psi_\bfk$. But it is still interesting to see a one to one correspondence between the emergent $c,\psi$ bands and the single particle $c_1,c_2$ bands in the $U\gg\gamma$ limit.  In the $U\gg\gamma$ limit, we also find a Kondo coupling between $c_1$ and the local $f$ moments. In the $U<\gamma$ limit, we believe there is also an effective Kondo coupling between the emergent $\psi$ band and the local moments in $\psi'$ layer. In this work we assume the spin moments in $\psi'$ layer are easier thermally fluctuating at finite $T$ or are quenched by an anti-Hund's coupling $J_A$ at $T=0$, so we can ignore the Kondo coupling. We can imagine that the effective Kondo coupling may be important to understand symmetry breaking (ferromagnetic phases) and superconductor phase, which we will investigate in a future work.

\subsection{Benchmark in the ancilla theory}
The above perturbation calculation only applies for $U\gg\gamma$ limit. Our ancilla theory should work in the whole range of $U$. Here we demonstrate that the same result can be obtained in the $U\gg\gamma$ limit of the ancilla theory. 

In the presence of ancilla fermion $\psi$, the mean-field Hamiltonian of the charge sector is:
\begin{equation}\label{eqn:totalHwithancilla}
\begin{split}
    H=&H_0+\Phi\sum_{\mathbf{k},a,\tau,s} \left(f^\dagger_{\mathbf{k};a\tau s}\psi_{\mathbf{k};a\tau s}+\mathrm{H.c.}\right)\\
    &+\frac{\Delta_\psi}{2} (N_\psi-N_f)-\mu (N_f+N_\psi),
\end{split}
\end{equation}
where $\mathbf{k} \in \mathrm{MBZ}$ and $N_\psi$ is the total density operator of $\psi$ electrons, $\Delta_\psi$ is tuned to fix $\frac{1}{N_\mathrm{site}}\langle N_\psi\rangle=4-\nu$. In the large $\Phi$ limit, the dominant term of Eq.~\eqref{eqn:totalHwithancilla} is $\Phi\sum_{\mathbf{k},\alpha,\eta,s} \left(f^\dagger_{\alpha\eta s}(\mathbf{k})\psi_{\alpha\eta s}(\mathbf{k})+\mathrm{H.c.}\right)$, $\frac{\Delta_\psi}{2}( N_\psi-N_f)$ and $\mu_(N_f+N_\psi)$. Hence the density $\langle N_\psi\rangle$ can be calculated as:
\begin{equation}
   \langle N_\psi\rangle=8N_\mathrm{site} \frac{\left(-\Delta_\psi/2+\sqrt{\Phi^2+\Delta_\psi^2/4}\right)^2}{\Phi^2+\left(-\Delta_\psi/2+\sqrt{\Phi^2+\Delta_\psi^2/4}\right)^2}.
\end{equation}
By solving $\langle N_\psi\rangle=N_\mathrm{site}(4-\nu)$, we can obtain:
\begin{equation}\label{eqn:Delta_psi}
    \Delta_\psi=\frac{2\nu}{\sqrt{16-\nu^2}}\Phi.
\end{equation}

If we use the notation of $\Delta_{\mathrm{Mott}}=2\sqrt{\Phi^2+\frac{\Delta_\psi^2}{4}}$, we find $\Delta_\psi=\frac{\nu}{4}\Delta_{\mathrm{Mott}}$ as found in Appendix.~\ref{app:ancilla}.

Then we can include $c_1$ electrons at $\Gamma$ point to estimate the energy shift of $c_1$ electrons. The free term is still Eq.~\eqref{eqn:freec1gamma}. Then the Hamiltonian of $c_1,f,\psi$ can be represented as:
\begin{equation}
 \sum_{a,\tau,s}\begin{pmatrix}
        c^\dagger_{1,\Gamma;a\tau s} & f^\dagger_{\Gamma;a\tau s} & \psi^\dagger_{\Gamma;a\tau s}
    \end{pmatrix}
    h
    \begin{pmatrix}
        c_{1,\Gamma;a\tau s} \\
        f_{\Gamma;a\tau s} \\
        \psi_{\Gamma;a\tau s}
    \end{pmatrix},
\end{equation}
where $h$ is:
\begin{equation}
    \begin{bmatrix}
        0 & \gamma & 0 \\
        \gamma & ·-\frac{\Delta_\psi}{2}-\mu & \Phi \\
        0 & \Phi &\frac{\Delta_\psi}{2}-\mu
    \end{bmatrix}.
\end{equation}
And then by simply second order perturbation theory, we can get the energy shift for $c_1$ band as:
\begin{equation}
    \delta E_{c_1}=\frac{\nu\Delta_\mathrm{Mott}-8\mu}{2\left(\Delta_{\mathrm{Mott}}^2-4\mu^2\right)}.
\end{equation}
The result agrees with that from the $\gamma^2/U$ perturbation calculation if we set $\Delta_{\mathrm{Mott}}=U,\mu=\mu_f-\nu U$. If we choose $\mu=0$, then there is:
\begin{equation}
    \delta E_{c_1}=\frac{\gamma^2\nu}{2\Delta_\mathrm{Mott}}.
\end{equation}

\section{Physical meaning of the ancilla fermion}

We have seen that the lower Hubbard band at $\nu=-2$ is dominated by the ancilla fermion $\psi$ and the small Fermi surface at $\nu=-2-x$ is formed by $\psi$. Here we show that the ancilla fermion $\psi$ represents a many-body excitation and should be interpreted as a composite polaron close to $\bfk=0$. Suppose we have a many body ground state $\ket{G_c}=P_s \ket{G_0}$, where $\ket{G_0}$ is the state in the enlarged Hilbert space with $c,\psi,\psi'$,  $\psi_{\bfk}$ can be related to a physical operator $O_{\bfk}$ from the equation $P_S \psi_{\bfk;\alpha} \ket{G_0}=O_{\bfk;\alpha} P_S \ket{G_0}$. It is easy to see that $O_{\bfk}$ has the same quantum number as $\psi_k$, but may be a complicated composite operator, which actually depends on  the many body state $\ket{G_c}=P_S \ket{G_0}$. Let us restricted to the active bands. Simply from symmetry, we can expand $O_\bfk$ in the form: 
\begin{equation}
    O_{\bfk;\alpha}=\sqrt{Z_\bfk} c_{\bfk;\alpha}+\sum_{\bfk',\mathbf q}F(\mathbf k',\mathbf q)c_{\bfk+\mathbf q;\beta}c^\dagger_{\bfk'+\mathbf q;\beta}c_{\bfk';\alpha'}+...
\end{equation}
Here $\alpha,\beta$ is the SU(8) flavor index.  When $\bfk=0$, $Z_\bfk=0$ because of $C_{3z}$ symmetry. Then the leading order of $O_{\bfk;\alpha}$ is a trion operator. Because most of the states in $\bf k$ space is from the $f$ band in THFM, we can expand $O_{\bfk=0}$ in terms of $f$ orbital as:
\begin{align}
    O_{\bfk=0;\alpha}&= P_0\frac{1}{\sqrt{N_\mathrm{site}}}\sum_i\sum_{\delta \mathbf r_1, \delta \mathbf r_2}Z(\delta \mathbf r_1, \delta \mathbf r_2) f_{\mathbf R_i+ \delta \mathbf r_1;\beta} \notag \\ &f^\dagger_\beta(\mathbf R_i+\delta \mathbf r_2)f_{\alpha}(\mathbf R_i-\delta \mathbf r_1-\delta \mathbf r_2) e^{-i \bf k \cdot \mathbf R_i}P_0+...
\end{align}
where  $P_0$ is the projection operator to the active bands. $i+ \delta$ labels the nearest neighbor site of $i$. $Z(\delta \mathbf r_1, \delta \mathbf r_2)$ is the wavefunction of the trion around the site $i$ and should be s-wave. Here we ignore spin dependence in $Z$ for simplicity.   In terms of the active band electron operator, close to $\bf k=0$, $O_\bfk$ is in the form:

\begin{align}
    O_{\bfk;\alpha}&\sim\frac{1}{N_\mathrm{site}} \sum_{\bfk',\mathbf q}Z(\mathbf k',\mathbf q)\mu_{f;\beta}(\bfk+\mathbf q)\mu_{f;\alpha}(\bfk')\mu_{f;\beta}^*(\mathbf k'+\mathbf q) \notag \\ 
    & c_{\beta;\bfk+\mathbf q}c^\dagger_{\beta;\bfk'+\mathbf q}c_{\bfk';\alpha}
\end{align}
where  $\mu_{f;\alpha}(\bf k)$ is the $f$ component of the Bloch wavefunction of the active band in the flavor $\alpha$. We expect $\mu_f(\mathbf k)\sim k_x \pm ik_y$, thus the two holes and one electron are spread in the whole MBZ except around the $\Gamma$ point. $Z(\mathbf k',\mathbf q) \sim\sum_{\delta \mathbf r_1, \delta \mathbf r_2} Z(\delta \mathbf r_1, \delta \mathbf r_2) e^{-i(\mathbf k'-\mathbf q)\cdot \delta \mathbf r_1} e^{-i(\mathbf k'+\mathbf q)\cdot \delta \mathbf r_2}$ is the Fourier transformation of $Z(\delta \mathbf r_1, \delta \mathbf r_2)$ and is s-wave.  It is easy to check that $O_{\bfk;\pm}$ has angular momentum $L=\pm  1$ around $\bf k=0$, so it can not mix with $c_{\bfk}$ at $\bfk=0$.

\subsection{Gauge theory description}

Lastly, we point out a low energy field theory description of the Mott phase with $\Phi \neq 0$. The constraint from the projection operator $P_S$ can be imposed by a gauge theory.  In this formalism, the ancilla $\psi$ couples to an U(1) gauge field $a_\mu$ and also an SU(8) gauge field $\alpha_\mu$\cite{Zhang2020}. In this sense the ancilla fermion $\psi_{\alpha}$ is really a ghost fermion and is not detectable by any physical probe.

However, the internal gauge fields are higgsed once we have a finite $\Phi$.  For simplicity, let us focus only on the U(1) gauge field. We also introduce a probe field $U(1)$ $A_\mu$, which provides the physical electro-magnetic field.  Thus, the physical electron $c$ couples to $A_\mu$.  Then from the hybridization $\Phi c^\dagger \psi$, we know that the boson $\Phi$ is coupled to $A_\mu-a_\mu$. When we have $\Phi \neq 0$, the condensation higgses $a_\mu-A_\mu$, thus the internal U(1) gauge field is always locked to the physical probing field.  In other words, we always have $a_\mu=A_\mu$ at low energy inside the Mott phase with $\Phi \neq 0$. As a result, the ancilla fermion $\psi$ now couples to $A_\mu$, just like the physical electron.  Although $\psi$ has zero spectral weight, its response to electro-magnetic field is the same as an ordinary electron inside the $\Phi \neq 0$ phase. As discussed above, the ancilla fermion $\psi(\bf k)$ corresponds to a composite  operator at $\bf k=0$ limit. But in the low energy effective theory, we can ignore its internal structure and just treat it as a point-like particle.

\section{Variational wavefunctions at $\nu=0, -2, -2-x$}

\begin{figure*}[ht]
    \centering
    \includegraphics[width=0.95\linewidth]{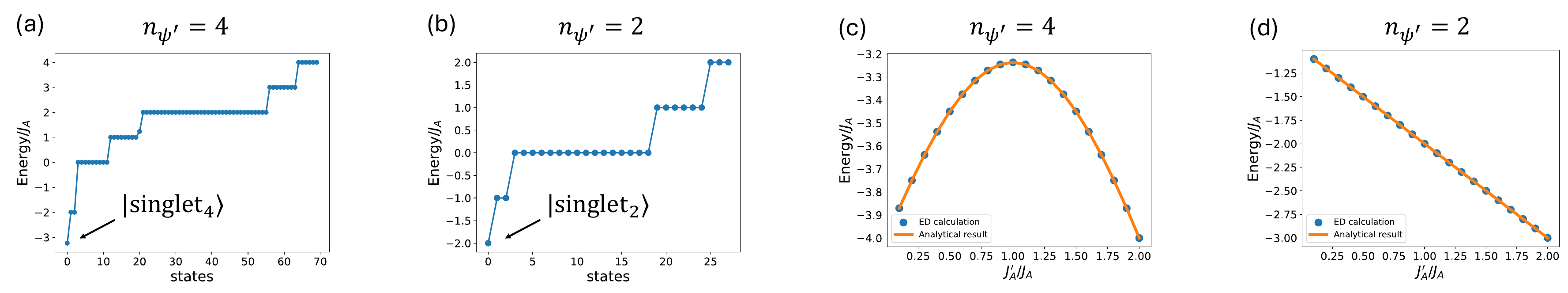}
    \caption{In (a) and (b), the plot is the ED spectrum for the spin Hamiltonian $H_J$ at single site. The number of $\psi^\prime$ is $4$ (a) and $2$ (b) respectively. We choose $J_A=J_A^\prime$ in the numerical calculation. In (c) and (d), we change $\frac{J^\prime_A}{J_A}$ and compare the ground state energy calculated obtained from ED with analytical calculations. The number of $\psi^\prime$ is $4$ (c) and $2$ (d) respectively. }
    \label{fig:ED_spin_singlet}
\end{figure*}

In the main text we mainly focus on the charge sector formed by $c,\psi$. To provide a full model wavefunction, we also need to specify the spin state for $\psi'$, which represents the local moments in the Mott phase. For $\nu=0,2$, we will show that a symmetric and featurless gapped spin state is possible for the local moments when there is a sizable anti-Hund's coupling $J_A$. In this case we can provide a symmetric wavefunction for the Mott state at $\nu=0,-2$ and the symmetric pseudogap metal at $\nu=-2-x$.   Note our wavefunction can be easily formulated within the subspace including only the active bands.

For the symmetric pseudogap metal, it violates the perturbative Luttinger theorem, but is actually consistent with the non-perturbative Oshikawa-Luttinger theorem\cite{oshikawa2000topological}, which  allows two different Fermi liquids for the unusual $U(1)\times U(1) \times SU(2)$ symmetry with separate charge conservation within each valley\cite{zhang2020spin}. One of them is at the non-interacting fixed point, and the second one corresponds to an intrinsically strongly interacting fixed point with Fermi surface volume per flavor smaller by $1/2$ of BZ. Our state corresponds to the second Fermi liquid (sFL),  which is also discussed in trivial moir\'e bands\cite{zhang2020spin} and  in bilayer nickelate\cite{yang2024strong}.  Especially Ref.~\onlinecite{zhang2020spin} already pointed out the possibility of a similar symmetric pseudogap metal when there is a large anti-Hund's coupling $J_A$ from electron-phonon coupling.  For the trivial moir\'e band and the bilayer nickelate, one can describe this second Fermi liquid state using conventional slave boson theory. But for TBG we need to use the ancilla theory.  The ancilla wavefunction constructed below can also be applied to the sFL phase for the trivial moir\'e band and bilayer nickelate.

\subsection{Anti-Hund's spin coupling}

 We assume that the dominant spin interaction is given by the anti-Hund's coupling term, expressed as follows~\cite{Wang2024THFM}:
\begin{equation}\label{eqn:antiHundH}
\begin{split}
    H_J=&\frac{J_A}{4}\sum_{i,\mu,\rho}\left({\psi^\prime}^\dagger_{i;K}\left(\sigma_\mu s_\rho\right)\psi^\prime_{i;K} \right)\left({\psi^\prime}^\dagger_{i;K^\prime}\left(\sigma_\mu s_\rho\right)\psi^\prime_{i;K^\prime}\right)\\
    &+\frac{J^\prime_A}{4}\sum_{i,\rho}
    \left({\psi^\prime}^\dagger_{i;K}\left(\sigma_0 s_\rho\right)\psi^\prime_{i;K}\right)\left({\psi^\prime}^\dagger_{i;K^\prime}\left(\sigma_0 s_\rho\right)\psi^\prime_{i;K^\prime}\right)\\
    &-\frac{J^\prime_A}{4}\sum_{i,\rho}\left({\psi^\prime}^\dagger_{i;K}\left(\sigma_z s_\rho\right)\psi^\prime_{i;K}\right)\left({\psi^\prime}^\dagger_{i;K^\prime}\left(\sigma_z s_\rho\right)\psi^\prime_{i;K^\prime}\right)
\end{split}    
\end{equation}  
where $J_A,J_A^\prime>0$. $\sigma_\mu$, $s_\rho$ are Pauli matrices acting on the subspace $(+,-)$ and $(\uparrow,\downarrow)$ respectively. $\mu,\rho\in\{0,x,y,z\}$. $n^\prime_{i;\tau}=\sum_{as}{\psi^\prime}^\dagger_{i;a\tau s}\psi^\prime_{a\tau s}$ representing the density at site $i$, valley $\tau$. The Hamiltonian respects $U(1)_O\times U(1)_V\times SU(2)_S$ symmetry, where generators are $\{\sigma_z,\tau_z, s_\rho\}$. 

In the following calculations, we demonstrate that the second ancilla layer
$\psi^\prime$ forms as spin singlets at each AA site for even filling $\nu$. We consider two representative cases: $\nu=0$ and $\nu=-2$. We further provide the final ancilla wavefunction at $\nu=0$, $-2$ and $-2-x$.
\subsection{$\nu=0$}
At filling $\nu=0$, there are 4 $\psi^\prime$ fermions per site. Performing an exact digonalization (ED) calculation for a single site yields the energy spectrum shown in Fig.~\ref{fig:ED_spin_singlet} (a). The lowest energy state shows no degeneracy, hence the original symmetry is preserved.  We label this state as $\lvert\mathrm{singlet}_4\rangle$. This state can also be solved analytically. We focus on the sector that $\sigma_z=\tau_z=s^2=0$. The basis for this sector is:
\begin{equation}
    \begin{split}
        \lvert1\rangle=&\frac{1}{\sqrt{3}}\left(G^\dagger_{K;1}G^\dagger_{K^\prime;-1}-G^\dagger_{K;0}G^\dagger_{K^\prime;0}+G^\dagger_{K;-1}G^\dagger_{K^\prime;1}\right)\lvert0\rangle,\\
        \lvert2\rangle=&F^\dagger_{K;1}F^\dagger_{K^\prime;-1}\lvert0\rangle,\\
        \lvert3\rangle=&F^\dagger_{K;0}F^\dagger_{K^\prime;0}\lvert0\rangle,\\
        \lvert4\rangle=&F^\dagger_{K;-1}F^\dagger_{K^\prime;1}\lvert0\rangle,
    \end{split}
\end{equation}
where $F^\dagger$ and $G^\dagger$ are defined as:
\begin{equation}
    \begin{split}
        F^\dagger_{\tau;1}=&{\psi^\prime}^\dagger_{+\tau\uparrow}{\psi^\prime}^\dagger_{+\tau\downarrow},\\
        F^\dagger_{\tau;0}=&\frac{1}{\sqrt{2}}\left({\psi^\prime}^\dagger_{+\tau\uparrow}{\psi^\prime}^\dagger_{-\tau\downarrow}+{\psi^\prime}^\dagger_{-\tau\uparrow}{\psi^\prime}^\dagger_{+\tau\downarrow}\right),\\
        F^\dagger_{\tau;-1}=&{\psi^\prime}^\dagger_{-\tau\uparrow}{\psi^\prime}^\dagger_{-\tau\downarrow},\\
        G^\dagger_{\tau;1}=&{\psi^\prime}^\dagger_{+\tau\uparrow}{\psi^\prime}^\dagger_{-\tau\uparrow},\\
        G^\dagger_{\tau;0}=&\frac{1}{\sqrt{2}}\left({\psi^\prime}^\dagger_{+\tau\uparrow}{\psi^\prime}^\dagger_{-\tau\downarrow}+{\psi^\prime}^\dagger_{+\tau\downarrow}{\psi^\prime}^\dagger_{-\tau\uparrow}\right),\\
        G^\dagger_{\tau;-1}=&{\psi^\prime}^\dagger_{+\tau\downarrow}{\psi^\prime}^\dagger_{-\tau\downarrow}.
    \end{split}
\end{equation}
The Hamiltonian Eq.~\ref{eqn:antiHundH} can be projected into the subspace $(\lvert1\rangle,\lvert2\rangle,\lvert3\rangle,\lvert4\rangle)$. After a transformation of basis for convenience:
\begin{equation}
    \begin{split}
        \lvert\tilde{1}\rangle=&\frac{1}{\sqrt{2}}\left(-\lvert1\rangle+\frac{1}{\sqrt{3}}\left(\lvert2\rangle-\lvert3\rangle+\lvert4\rangle\right)\right),\\
        \lvert\tilde{2}\rangle=&\frac{1}{\sqrt{2}}\left(\lvert1\rangle+\frac{1}{\sqrt{3}}\left(\lvert2\rangle-\lvert3\rangle+\lvert4\rangle\right)\right),\\
        \lvert\tilde{3}\rangle=&\frac{1}{\sqrt{6}}\left(\lvert2\rangle+2\lvert3\rangle+\lvert4\rangle\right),\\
        \lvert\tilde{4}\rangle=&\frac{1}{\sqrt{2}}\left(\lvert2\rangle-\lvert4\rangle\right),
    \end{split}
\end{equation}
The projected Hamiltonian in the new basis can be written as:
\begin{equation}
    \begin{bmatrix}
        -4J_A+\frac{4J^\prime_A}{3} & \frac{4J^\prime_A}{3} & -\frac{4J^\prime_A}{3} & 0\\
        \frac{4J^\prime_A}{3} & 2J_A-\frac{2J^\prime_A}{3} & \frac{4J^\prime_A}{3} & 0\\
        -\frac{2J^\prime_A}{3} & \frac{4J^\prime_A}{3} & 2J_A+\frac{4J^\prime_A}{3} & 0 \\
        0 & 0& 0 & 2J^\prime_A
    \end{bmatrix}.
\end{equation}
By diagonalizing the $4\times 4$ matrix, we can obtain the lowest energy state as:
\begin{equation}
    \begin{split}
    \lvert\mathrm{singlet}_4\rangle=&\frac{1}{\mathcal{N}}\Bigg(\left(-9+4\frac{J^\prime_A}{J_A}-3\sqrt{9-\frac{8J_A^\prime}{J_A}+\frac{4{J^\prime_A}^2}{J_A^2}}\right)|\tilde{1}\rangle\\
    &+\frac{4J^\prime_A}{J_A}\lvert\tilde2\rangle-\frac{2J^\prime_A}{J_A}\lvert\tilde3\rangle\Bigg),
    \end{split}
\end{equation}
where $\mathcal{N}$ is the normalizatioin factor. The corresponding energy is $E=-J_A-\sqrt{9J_A^2-8J_AJ^\prime_A+4{J^\prime_A}^2}$. Its comparison with the ED calculation is shown in Fig.~\ref{fig:ED_spin_singlet}(c). The wavefunction for $\psi^\prime$ is:

\begin{equation}
    \lvert\Psi_{\psi^\prime} \rangle=\otimes_{i}\lvert\mathrm{singlet}_4\rangle_i
\end{equation}

For the charge sector, we use the Slater determinant given by the mean-field Hamiltonian Eq.~\ref{eqn:MFBM_app} (BM model) or Eq.~\ref{eq:THFM_ancilla_app} (THFM). We choose $\Delta_\psi$ and $\mu$ to make $\langle n_{i;c}\rangle=4$, $\langle n_{i;\psi}\rangle=4$.

The ancilla wavefunction can be written as:
\begin{equation}\label{eqn:final_ancilla_wavefunction}
    \begin{split}
        \lvert\psi_c\rangle=&P_S \left(\lvert \mathrm{Slater}[c,\psi]\rangle\otimes\lvert\Psi_{\psi^\prime}\rangle\right),\\
    \end{split}
\end{equation}
where $P_S$ is defined in Appendix.~\ref{app:ancilla}.
\subsection{$\nu=-2$}
At filling $\nu=-2$, there are 2 $\psi^\prime$ fermions per site. The ED calculation for a single site shows that the lowest energy state is non-degenerate, as illustrated in Fig.~\ref{fig:ED_spin_singlet} (b). We denote this state as $\lvert\mathrm{singlet}_2\rangle$. This state can also be solved analytically within the sector $\sigma_z=\tau_z=s^2=0$. The basis for this sector is:
\begin{equation}
    \begin{split}
        \lvert1\rangle=&\frac{1}{\sqrt{2}}\left({\psi^\prime}^\dagger_{+K\uparrow}{\psi^\prime}^\dagger_{-K^\prime\downarrow}-{\psi^\prime}^\dagger_{+K\downarrow}{\psi^\prime}^\dagger_{-K^\prime\uparrow}\right)\lvert0\rangle,\\
        \lvert2\rangle=&\frac{1}{\sqrt{2}}\left({\psi^\prime}^\dagger_{-K\uparrow}{\psi^\prime}^\dagger_{+K^\prime\downarrow}-{\psi^\prime}^\dagger_{-K\downarrow}{\psi^\prime}^\dagger_{+K^\prime\uparrow}\right)\lvert0\rangle.
    \end{split}
\end{equation}
The Hamiltonian Eq.~\ref{eqn:antiHundH} projected into this $2-$particle basis can be written as:
\begin{equation}
    \begin{bmatrix}
        -J^\prime_A & -J_A \\
        -J_A & -J^\prime_A
    \end{bmatrix}.
\end{equation}
Then we can obtain the lowest energy state as:
\begin{equation}
    \lvert\mathrm{singlet}_2\rangle=\frac{1}{\sqrt{2}}\left(\lvert1\rangle+\lvert2\rangle\right).
\end{equation}
The corresponding energy is $E=-J_A-J^\prime_A$. Its comparison with the ED calculation is shown in Fig.~\ref{fig:ED_spin_singlet}(d). The wavefunction for $\psi^\prime$ is:
\begin{equation}\label{eqn:singlet2}
    \lvert\Psi_{\psi^\prime} \rangle=\otimes_{i}\lvert\mathrm{singlet}_2\rangle_i,   
\end{equation}

We determine the wavefunction of the charge sector and the final ancilla wavefunction similarly to Eq.~\ref{eqn:final_ancilla_wavefunction}, but with parameters $\Delta_\psi$ and $\mu$ chosen to satisfy $\langle n_{i;c}\rangle=2$ and $\langle n_{i;\psi}\rangle=6$.

\subsection{$\nu=-2-x$: symmetric pseudogap metal}
At filling $\nu=-2-x$, each AA site still hosts 2 $\psi^\prime$ fermions. Therefore, we retain the wavefunction of $\psi^\prime$ given by Eq.~\ref{eqn:singlet2} and use Eq.~\ref{eqn:final_ancilla_wavefunction} to construct the final ancilla wavefunction. The parameters $\Delta_\psi$ and $\mu$ are selected to make $\langle n_{i;c}\rangle=2-x$, $\langle n_{i;\psi}\rangle=6$. For $\lvert \mathrm{Slater}[c,\psi]\rangle$, we can easily project the electron $c$ in the subspace with only the active bands.  In the end, we obtain a wavefunction within the active bands subspace. This state has an emergent heavy fermion structure: itinerant sector from $c,\psi$ and localized sector from $\psi'$. But the heavy fermion physics is purely emergent at low energy and does not rely on the existence of remote band and a microscopic heavy fermion model.  The spirit is similar to a previous theory of a similar pseudogap metal phase for underdoped cuprate, formulated purely in a one-band model\cite{Zhang2020}.

\end{document}